\documentclass{elsart}
\usepackage{epsfig}
 
\newcommand{\req}[1]{(\ref{#1})}              
\newcommand{\dd}{{\rm d}}                     
\newcommand{\be}{\begin{equation}}
\newcommand{\ee}{\end{equation}}

\newcommand{\rf}{r_{\rm ff}}                 
\newcommand{\rd}{r_{\rm d}}                  
\newcommand{\zd}{d_{\rm d}}                      
\newcommand{\df}{d_{\rm ff}}         
\newcommand{\ds}{d_{\rm sub}}         
\newcommand{\Qt}{Q_{\rm tot}}               
\newcommand{\ri}{r_{\rm i}}             
\newcommand{\mic}{$\mu$m}                                

\newcommand{\Ka}{K$\alpha$ }
\newcommand{\Kb}{K$\beta$ }
\newcommand{\La}{L$\alpha$ }

\begin{document}
\begin{frontmatter}
\title{Simulating CCDs for the {\em Chandra} Advanced CCD Imaging Spectrometer}

\author{L.K. Townsley},
\ead{townsley@astro.psu.edu}
\author{P.S. Broos},
\author{G. Chartas},
\author{E. Moskalenko\corauthref{posth}},
\corauth[posth]{posthumous}
\author{J.A. Nousek}, \and
\author{G.G. Pavlov}

\address{Department of Astronomy \& Astrophysics, The Pennsylvania State 
University, 525 Davey Lab, University Park, PA 16802\\
http://www.astro.psu.edu}


\begin{abstract}

We have implemented a Monte Carlo algorithm to model and predict the
response of various kinds of CCDs to X-ray photons and
minimally-ionizing particles and have applied this model to the CCDs in
the Chandra X-ray Observatory's Advanced CCD Imaging Spectrometer.
This algorithm draws on empirical results and predicts the response of
all basic types of X-ray CCD devices.  It relies on new solutions of
the diffusion equation, including recombination, to predict the radial
charge cloud distribution in field-free regions of CCDs.  By adjusting
the size of the charge clouds, we can reproduce the event grade
distribution seen in calibration data.  Using a model of the channel
stops developed here and an insightful treatment of the insulating
layer under the gate structure developed at MIT, we are able to
reproduce all notable features in ACIS calibration spectra.

The simulator is used to reproduce ground and flight calibration data
from ACIS, thus confirming its fidelity.  It can then be used for a
variety of calibration tasks, such as generating spectral response
matrices for spectral fitting of astrophysical sources, quantum
efficiency estimation, and modeling of photon pile-up.

\end{abstract}


\end{frontmatter}

\section{Introduction} \label{sec:intro}

\subsection{Motivation}

The July 1999 launch of the Chandra X-ray Observatory ({\em Chandra}),
NASA's latest Great Observatory, heralded a new era for observational
X-ray astronomy.  With half-arcsecond image quality and broad (0.1--10
keV) spectral sensitivity, the {\em Chandra} mirrors provide X-ray
imaging on a par with the best ground-based visual and near-infrared
telescopes and finally allow straightforward comparisons between images
in X-rays and those in longer wavebands.  {\em Chandra}'s most frequently
used focal-plane instrument is the Advanced CCD Imaging Spectrometer (ACIS),
a CCD camera that operates in photon-counting mode to record both
spatial and spectral information from celestial X-rays.  Each photon
that is stopped by photoabsorption produces a cloud of secondary
electrons; this charge cloud is pixelized and recorded as an ``event''
with well-defined characteristics, including position, amplitude, and a
measure of its spatial concentration (``grade'').
		
Our Monte Carlo simulations are used to augment ACIS calibration data;
once the simulator is tuned to reproduce monochromatic calibration
data, it can be used to predict the instrument's spectral response at
energies not covered by the calibration sampling.  These predictions
are used to generate the ACIS response matrix, the mathematical
representation of the ACIS spectral redistribution function.  The tool
can also be used to simulate photon pile-up\footnote{Photon ``pile-up''
is the condition where two or more photons strike the CCD during a
single integration, in the same or neighboring pixel.  The overlapping
charge clouds, if they are even recognized as an event, will have an
energy and spatial pixel distribution that differs dramatically from
events induced by single photons.  This aliasing corrupts both the
spectrum and image of bright astrophysical sources observed with ACIS
and should be avoided whenever possible by careful observation
planning.}, which can occur when bright sources are observed.  It also
incorporates a model for charge transfer inefficiency (CTI), a
phenomenon that spectrally and spatially degrades events from ACIS
detectors.

Our goal is to use the simulator to predict the spatial and spectral
response of ACIS in order to choose appropriate targets for
observation, to configure the instrument to yield the best possible
data, and to interpret the data after observation.  As a check on the
fidelity of the response matrices and data modeling, the observer can
run a model spectrum and spatial distribution of photons through the
simulator and assess the model's ability to reproduce the data.  For
sources suffering from pile-up or CTI, iterative modeling with the
simulator may be the only way to recover realistic source spectra.
This paper is intended for novice ACIS users interested in applying
device physics to obtain astrophysically relevant results, so it
contains substantial introductory material on ACIS and charge cloud
propagation in CCDs.


\subsection{Background}

A brief overview of {\em Chandra} and its calibration is given by
O'Dell and Weisskopf \cite{odell98} and references therein.  The
on-orbit performance of the telescope is summarized by Weisskopf {\em
et al.} \cite{weisskopf00}.  A detailed description of ACIS will be
given in an upcoming paper \cite{garmire01}.  Extensive online
documentation for {\em Chandra} and its instruments is available from
the Chandra X-ray Center (CXC), at chandra.harvard.edu \cite{pog}.  
The use of CCDs in X-ray astronomy is described by Lumb
{\em et al.} \cite{lumb91}.

ACIS employs eight bulk front-illuminated (FI) charge-coupled devices
and two back-illuminated (BI) devices, all manufactured at Lincoln
Laboratories.  Details of these devices are given in Burke {\em et al.}
\cite{burke97}, Prigozhin {\em et al.} \cite{prigozhin98a} (hereafter
PRBR98) and Pivovaroff {\em et al.} \cite{pivovaroff96}.  The extensive
subassembly calibration of individual chips, carried out at MIT's
Center for Space Research (MIT/CSR) and at various synchrotron
facilities, is outlined in Bautz {\em et al.} \cite{bautz98} and
references therein.  Extensive documentation of ACIS calibration,
hardware, and operating modes is available via the ACIS Operations
Handbook \cite{sop01} and the ACIS Calibration Report
\cite{calreport}.

Recent results of MIT/CSR modeling of the ACIS CCDs, focusing primarily
on the FI devices, are presented by Bautz {\em et al.} \cite{bautz99} and
Prigozhin {\em et al.} \cite{prigozhin00}.  The X-ray Imaging Spectrometers
on the Japanese satellite Astro-E used the same type of device; calibration
and modeling results for these devices is described by Hayashida
{\em et al.} \cite{hayashida00}.

At the beginning of science operations with {\em Chandra}, the ACIS FI
detectors were damaged in the extreme environment of the Earth's
radiation belts, due to the unanticipated forward scattering of charged
particles by the Chandra mirrors.  This resulted in substantial CTI in
the parallel registers of FI devices, causing the gain, quantum
efficiency, energy resolution, and grades to exhibit row-dependent
effects.  The ACIS team has performed extensive laboratory tests to
determine the nature and exact cause of the damage.  This is summarized
in recent papers by Prigozhin {\em et al.} \cite{prigozhin00b} and Hill
{\em et al.} \cite{hill00}.

As a result of damage intrinsic to the manufacturing process, the ACIS
BI devices have always shown modest CTI effects, but the protection
afforded by the depletion region in front of the gates prevented the
on-orbit radiation damage seen in the FI devices.  Since BI CTI was
known long before launch, we developed a Monte Carlo model that
simulates CTI in conjunction with our CCD simulator.  These ongoing
efforts to mitigate the BI CTI by modeling and post-processing the data
also allowed us to address the new problem of FI data affected by CTI.
A preliminary description of the method, emphasizing the FI devices,
was given in Townsley {\em et al.} \cite{townsley00}.  A more complete
description of our techniques, adding details for both FI and BI
devices, is presented as a companion paper \cite{townsley01b}.
Hardware and flight software mitigation efforts are underway at MIT/CSR
and show promise for further removing the effects of radiation damage
on ACIS data; see Prigozhin {\em et al.} \cite{prigozhin00b} for
details.
  
The basic algorithms used in our simulator originated with the X-ray
group at Leicester University, which continues independent model
development ({\em e.g.} \cite{hill99,mccarthy95,holland95}).  Lumb and
Nousek \cite{lumb92} developed the first Penn State simulator.  The
current model relies on new solutions of the diffusion equation to
predict the radial charge cloud distribution in field-free regions of
CCDs (Pavlov and Nousek \cite{PN99}, hereafter PN99).

The current version of the code attempts to reproduce the detected CCD
event spectrum, the quantum efficiency, and the event grade
distribution resulting from a monochromatic incident
X-ray flux.  Our model for channel stops and our interpretation of the
MIT group's model for the insulating layer under the gate structure
\cite{prigozhin00} are described here; these enhancements were
necessary to explain subtle redistribution features in the spectra.  We
detail our model of charge transfer inefficiency (CTI) in an
accompanying paper \cite{townsley01b}, but note the effects of CTI
here.

\section{The Generic CCD Simulator} \label{sec:generic}


\subsection{CCD Geometry}

Our Monte Carlo algorithm simulates the response of three basic types
of CCD devices:  epitaxial front-illuminated, bulk front-illuminated,
and back-illuminated.  Each type of device is assumed to consist of a
stack of slabs with different properties.  The fundamental structure of
the algorithm follows that described by McCarthy {\em et al.}
\cite{mccarthy95}.

\begin{figure}[htb]
\centerline{\epsfig{file=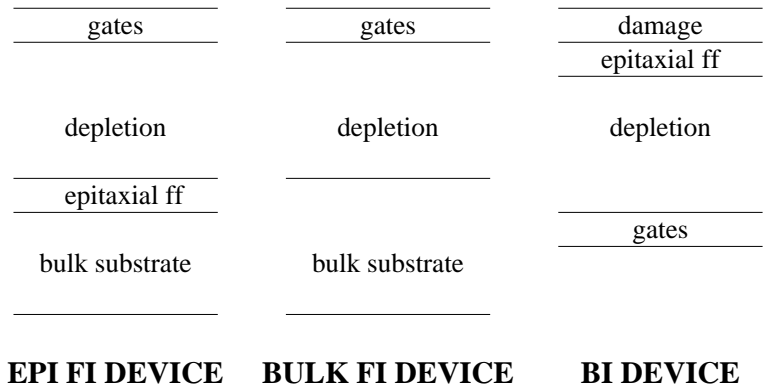,height=2.0in}}
\caption{\protect \small The generic structure of epitaxial FI, bulk FI,   
and BI devices.  Photons come from the top in this diagram.  Slabs are
not to scale.  The label ``ff'' stands for ``field-free.'' ACIS contains BI and bulk FI devices.}
\normalsize
\label{fig:ccdgeoms}
\end{figure}

The three types of CCDs are modeled by arranging the component slabs in
appropriate order, from the viewpoint of the impinging photon (see
Figure~\ref{fig:ccdgeoms}).  The user specifies the device by
supplying the thickness of each layer and various other important
parameters (such as the dimensions of rectangular pixels, the operating
temperature, and the acceptor concentration).  Both ACIS BI and FI
(bulk type) chips have been modeled using this technique.

In an epitaxial FI device, the top section is the gate structure, which
can produce secondary fluorescent photons that can interact in an
active part of the device and contribute substantially to the Si~\Ka
peak in the spectrum \cite{prigozhin98a}.  Beneath this is the depletion
(``field'') region, interrupted near the surface by the channel stops
that define the pixel boundaries, and the buried channel, where charge
is held until it is clocked out through the gates.  Adequate simulation
of the FI spectral redistribution function required inclusion of a
model for charge interaction in these channel stops (detailed below).
The depletion region is the main active section of the device; photons
interacting in this slab are efficiently converted to electron charge
clouds that are swept towards the gates by a nearly-uniform electric
field.  The depletion region is followed by an epitaxial field-free
region, then a bulk silicon substrate.  A bulk FI device (such as the
ACIS FI chips) has the same structural elements down through the
depletion layer as in an epitaxial device, simply followed by a bulk
field-free substrate several hundred microns thick.

In a BI device, the top layer is a ``damage'' or surface layer (often
$SiO_2$), left after the bulk substrate on which the device was formed
has been etched away.  Below this is a thin epitaxial field-free layer,
which acts as a reflecting layer and prevents charge from the depletion
region from leaking out into the damage layer and generating dark
current.  Since this epitaxial layer has no electric field across it,
charge clouds suffer more spreading as they traverse it than in field
regions.  The depletion layer lies under this epitaxial layer and makes
up most of the thickness of the device; below it is the buried channel,
channel stops, and gate structure at the bottom of the device.  It is
important to note that ACIS BI devices are extremely thin compared to
their FI counterparts.  This makes it difficult to predict the spectral
redistribution function of these devices as a function of photon energy,
since both top and bottom surface layers can come into play.


\subsection{Photon Interactions}

The X-ray source model (usually generated by SAOsac \cite{jerius95} or
MARX\cite{wise97} for celestial sources or by our own code for uniform
illumination) specifies the (x,y) position where each X-ray photon
enters the upper-most slab of the CCD and specifies the trajectory
(direction cosines) of the photon.  The simulation code follows that
trajectory, iteratively modeling the photon's propagation through each
slab it encounters.  The propagation length of photons in a single slab
is characterized by the energy-dependent linear absorption coefficient
of the material, $\alpha$, used to parameterize the bulk transmittance,
$T$, of a slab of uniform material in the usual way: $T(E) =
\e^{-\alpha(E) d}$, where $E$ is the photon energy and $d$ is the
thickness of the slab.

When the photon enters a slab, the code computes the distance that the
photon would travel in that slab's material before being absorbed: 
\be
z = -\frac{1}{\alpha}\ln(\mathcal{R}_{U})\,
\label{depth}
\ee
where $\mathcal{R}_{U}$ is a
uniform random number.  If $z > d$, then the photon is considered to
have traveled through the slab and the calculation is repeated for the
adjacent slab (which is composed of a different material).

The simulation accounts for the possibility of secondary fluorescent
photon generation (both \Ka and \Kb) from silicon.  A random
propagation direction is generated and a new trajectory for the
fluorescent photon is calculated, starting from the interaction
position of the original photon.  Fluorescent photons from other
elements in the device (such as oxygen) may occur, but with a very low
probability, so they are not included in the current simulator.

As the charge cloud propagates through the field-free regions of the
device, some of the charge may recombine and be lost.  Charge spreading
and recombination have been modeled by solving the diffusion equation
for each slab of the device, using the new solution of PN99 in
field-free layers.


\subsection{Charge Transport Theory}

The photon interacts with the device via the photoelectric effect and
produces a cloud of charge.  This charge spreads through the layers of
the device, with the spreading rate and the charge reflection and
absorption dependent on the properties of each layer.  We are careful
to account for the complicated geometry arising when a photon interacts
close enough to the boundary between any two layers that the initial
charge cloud spreads across that boundary.  Once the charge reaches the
buried channel, it is ``detected'' by recording an appropriate number
of electrons in each pixel over which the charge spread.  The degree to
which a given photon's charge cloud is split across pixels depends on
the photon's energy, its interaction depth, and the proximity of the
interaction to pixel boundaries.

We must predict the pixelized charge distribution resulting from each
X-ray interaction.  Our model follows the simple formulations of charge
generation and transport developed by other groups interested in using
CCDs for X-ray astronomy ({\em e.g.}
\cite{hopkinson83,janesick85,janesick87,hopkinson87}).  PN99 summarize
the work of these authors -- see that paper and references therein for
a detailed introduction.  The following subsections describe how we
apply the previous work in our simulator.  A concise description of the
X-ray detection process in a CCD is given in Chapter~4 of the ACIS
Calibration Report \cite{calreport}.

The empirical equations developed to describe charge cloud generation
and evolution in silicon CCDs do not always perfectly reproduce our
ACIS data.  To account for this incomplete modeling, we include
energy-dependent tuning parameters in these equations.  These tuning
parameters are adjusted iteratively and in parallel to improve the
model's ability to reproduce calibration data.  They are described in
detail in Section~\ref{sec:tuningparams};
Figures~\ref{fig:maintuning}--\ref{fig:cstuning} give the values for
the tuning parameters at all calibration energies for both FI and BI
devices.


\subsubsection{The Initial Charge Cloud}

Each photon is allowed to interact with the CCD at a random depth
as described in Equation~\req{depth}.  
To estimate the amount of charge contained in the resultant electron cloud,
Fano noise is added to the incident photon's energy $E_{i}$ (employing a
tuning parameter $\mathcal L$ to get the linewidth to match the data for
monochromatic photons); then this energy $E_{f}$ is converted into an initial charge $Q_{0}$: 
\be
E_{f} = E_{i} + \mathcal{R}_{N}(0)\sqrt{FwE_{i}{\mathcal L}}\,
\label{fano}
\ee

\be 
Q_{0} = E_{f} / w
\label{initchg}
\ee
where $w=3.71$eV e$^{-1}$ is the energy necessary to liberate an
electron-hole pair from the silicon at temperature 153K and $F = 0.115$
is the Fano factor.  The notation $\mathcal{R}_{N}(0)$ means that, for
each photon, we sample a normal random distribution of mean 0 and
standard deviation 1.  Readout noise will be applied later, after the
charge is mapped onto the pixels.

The initial charge cloud is assumed to have a Gaussian radial profile
with an energy-dependent radius (PN99, p.~350) of:
\be
r_{i} \simeq 0.0062E_{f}^{1.75}
\hspace{0.2in}
~\mu{\rm m}~,
\label{rinitial}
\ee 
where $E_{f}$ is in~keV.  This characteristic size of the initial charge
cloud differs from the PN99 value by a factor of two for historical
reasons; in practice, a tuning parameter (detailed below) is used to tailor the
charge cloud radius to the data.

For interactions in the oxide layers, we assume that the initial charge cloud
radius is $r_{i}$ as given above, modified by a tuning parameter
${\mathcal I}$:  $r_{i,oxide} = r_{i}{\mathcal I}$.  
Equation~\req{rinitial} is empirical and suited only for photon
interactions in $Si$, not $SiO_2$, so it is not surprising that some
adjustment to this value is necessary.  Prigozhin {\em et al.}
\cite{prigozhin00} review this problem in detail.

If the incident photon has enough energy ($\ge$ 1839~eV) it has a
chance (4.3\%) of producing a Si \Ka fluorescent photon, which we assume
for simplicity is emitted from the original interaction point in a
random direction.  It propagates a distance $z$ given by Equation~\req{depth} and a new charge cloud is produced and treated
independently of any remaining charge from the original photon.  The
original charge cloud retains its size as determined by the energy of
the original photon and the remaining charge left in it is propagated
from the original interaction point.


\subsubsection{Interaction in the Depletion Region}

A photon interacting at a depth $z_0$ in the depletion region
(thickness $\zd$) will have a charge cloud radius after drifting
through the depletion region given by Hopkinson's Equation~7
\cite{hopkinson87}, presuming the linear regime where the drift
velocity is proportional to the electric field:
\be
\rd^{\rm lin} = 1\times 10^{4} \left( \frac{4D\epsilon}{\mu eN_{a}}\, \ln\frac{\zd}{\zd-z_0}\right)^{1/2}
\hspace{0.2in}
~\mu{\rm m}~,
\ee
where $\epsilon$ is the electric permittivity of silicon ($1.044\times
10^{-12}$~F~cm$^{-1}$), $\mu$ is the electron mobility, $e$ is the
electronic charge in Coulombs, and $N_a$ is the concentration of
acceptors in number~cm$^{-3}$.  Following the prescription of Hopkinson
(\cite{hopkinson87}), we restrict photons from interacting too close to
the depletion boundary ($z_0 \neq \zd$) so that the logarithm in
$\rd^{\rm lin}$ does not diverge.

This radius is increased to take into account that 
we may be working in the saturated regime, where the electric field
is strong enough that the drift velocity of the charge carriers has
approached some terminal value.  This is likely to be true for only
certain parts of each pixel; saturation depends on the detailed
structure of the electric field, both in depth and laterally across
each pixel.  In practice, this uncertainty is taken up by a tuning
parameter in the code.  From PN99's Equation~17, we add another
radius term in quadrature with the term from the linear regime,
yielding
\be
\rd^{\rm sat} = \left( \left(\rd^{\rm lin}\right)^2 + 1\times 10^{8} \cdot \frac{4D\epsilon}{\mu eN_{a}}\frac{z_0}{\zd}\mathcal{S} \right)^{1/2}
\hspace{0.2in}
~\mu{\rm m}~,
\label{rdsat}
\ee
where $\mathcal{S}$ is the tuning parameter.  
Then, finally,
\be
\rd = \sqrt{r_{i}^2 + \left(\rd^{\rm sat}\right)^2}
\label{totalrd}
\ee
is the total charge cloud radius in the depletion region.								
In the above formulation we include an explicit temperature dependence
for the electron mobility $\mu$ as given by Equation~5 of Jacoboni {\em
et~al.} \cite{jacoboni77}:
\be
\mu = A_{J}T^{-\gamma}
~{\rm cm^2V^{-1}s^{-1}}~,
\label{muoft}
\ee
where $A_{J} = 1.43\times10^{9}$, $\gamma = 2.42$, and T is the
temperature in Kelvins.  Jacoboni {\em et al.} say this equation holds
``around room temperature'' and for high-purity silicon.  It appears
from their Figure~3 that this equation adequately fits the data down to
roughly 120K.  This constrains us to model only those devices with
acceptor concentrations less than $\sim 10^{16}$cm$^{-3}$ (PN99).  The
ACIS devices satisfy this criterion except in the bulk substrate of the
FI device.

The other important quantity for simulating photons interacting in the
depletion region is the pixelization of the charge collected at the
gates.  Defining pixel $(0,0)$ as the one in which the photon
interacted, the charge in pixel $(i,j)$ is given by PN99's
Equation~61:
\be
Q_{ij}(x_0,y_0)=\frac{Q_0}{4}\left[{\rm erf}\left(\frac{a_{i+1}-x_0}{\sqrt{\rd^2
+\ri^2}}\right)
-{\rm erf}\left(\frac{a_{i}-x_0}{\sqrt{\rd^2+\ri^2}}\right)\right]
\left[{\rm erf}\left(\frac{b_{j+1}-y_0}{\sqrt{\rd^2+\ri^2}}\right)
-{\rm erf}\left(\frac{b_{j}-y_0}{\sqrt{\rd^2+\ri^2}}\right)\right]
\label{Qij-depl}
\ee
where erf$(x)$ is the error function, $a$ and $b$ are the pixel sizes
in two dimensions, $a_i=-a/2+ia$, $b_j=-b/2+jb$, and $\rd$ is given by
Equation~\req{totalrd}.  The initial charge $Q_0$ is given by
Equation~\req{initchg}.  The quantities $x_0$, $y_0$ are the photon
interaction coordinates in the reference frame with its origin at the
center of the pixel under which the photon is absorbed ($-a/2<x_0<a/2$,
$-b/2<y_0<b/2$).


\subsubsection{Interaction in a Field-Free Region}

For a photon interacting in the field-free zone (either bulk or
epitaxial) just below the depletion region, charge diffusion and
recombination become important and it is likely that only a fraction of
the initial charge will be collected at the gates.  PN99 give an
expression (their Equation~29) for the total charge collected in a
field-free region:
\be
\Qt=
Q_0\, \frac{\gamma\, \cosh\gamma (1-\zeta) + \beta\, \sinh\gamma (1-\zeta)}
{\gamma\, \cosh\gamma + \beta\, \sinh\gamma}
\label{qt}
\ee
where $\gamma=\df/L$, $\zeta=(z_{0}-\zd)/\df$,
$\beta=(\df/L)(\mathcal{T}/\mathcal{R})$, $\df$ is the thickness of the
field-free zone, $L$ is the diffusion length, and $\mathcal{T}$ and
$\mathcal{R}$ are the transmission and reflection coefficients at the
boundary.  This is equivalent to Hopkinson's Equation~8
\cite{hopkinson87}.  We use this quantity in a calculation of the
recombination noise that must be incorporated into the final collected
charge fraction.  It is simplified by the fact that $\mathcal{T}=0$,
$\mathcal{R}=1$ for an epitaxial region and $\mathcal{T}=1$,
$\mathcal{R}=0$ for a bulk region.

The shape of the charge cloud in the field-free region is not Gaussian,
as explored in detail by PN99.  Their Equation~66 gives the charge
collected in pixel $(i,j)$ after creation in a field-free region and
diffusion through that region and the depletion region:
\be
Q_{ij}(x_0,y_0)=\frac{Q_0}{4\pi^{1/2}} \int_0^\infty \frac{\dd\tau}{\tau^{3/2}}
\,\, G_i(\alpha,\xi_0,\tau)\,\, G_j(\beta,\eta_0,\tau)\,\, S(\zeta,\tau)\,\,
\exp\left(-\frac{\tau}{4\Lambda^2}\right)~,
\label{qij}
\ee
where 
\be
G_i(\alpha,\xi_0,\tau)=
{\rm erf}\left(\frac{\alpha_{i+1}-\xi_0}{\sqrt{\tau+\tau_{\rm d}}}\right) 
-{\rm erf}\left(\frac{\alpha_{i}-\xi_0}{\sqrt{\tau+\tau_{\rm d}}}\right)~,  
\label{gi}
\ee
\be
S(\zeta,\tau)=\sum_{m=-\infty}^\infty C_m\, (\zeta-2m)\,
\exp\left[-\frac{(\zeta-2m)^2}{\tau}\right],
\label{sum}
\ee
$\alpha=a/\df$, $\beta=b/\df$, $\xi_0=x_0/\df$, $\eta_0=y_0/\df$,
$\tau_{\rm d}=(\rd^2+\ri^2)/\df^2$, $\zeta= (z_0-\zd)/\df$,
$\alpha_i=(-a/2+ia)/\df$, $\beta_j=(-b/2+jb)/\df$, $\Lambda=L/\df$, and
$\tau=4Dt/\df^2$.  As before, $a$ is the pixel size in $x$, $b$ is the
pixel size in $y$, and $(x_0,y_0,z_0)$ is the interaction site; $C_m =
(-1)^m$ for a completely reflective (epitaxial device) and $C_m=1$ for
a completely transparent (bulk device) lower boundary of the field-free
zone.  We follow PN99's recommendation to integrate over the
logarithmic variable $u=\ln\tau$ ($\dd\tau=\e^u\dd u$).  This changes
the formal limits of integration to $\pm\infty$.  In practice, these
limits are determined empirically; the range [$-$19,5] is adequate for
all practical cases.


\subsubsection{The Special Case of Substrate Interaction in an Epitaxial Device}
	
For FI epitaxial devices in the instance where a photon interacts in
the bulk silicon below the epitaxial layer, the situation becomes more
complicated.  Typically this layer has high acceptor concentrations
($N_a \sim 10^{18}$cm$^{-3}$), so the diffusion length is short ($L
\sim 10$\mic) \cite{hopkinson87} and our expression \req{muoft} does
not hold.  We currently just combine the charge cloud radius from the
bulk region with those from the other regions by adding them in
quadrature, which presumes they are Gaussian when they aren't quite.
We ignore this inconsistency because it affects few
events for reasonable epitaxial device geometries.

The radial charge cloud profile in the substrate is given by
Hopkinson's Equation 19 \cite{hopkinson87}:
\be
q_{\rm sub}(r) = \frac{Q_0\ds}{2\pi v^3}\left(1+\frac{v}{L}\right)
\exp\left(-\frac{v}{L}\right)
\label{hop19}
\ee
where $v=\sqrt{r^2 + \ds^2}$ and $\ds$ is the depth of the interaction
in the substrate.  We can calculate the maximum charge fraction
available from a photon interaction in the substrate by taking this
equation in the limit $r \rightarrow 0$:
\be
q_{\rm sub}(r=0) = \frac{Q_0}{2\pi \ds^2}\left(1+\frac{\ds}{L}\right)
\exp\left(-\frac{\ds}{L}\right).
\ee
We then calculate the relevant 1-$\sigma$ charge cloud radius for the
substrate, $r_{\rm sub}$, by stepping out in radius from $r=0$ until the charge
fraction (Equation \req{hop19}) is 67\%.  

We estimate the charge cloud
radius in the epitaxial region, $\rf$, by PN99's Equation~53 (with
$z_0-\zd=\df$):
\be
\rf^2=2\df L \tanh\frac{\df}{L}
\hspace{0.2in}
~\mu{\rm m}^2~.
\ee
The total charge cloud radius at the gates is then found by combining
these in quadrature with the depletion radius and the initial radius:
\be
r_{\rm total} = \sqrt{\ri^2 + \rd^2 + \rf^2 + r_{\rm sub}^2}.		
\ee
Then equation \ref{Qij-depl} is used to pixelize the charge cloud,
substituting $r_{\rm total}$ for $\sqrt{\rd^2 + \ri^2}$.

\section{Specifics of the ACIS CCDs} \label{sec:acisspecs}

There are several characteristics of ACIS data that are due to the
specific geometry of the ACIS devices.  This is particularly true for
the BI chips, which suffer substantial serial as well as parallel
charge transfer inefficiency (CTI), that by its nature is
position-dependent.  Thus we review here the exact definition of ACIS
events and the readout process for the CCDs. 


\subsection{ACIS Events}

Photons interact with the ACIS CCDs to produce ``events,'' islands of
pixelized charge with characteristic patterns which differ from the
patterns produced by charged particle background interactions.
Due to telemetry limitations, the information sent from ACIS to the
ground mainly consists of a list of these events rather than the
full-frame CCD exposures familiar from visual astronomy.

\begin{figure}[htb]
\centerline{\mbox{
	\epsfig{file=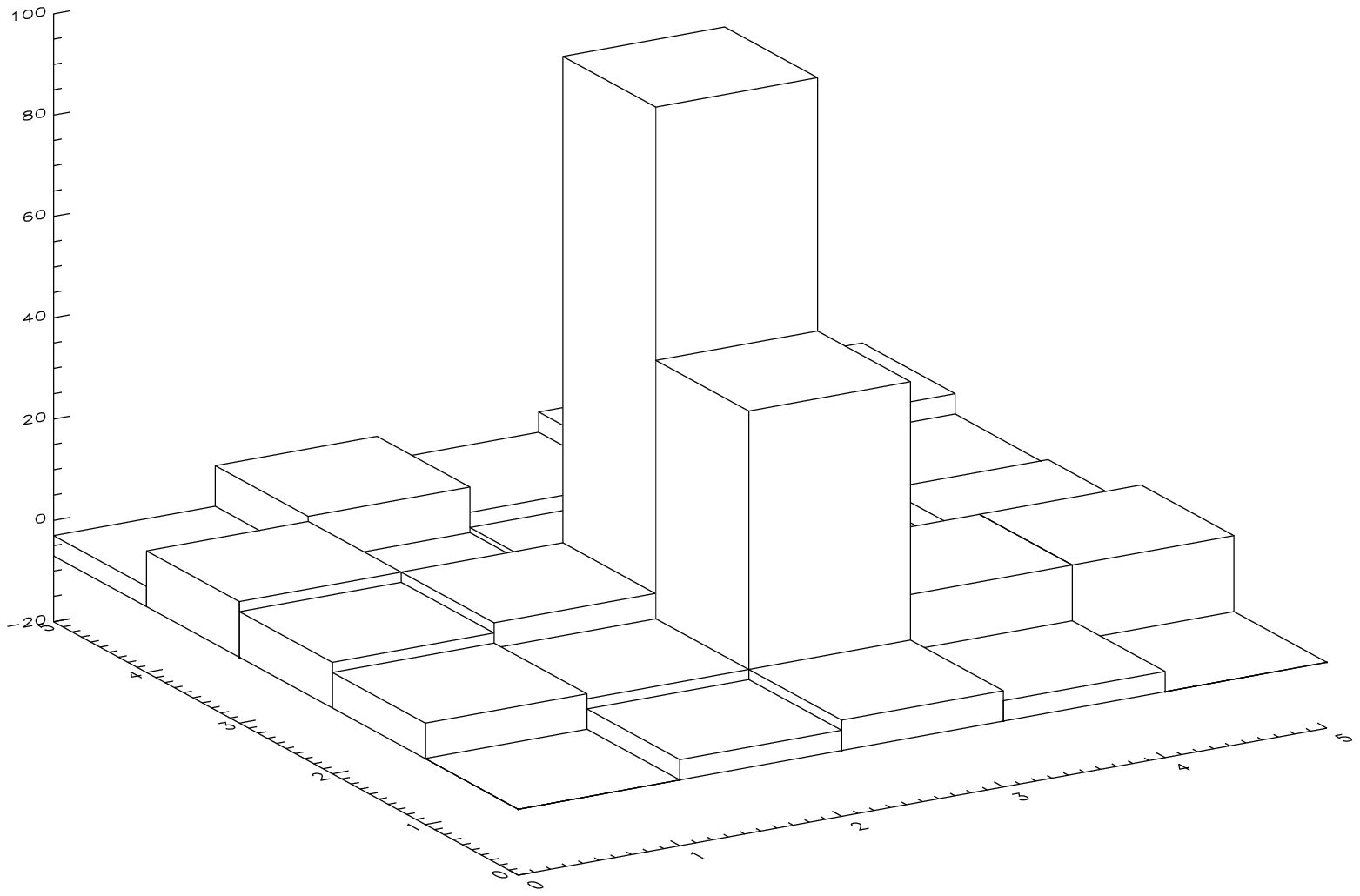,height=3.0in}
	\epsfig{file=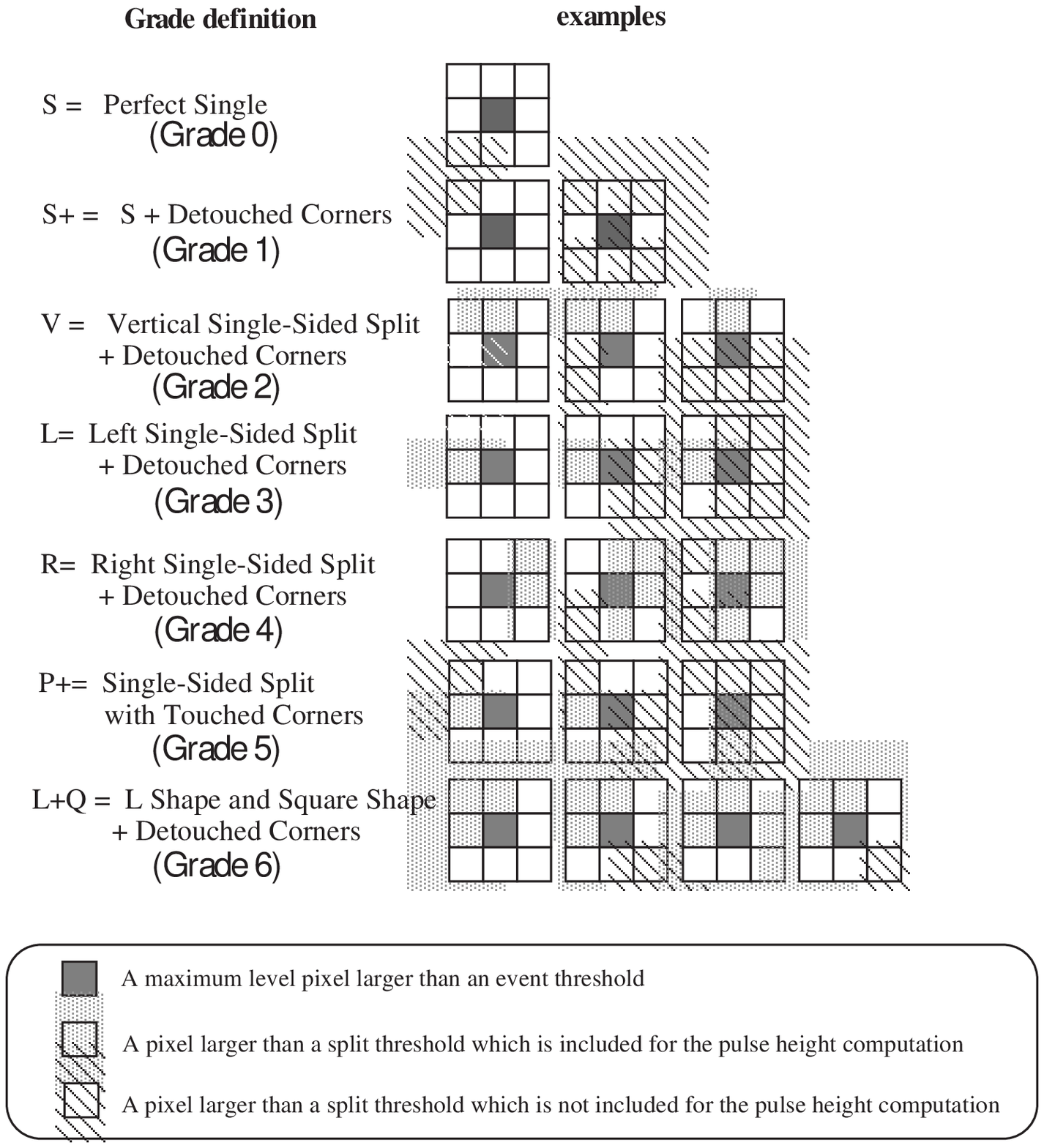,height=4.0in}}}
\caption{\protect \small  The left panel shows the neighborhood of 
pixels defining an ACIS event.  The vertical axis is the pixel amplitude
in digital numbers (DN).  The right panel illustrates the ASCA
grade definitions, reproduced with permission from Yamashita \cite[page
47]{yamashita}.  According to these definitions, the event in the
left panel would have ASCA grade 2.}

\normalsize
\label{fig:lego_event}
\end{figure}

The left panel of Figure~\ref{fig:lego_event} shows an ACIS event, made
up of a 3 $\times$ 3 neighborhood of pixels (here embedded in a 5
$\times$ 5 subarray representing a small area on the CCD) where the
central pixel is by definition the one containing the most charge.  In
order for the flight software to recognize a pixelized charge cloud as
an event, this central pixel must be larger than the ``event detection
threshold,'' set at 20 digital numbers (DN) for BI chips and 38 DN for
FI chips.  These values are based on the intrinsic noise levels in the
chips and their readout amplifiers.  A second, lower threshold is
applied to the 8 pixels surrounding the central pixel; pixels larger
than this ``split event threshold'' (13 DN for all ACIS devices) are
considered to contain real charge (as opposed to noise).

The event grade is determined by noting which of these neighboring
pixels are above the split event threshold.  The right panel of
Figure~\ref{fig:lego_event} shows the grading scheme used on the ASCA
X-ray satellite, a recently-completed Japanese mission.  Following the
ASCA grade definitions, the event in the left panel would have ASCA
Grade 2 (which is composed of both ``up'' and ``down'' singly-split
events -- our example shows a ``down'' split).

Although ACIS uses a different grading scheme internally, the ASCA
grades are familiar to many users and are often used by analogy in ACIS
data analysis.  Although the two missions calculate event energies
using a slightly different algorithm, there is a straightforward
mapping of ACIS event grades into ASCA grades.  Generally speaking, the
same event grades useful for ASCA spectral analysis are useful for ACIS
spectral analysis, so the standard ASCA grade filtering is employed on
ACIS event lists as well.

\subsection{ACIS Readout Geometry}

\begin{figure}[htb]
\centerline{\epsfig{file=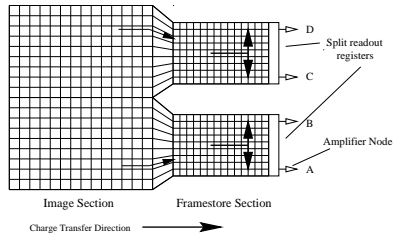,height=2.0in,angle=-90}}
\caption{\protect \small Geometry of an ACIS CCD.  Note the
``handedness'' in the serial readouts.  Image array pixels are
24$\times$24 \mic.  Framestore pixels are 13.5$\times$21 \mic.  This
figure is borrowed from Chapter 4 of the ACIS Calibration Report
\cite{calreport}.}

\normalsize
\label{fig:layout}
\end{figure}

Figure~\ref{fig:layout} shows the layout of an ACIS CCD (see Chapter 4
of the ACIS Calibration Report \cite{calreport} for details).  It
consists of an ``image array,'' the 1024 $\times$ 1026 pixel collecting
area for photon events, a ``framestore array,'' a holding area for
photon events, also 1024 $\times$ 1026 pixels (but with a different
pixel size), that is shielded from X-rays, and four readout
amplifiers.  After an exposure, charge in the image array is quickly
swept into the framestore array (via fast parallel transfers) then
slowly read out through the amplifiers (via slow parallel and serial
transfers).  Note that these four amplifiers divide the chip into four
distinct regions; serial readouts for amplifier nodes A and C shuttle
charge in the same direction, while serial readouts for amplifier nodes
B and D shuttle charge in the opposite direction.  When a chip exhibits
measurable serial CTI , this ``handedness'' in the readout becomes
noticeable.

\subsection{ACIS Absorption Coefficients and Gate Structure}
\label{sec:gates}

Prigozhin\cite{prigozhin98b} used a synchrotron to measure the
transmission of phosphorus-doped polycrystalline silicon and $SiO_2$
made via the same processes used in ACIS devices, and a $SiO_2 +
Si_3N_4 + SiO_2$ sandwich that makes up the insulating layer under the
gate structure.  He measured EXAFS (extended X-ray absorption fine
structure) around the absorption edges of nitrogen, oxygen, and
silicon, effects that can substantially alter the depths at which
photons interact in the CCDs and hence the character (recovered charge
and grade) of the events produced.  Our model uses these new absorption
coefficients; transmission values are taken from the Henke data
\cite{henke93} when no synchrotron data were available
\cite{prigozhin98b}.

These transmission data were used to generate the $\alpha(E)$ values
used in Equation~\req{depth} to simulate the photon interaction
depths.  The complex geometry of the gates and the materials used in
the gates and gate insulating layer contribute to the complicated
low-energy quantum efficiency (QE) of the ACIS FI devices.  The $SiO_2$
damage layer that photons first encounter in the ACIS BI devices
similarly complicates their QE.  These ``surface'' layers also
contribute to the CCDs' spectral redistribution functions.

Given the importance of these surface layers, we have added them to the
CCD model.  The damage layer of the BI device is fairly easy to
include, as we can simply assume that it is just another slab of
$SiO_2$ in our model, although charge liberated in the damage layer is
not efficiently propagated and is largely lost to recombination.  The
fraction of charge propagated was left as a tuning parameter in our
modeling.  The gate structure is more complicated -- the true geometry
is quite complex due to three different gate configurations with
non-uniform polysilicon gate structures surrounded by non-uniform
$SiO_2$ (see Figure~1 in PRBR98).  We have not attempted to incorporate
the exact geometry of the gates; instead, we have modeled the gate
structure as a series of uniform-thickness slabs as well.

\begin{figure}[htb]
\centerline{\mbox{
	\epsfig{file=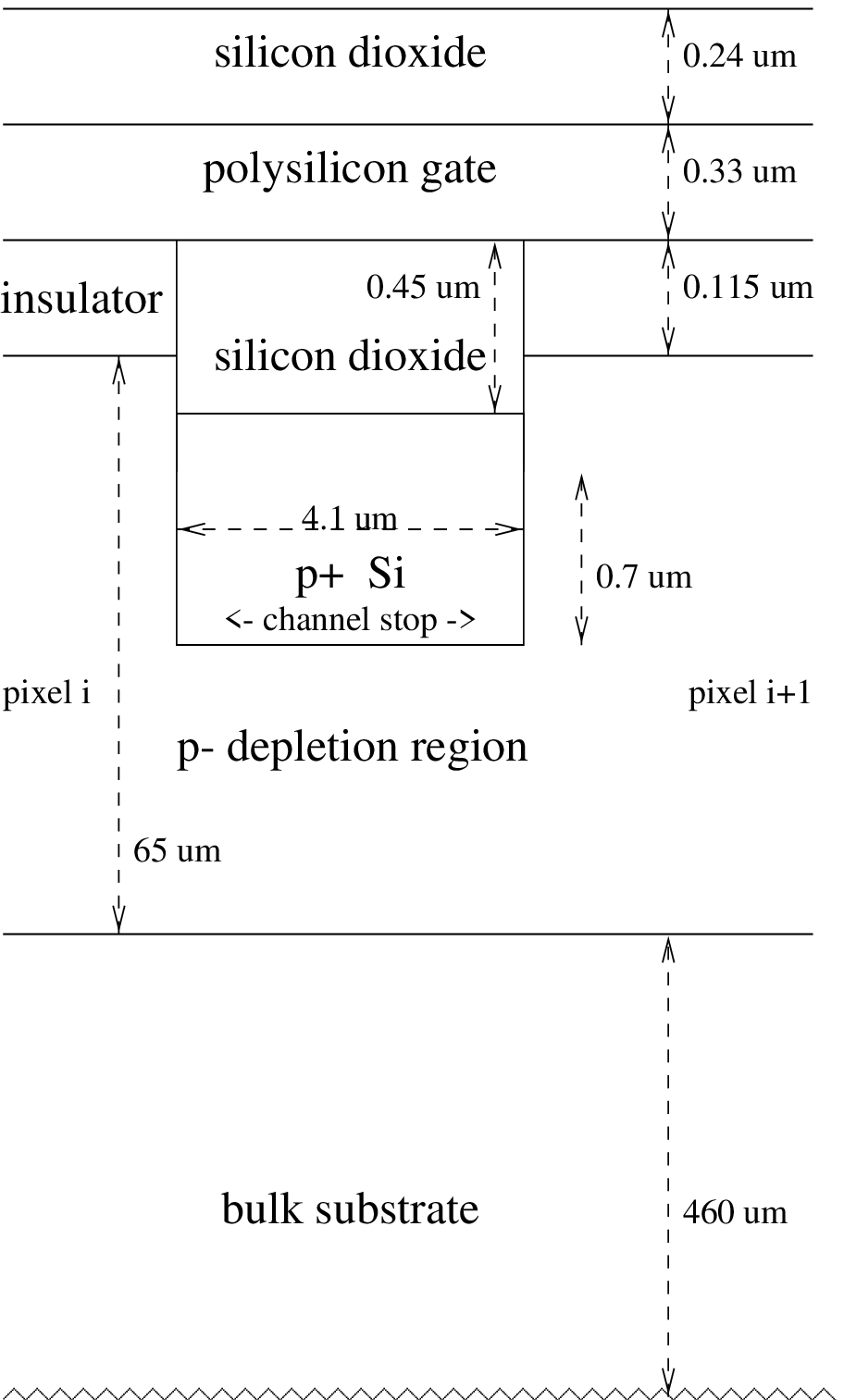,height=3.0in}
	\hspace{1.0in}
	\epsfig{file=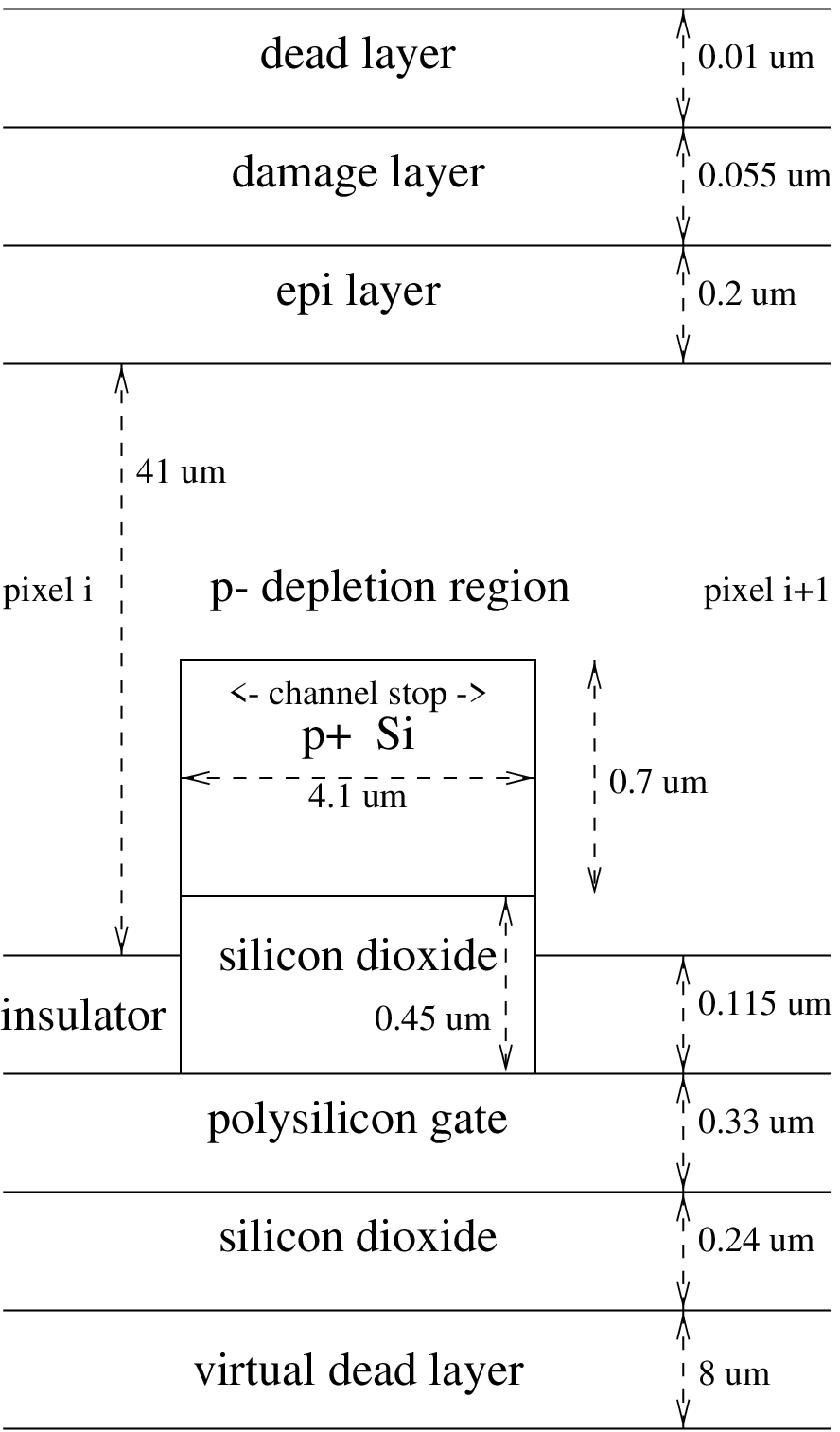,height=3.0in}}}
\caption{\protect \small Idealized geometry of an ACIS FI device (left)
and an ACIS BI device (right) used in our model.  Note that the layer
thicknesses are not drawn to scale.  ``Insulator'' is 
the silicon dioxide/nitride sandwich that makes up the gate insulator.}
\normalsize
\label{fig:layers}
\end{figure}

Exact layer thicknesses used in simulating the ACIS devices are shown
in Figure~\ref{fig:layers} (not to scale).  The layer thicknesses were
determined by MIT/CSR \cite{calreport} and are used here without
revision, except for the polysilicon layer in the channel stop and the
extra ``virtual dead layer'' in the BI device (described below).  We
required the channel stop polysilicon to be thicker than the MIT model
(0.7$\mu$m vs.\ 0.45$\mu$m) to reproduce the amplitude of a certain
spectral feature we call the ``soft shoulder'' (described in
Section~\ref{sec:specfeatures} below).  Perhaps this is necessary
because we simplified the geometry of the channel stops, again
representing them as simple slabs.  PRBR98 show the actual geometry of
the channel stops in their Figure 2 and include it, as well as the
actual geometry of the gate structure, in their model.  The gate
insulator is the $SiO_2 + Si_3N_4 + SiO_2$ sandwich; the components
have thicknesses of 0.015\mic, 0.04\mic, and 0.06\mic, respectively,
with the 0.06$\mu$m $SiO_2$ layer closest to the depletion region.

The ``virtual dead layer'' is an imaginary construct used to try to
increase the number of $Si$-fluorescence photons produced by the model
for high photon energies.  It is placed at the bottom of the BI device
in the model so that it won't affect the device's quantum efficiency.
We tried this technique because without it the amplitude of this peak
is underestimated for the BI device.  The actual source of these
fluorescent photons is unknown to us, although a possible explanation
could be the vertical rim of $Si$ left around the edge of the BI device
after the bulk $Si$ is etched away in the thinning process
\cite{burke97}.  This rim extends several hundred microns (the
equivalent of $\sim$20 pixels) above the surface of the BI device, so
$Si$ fluorescent photons generated there could propagate through the
vacuum above the device and interact with it far from their generation
sites.

As mentioned above, each layer has its own energy-dependent linear
absorption coefficients that are used to generate interaction depths
for that layer.  The gate structure model is the same for FI and BI
devices.  This idealized gate structure greatly simplifies the
simulator code and run-time but may limit our ability to reproduce the
event grades and subtle spectral features exactly.

Following PRBR98, we assume that photon interactions in the polysilicon
gate layer can produce $Si$ fluorescent photons that are detected if
they propagate into an active area in the device.  Any leftover charge
interacting in the polysilicon layer is presumed lost.  Photons
interacting in all the other gate structure layers are also lost,
except for those interacting in the $SiO_2$ layer closest to the
depletion region (see Section~\ref{sec:specfeatures} for details).

\section{The Calibration Data} \label{sec:caldata}

\subsection{Before Launch}

The simulator was tuned against calibration data obtained in Spring 1997
from the X-ray Calibration Facility (XRCF) at NASA's Marshall Space
Flight Center.  Details of the experimental setup are described in
O'Dell and Weisskopf \cite{odell98}.  Some of the ACIS tests performed
at XRCF are described in Nousek {\em et al.} \cite{nousek98} and in the
Calibration Report \cite{calreport}.  XRCF data were collected in time
intervals called ``phases.'' Only Phase H and I data were used for
tuning the simulator.

Phase H calibration involved the {\em Chandra} mirrors and ACIS
together; data from that phase useful for our tuning effort generally
consist of defocused images intended to measure the system's effective
area.  Even though these measurements were configured to minimize
photon pile-up, the spectra are still substantially affected by it --
this makes spectral features in most Phase H data less reliable for
tuning than the Phase I data.

In Phase I, the {\em Chandra} mirrors had been removed, so the source
photons illuminated the ACIS focal plane nearly uniformly.  Pile-up is
much less of a problem for these data.  This XRCF dataset is much
smaller than the MIT lab dataset; the PSU model is thus tuned to fewer
test events than the MIT model in all cases, sometimes to very few ($<
10000$) events.  MIT's results from subassembly calibration (such as
the depletion depth and the pixel size) were used as defaults for our
simulator; we only deviated from values determined in subassembly
calibration when it was necessary to produce better agreement with our
data.  

To date the PSU simulator has been used for detailed modeling of the FI chip
in the ``I3'' focal plane position and the BI chip
in the ``S3'' focal plane position.  These devices contain the aimpoints
for the ACIS imaging and spectroscopy arrays, respectively.
The S3 BI chip is especially important to understand because the
zeroth-order (undispersed) image from the {\em Chandra} transmission
gratings falls on S3.

The XRCF data allowed us to tune the simulator at 21 energies, unevenly
sampling the range 277--9000~eV.  We chose to tune using the XRCF data
exclusively (including no subassembly data) because they were most
readily available to us and because they were obtained from the
fully-assembled ACIS, using flight electronics and with all CCDs
populating the flight focal plane.

After CTI correction \cite{townsley01b} was performed on the BI data,
all XRCF data were filtered to contain only ASCA Grade 0, 2, 3, 4, and
6 (shorthand:  ``g02346'') events prior to tuning the simulator to
match them.  This helped to remove contamination in the data from
pile-up, cosmic rays, and some of the source continuum spectrum.
Source continuum ({\em i.e.} non-monochromatic incident flux) is still
present in some of these filtered spectra, especially at energies below
that of the main peak (examples are given in
Section~ref{sec:comparisons}).  We cannot guarantee that our tuning is
immune from corruption by the presence of these extraneous source
photons masquerading as the CCD energy redistribution function, but we
were aware of this potential error and tuned with it in mind.

\subsection{After Launch}

When {\em Chandra} passes through Earth's radiation belts, ACIS is
moved out of the lightpath of the mirrors to a stowed position that
protects it from damaging protons.  In this position, ACIS views the
External Calibration Source, which produces line emission at Al \Ka, Ti
\Ka and \Kb, and Mn \Ka and \Kb that illuminates the focal plane nearly
uniformly.  The resulting complex spectrum from this observing
configuration, including instrumental lines, pile-up peaks, and CCD
spectral redistribution features, was carefully characterized on the
ground \cite{calreport}.

Since these calibration data are collected at least once per orbit (for
the purpose of monitoring CTI), combining these observations resulted
in a large database useful for checking the simulator tuning.  There
are $\sim 1 \times 10^7$ events per amp on the BI chip S3 and $\sim 2
\times 10^6$ events per amp on the FI chip I3.  We checked that the CTI
did not vary significantly across the several-month timespan of the data then
corrected the data for CTI; an additional noise term was added to the
simulator and tuned to match the row-dependent energy resolution
degradation seen in the FI devices \cite{townsley01b}. We found that
some tuning parameters required adjustment to match these on-orbit
data, which is not surprising since CTI has changed the character of
the devices substantially.

\section{Spectral Features}
\label{sec:specfeatures}

\subsection{Canonical Spectra}

\begin{figure}[htb]
\centerline{\mbox{
         \epsfig{file=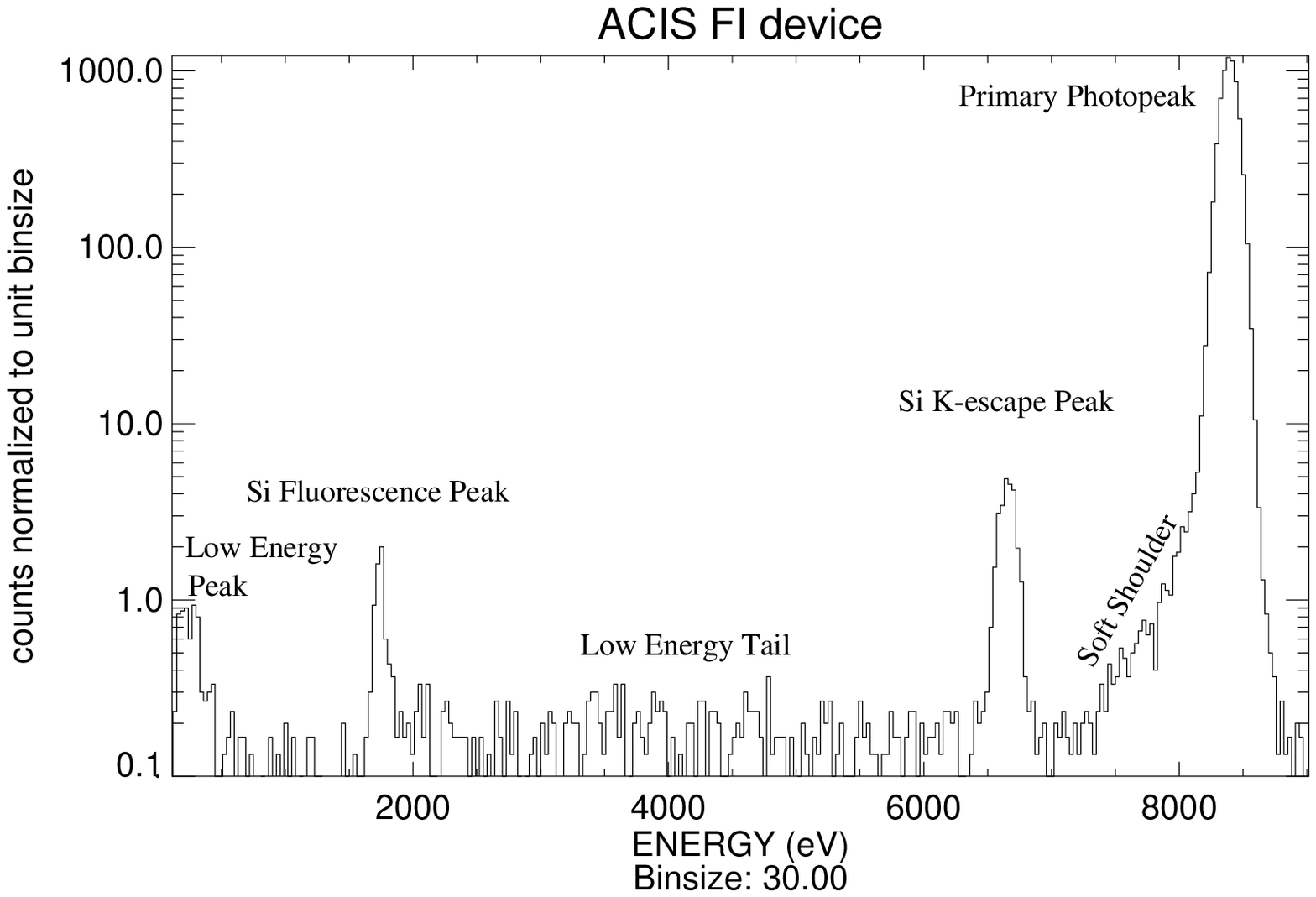,width=3.0in,angle=-90 }
         \hspace{0.5in}
         \epsfig{file=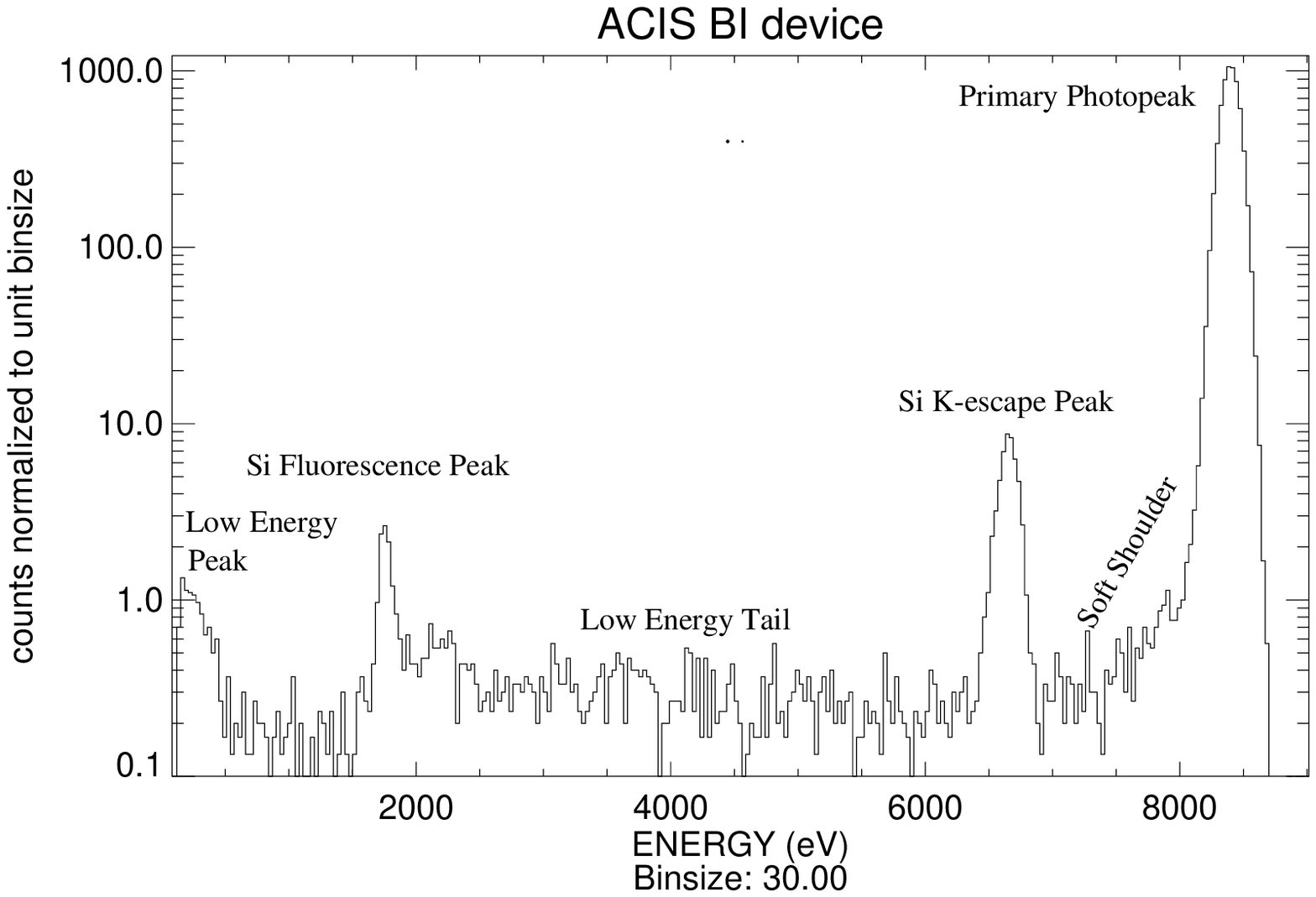,width=3.0in,angle=-90}}}
\caption{\protect \small Canonical FI and BI spectra from 8.398~keV DCM
data.  Spectral features used in tuning are marked.  Both datasets
contain $\sim 2 \times 10^5$ g02346 events.  The BI data were corrected
for CTI.}

\normalsize
\label{fig:canonical}
\end{figure}

Canonical monochromatic spectra for the I3 FI device and the S3 BI
device are shown in Figure~\ref{fig:canonical}.  These are XRCF Phase I
data at 8.398~keV, taken from all amplifiers of each CCD and obtained
using the Double Crystal Monochromator (DCM), which produces a clean
spectrum; all of the prominent features are due to the CCD spectral
redistribution (except the main peak, of course) rather than corruption
from the input source spectrum.  Data were obtained from rows 10--265
and 745--1000 on each device.  The BI data have been corrected for
CTI.  Thus these spectra illustrate the CCDs' spectral redistribution
functions as well as any XRCF dataset can. The spectral features used
to tune the simulator are indicated, following the nomenclature of
Bautz {\em et al.} \cite{bautz99}.  Both datasets contain the same
number of events and were filtered to keep only grades g02346.

The spectra are reassuringly similar.  The BI device has slightly worse
spectral resolution ($\sigma \simeq 69~eV$) than the pre-launch FI
device ($\sigma \simeq 61~eV$); this is probably due at least in part
to residual CTI effects.  The BI device shows slightly larger Si
fluorescence, Si K-escape, and low-energy peaks and a larger
low-energy tail.  This could be due to photons interacting in the dead
and damage ($SiO_2$) layers at the top of the BI device -- these layers
cause spectral redistribution similar to that in the gate structure.
Since the gate structures of the ACIS FI and BI devices are similar,
the added $SiO_2$ layers in the BI device might increase the relative
strength of these redistribution features. 

As mentioned in Section~\ref{sec:gates}, the thin insulator layer under
the gates is invoked to explain the low-energy peak and the low-energy
tail in the XRCF FI data, based on ideas from the MIT ACIS team
\cite{prigozhin00}.  We consider only the lowest $SiO_2$ layer in this
sandwich to be capable of producing events.  There, photons are
converted to electrons much less efficiently.  Prigozhin {\em et al.}
\cite{prigozhin00} suggest a conversion based on ACIS subassembly
calibration data of $w=17$~eV~e$^{-1}$ (see Equation~\req{initchg}) in
$SiO_2$, but that 65\% of the electrons generated there are lost,
making the apparent conversion factor $w \sim 52$~eV~e$^{-1}$.
Application of this conversion factor yields events that populate the
low-energy peak.

Prigozhin {\em et al.} \cite{prigozhin00} also introduced the idea of
``boundary'' events, where a photon interacts close enough to the
boundary between the gate insulator ($SiO_2$) and depleted $Si$ that
its initial charge cloud is split between these layers, each part then
interacting with the device according to the physics appropriate for
the material in each layer.  Photons interacting at a range of depths
have varying fractions of their initial charge clouds interacting in
the depletion region -- the resulting events have a range of energies
that generate the continuum seen in the spectral redistribution
function.  In our code, we extend this concept to boundaries between
other layers.  Our criterion for invoking this boundary splitting is
that the photon must interact within one initial charge cloud radius of
the boundary, on either side of that boundary.  Note that this initial
charge cloud radius is energy dependent.

For BI devices, the insulator layer between the depletion region and
the gates is included, but it is not the primary cause of continuum
counts or the low-energy peak.  Since these spectral features appear in
the BI data as well as the FI data, it made sense to treat the damage
layer, also made of $SiO_2$, in the same way as the gate insulator.
Since this is the first layer photons encounter, a substantial number
of photons interact in this layer and serve to populate the continuum
and low-energy peak in the spectrum.  Empirically, we determined that
the damage layer retains even less charge than normal $SiO_2$; in order
to reproduce the observed ratio of the low-energy peak to the primary
photopeak, we had to reduce the charge in the damage layer by a factor
of 2 (independent of energy).


\subsection{Channel Stops}

A model for the channel stops is often used to account for the soft
shoulder of the main peak apparent in the FI data (and at the highest
energies in the BI data).  PRBR98 employ a detailed model of the
channel stop geometry (see their Figure 2), supported by detection
efficiency maps with sub-pixel spatial resolution \cite{bautz99}, to
reproduce the energy and grade distribution of the soft shoulder.

We assume that photons interacting in the channel stop $SiO_2$ layer or
the p+ silicon suffer splitting of their charge clouds, that the charge
is quickly swept to the right or left side of the channel stop, and
that the fraction swept to each side depends on the interaction
distance from each side.  Each separated charge cloud is assumed to
propagate from a new x position (at the edge of our simplified channel
stop -- see Figure~\ref{fig:layers}) through the depletion region as a
normal photon would, using an initial charge cloud radius calculated
from the original photon's energy.

This model is motivated by the overabundance of right- and left-hand
split events (ASCA grades 3 and 4) present in the spectral soft
shoulders of the FI XRCF data.  We include a special location-dependent
loss mechanism and an extra noise term for this process, necessary to
reproduce the offset in the mean of the soft shoulder energy from the
main peak mean and to reproduce the width of the soft shoulder.  These
are incorporated into the code as energy-dependent tuning parameters
$\mathcal C$ and $\mathcal W$, respectively.

Then, defining $Q_{0}$ to be the initial charge, $w_{c}$ to be the width
of the channel stop in microns, and $x_{c}$ to be the photon
interaction position expressed as a distance from the left edge of the
channel stop (in microns), the charge to be propagated from the left
edge of the channel stop becomes:
\be
Q_{left} = \frac{w_{c} - x_{c}}{w_{c}}Q_{0} -  {\mathcal C}x_{c} +  \mathcal{R}_{N}(0){\mathcal W}\frac{x_{c}}{w_{c}}
\label{csleft}
\ee
while the charge to be propagated from the right edge of the channel stop is:
\be
Q_{right} = \frac{x_{c}}{w_{c}}Q_{0} - (w_{c} - x_{c}){\mathcal C} +  \mathcal{R}_{N}(0){\mathcal W}\frac{w_{c} - x_{c}}{w_{c}}~,
\label{csright}
\ee
where $\mathcal{R}_{N}(0)$ is a sample of a normal random distribution of
mean 0 and standard deviation 1, as before (different samples are used
for the left and right charge calculations).

Note that the channel stops are generally not responsible for the soft
shoulders seen in the BI XRCF data (except above $\sim$5~keV, where
some photons penetrate far enough into the device to interact in the
channel stops).  We do see soft shoulders in the BI data, but the grade
distribution of events in the soft shoulders is similar to that of
other spectral features; the soft shoulders are not dominated by grade
3 and 4 events.  This adds confusion to an already confusing device --
as usual, the culprit appears to be CTI (see the companion paper
\cite{townsley01b}).  We do include the channel stop model in the BI
device simulator to model the high-energy spectral redistribution
function and because this feature may be more important for future
modeling of thinner BI devices.

\section{Tuning Parameters} \label{sec:tuningparams}
 
As mentioned earlier, there are six energy-dependent tuning parameters
used in the simulator to get its output to match XRCF data.  They are
necessary because the simulator does not incorporate all the device
physics required to predict the data exactly; in effect, the existence
of a tuning parameter represents an uncertainty in the model.  The tuning
parameters were designed to be as independent of each other as possible,
so that adjusting the value of one parameter has minimal effect on the
others.


\subsection{Definitions}
 
The current tuning parameters can be grouped into those that affect the
main peak, those that adjust the sizes of the charge clouds, and those
that pertain to the channel stops.  Most were described in earlier sections
via the equations in which they appear.

\begin{itemize}
\item Main Peak Parameters
  \begin{description}
    \item[gain] $\mathcal G$ adjusts the mean of the main peak to match the data
    at a given energy -- compensates for the model's imperfect prediction
    of event amplitude.  It is applied to each simulated 3$\times$3 pixel
    neighborhood $\bf N$ as the last step in generating a simulated event:
      ${\bf N}_{\rm final} = {\mathcal G}{\bf N}$.
    \item[linewidth] $\mathcal L$ increases the width of the main peak to match 
    the data -- incorporates any unknown noise terms that broaden the
    spectral response.  See Equation~\req{fano}.
  \end{description}
\item Charge Cloud Radius Parameters
  \begin{description}
    \item[saturation] $\mathcal S$ increases the size of the charge cloud 
    in the depletion
    region to achieve the correct grade distribution;
    note that this parameter
    is tuned only so the simulation reproduces the correct percentage of g0
    events ({\em i.e.} no conscious effort is made to get the correct
    distribution of other grades).  See Equation~\req{rdsat}.
    \item[initial oxide] $\mathcal I$ adjusts the initial size of the 
    charge cloud in the $SiO_2$
    insulating layer under the gates and in the damage layer for BI devices;
    this affects the continuum level and the sharpness of the low-energy
    feature.  See Equation~\req{rinitial} and the notes following.
  \end{description}
\item Channel Stop Parameters
  \begin{description}
    \item[shoulder linewidth] $\mathcal W$ increases the width of the soft
    shoulder of the main peak -- accounts for additional uncertainty in 
    the amount of charge recovered for events generated in channel stops.
    See Equations~\req{csleft} and \req{csright}.
    \item[charge loss] $\mathcal C$ adjusts the mean energy of the soft 
    shoulder of the main peak -- accounts for charge loss in channel stops.
    See Equations~\req{csleft} and \req{csright}.
  \end{description}
\end{itemize}


\subsection{Tuning Parameter Values}

The following figures show the values of all the tuning parameters for
CCDs I3 and S3, as a function of energy.  The gain tuning parameter was
derived from the on-orbit External Calibration Source data, corrected for
CTI.  Since the monochromatic spectral redistribution features are
subtle and often confused by background or other line features in the
on-orbit data, tuning parameters relating to these features were
obtained using XRCF data.

The tuning was performed by hand, adjusting the tuning parameters one
at a time to match the data.  The main peak parameters were tuned
first, since they have the largest impact on the spectral
redistribution function.  Then the saturation parameter $\mathcal S$
was adjusted to match the number of g0 events.  Finally the other
parameters affecting smaller redistribution features were tuned. 
Since the tuning was performed interactively, checks were made at
each iteration to ensure that tuning one parameter did not adversely
affect regions of the spectrum governed by other tuning parameters. 

For simulating photons with energies not measured on-orbit or at XRCF,
the code performs a linear interpolation between the two closest
measured tuning parameter values.  When forced to extrapolate (for
simulated energies beyond the range measured at XRCF), the code uses
the tuning parameter values appropriate for the closest measured
energy.

Error bars are not given because the errors in these parameter
estimates are dominated by systematic uncertainties, due mainly to
corruption of the spectra by CTI, continuum photons from the XRCF
sources, or by pile-up.  For some energies we also have very limited
quantities of data; since only a few of the events are redistributed
out of the main peaks, it is difficult to constrain the tuning
parameters in these datasets.

\begin{figure}[htb] 
 
\centerline{\mbox{
         \epsfig{file=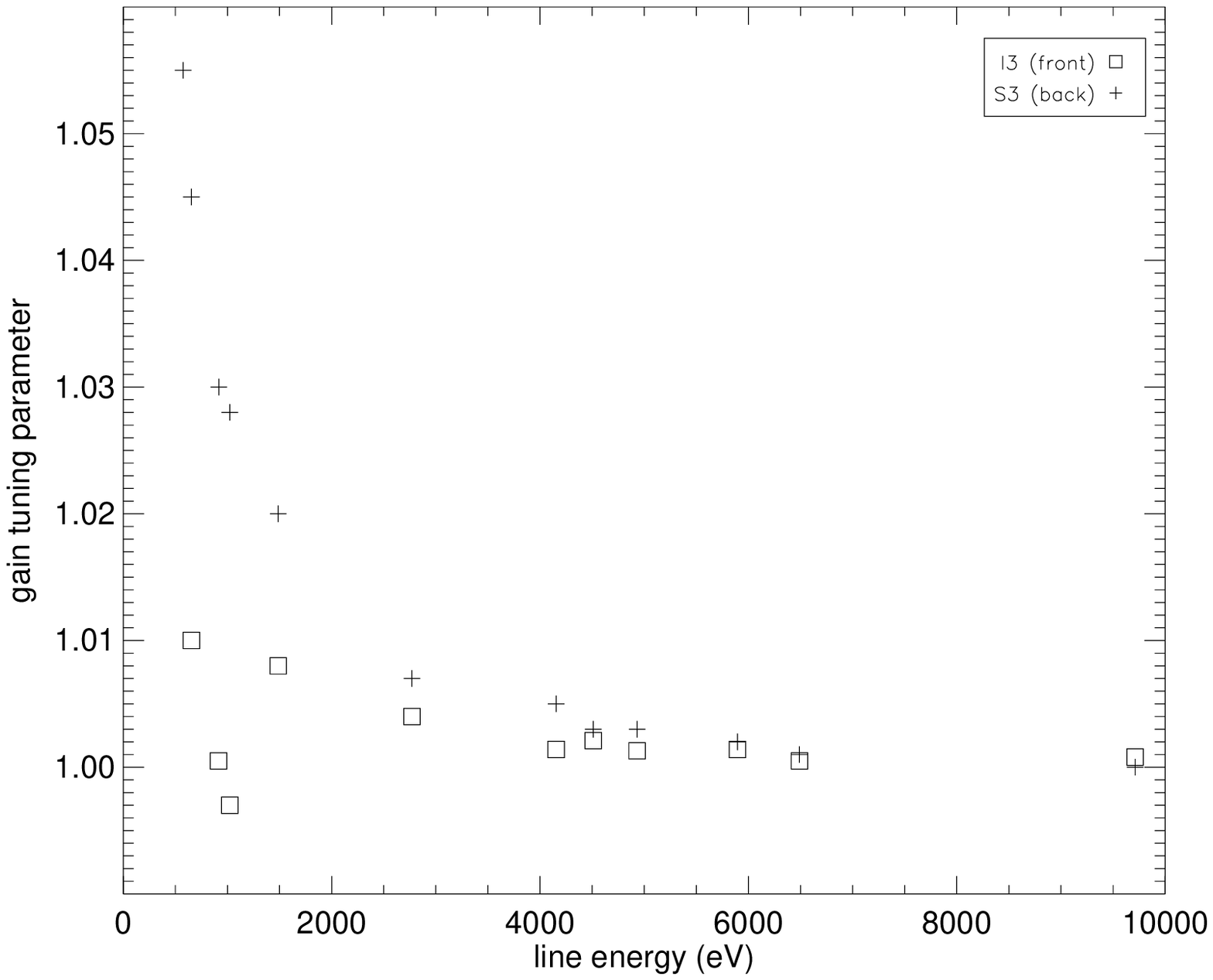,width=3.0in}
         \hspace{0.5in}
         \epsfig{file=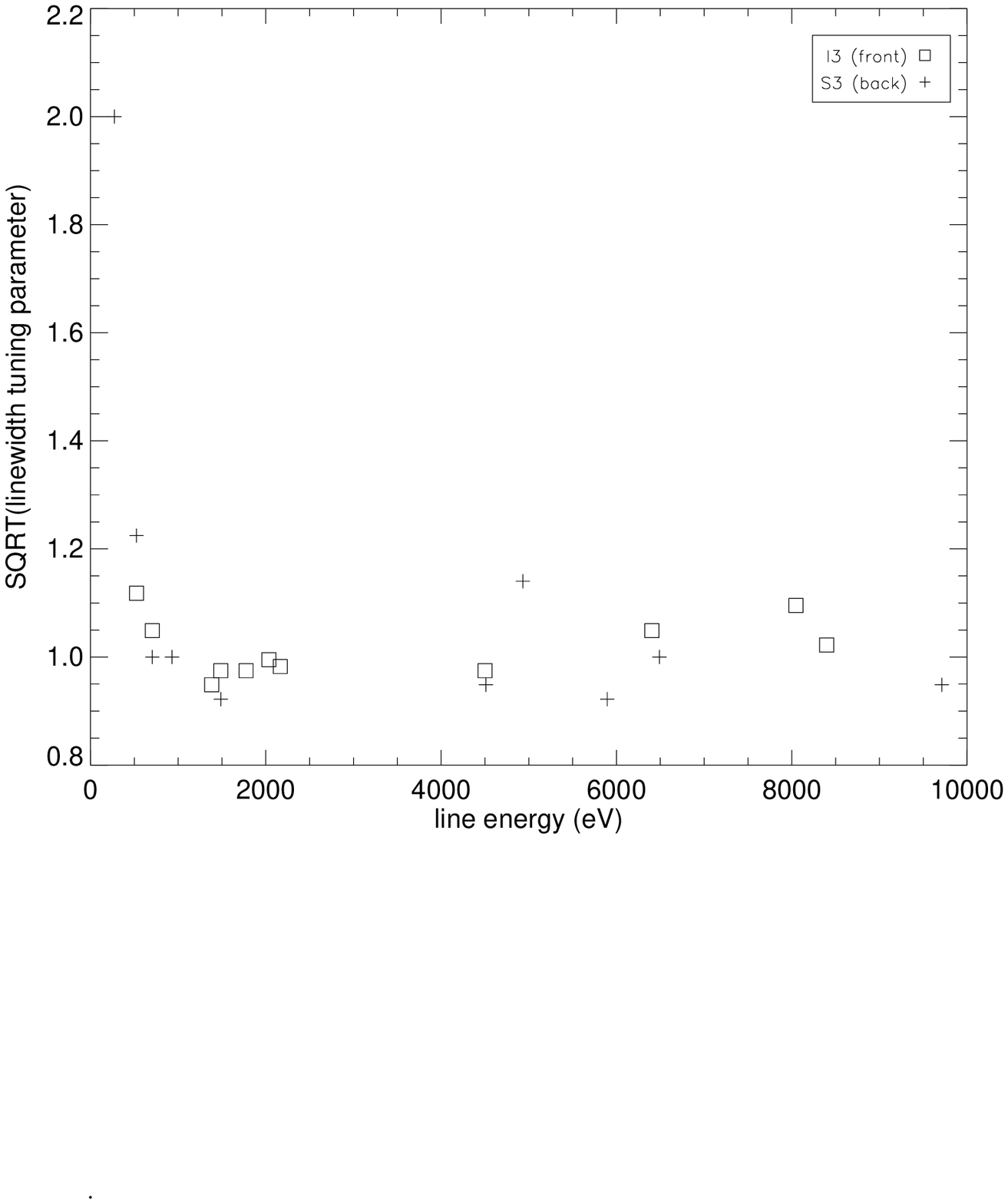,width=3.0in}}}
                  
\caption{\protect \small Main peak tuning parameters for the I3 and S3
chips.  The left panel shows $\mathcal G$ as a function of photon
energy and can be interpreted as a measure of gain non-linearity in
these devices.  The right panel shows $\sqrt{\mathcal L}$, the ratio of
measured width to theoretical width for each line.}


\normalsize
\label{fig:maintuning}
\end{figure}

The left panel in Figure~\ref{fig:maintuning} shows the gain
adjustments necessary to make simulated event energies reproduce the
data.  Only on-orbit data are plotted because there appears to be a
small gain change between XRCF and on-orbit data.  The FI adjustments
are quite small, but the BI device shows that substantial (several
percent) adjustments are necessary at low energies to make the
simulator reproduce the data.  The points below 1.4~keV are from
several observations of the oxygen-rich supernova remnant E0102-72.3,
taken at a variety of positions on the I3 and S3 devices.  The datasets
were combined and the line positions measured from the composite
spectra. 

The amplitude and strong energy-dependence of $\mathcal G$ below 2~keV
in the BI device reflects our limited understanding of the geometry and
charge cloud characteristics in and near its thin dead, damage, and
epitaxial layers.  We have no knowledge of the BI gain adjustment
necessary in the important soft spectral range 0.2--0.5~keV, due to a
lack of suitable calibration information.  As noted above, the
simulator uses the gain tuning parameter at the last measured energy
(0.574~keV) to simulate all lower-energy events.  Given the trend
suggested by Figure~\ref{fig:maintuning}, this could be a substantial
underestimation of the actual value needed.

The right panel of Figure~\ref{fig:maintuning} gives the ratio of
measured to theoretical linewidth.  For this parameter, I3 was tuned
using mostly XRCF data but S3 was tuned using mostly ECS data.  The
rise at low energies implies that there is some additional source of
line broadening not included in the model.  This could be the result of
photon interactions in the surface layers of the devices.  Since these
are complex layers consisting of silicon compounds with an electric
field structure different from that in depleted silicon, events
generated there may not follow the Fano noise described in
Equation~\req{fano} \cite{prigozhin00}.

\begin{figure}[htb] 
 
\centerline{\mbox{
         \epsfig{file=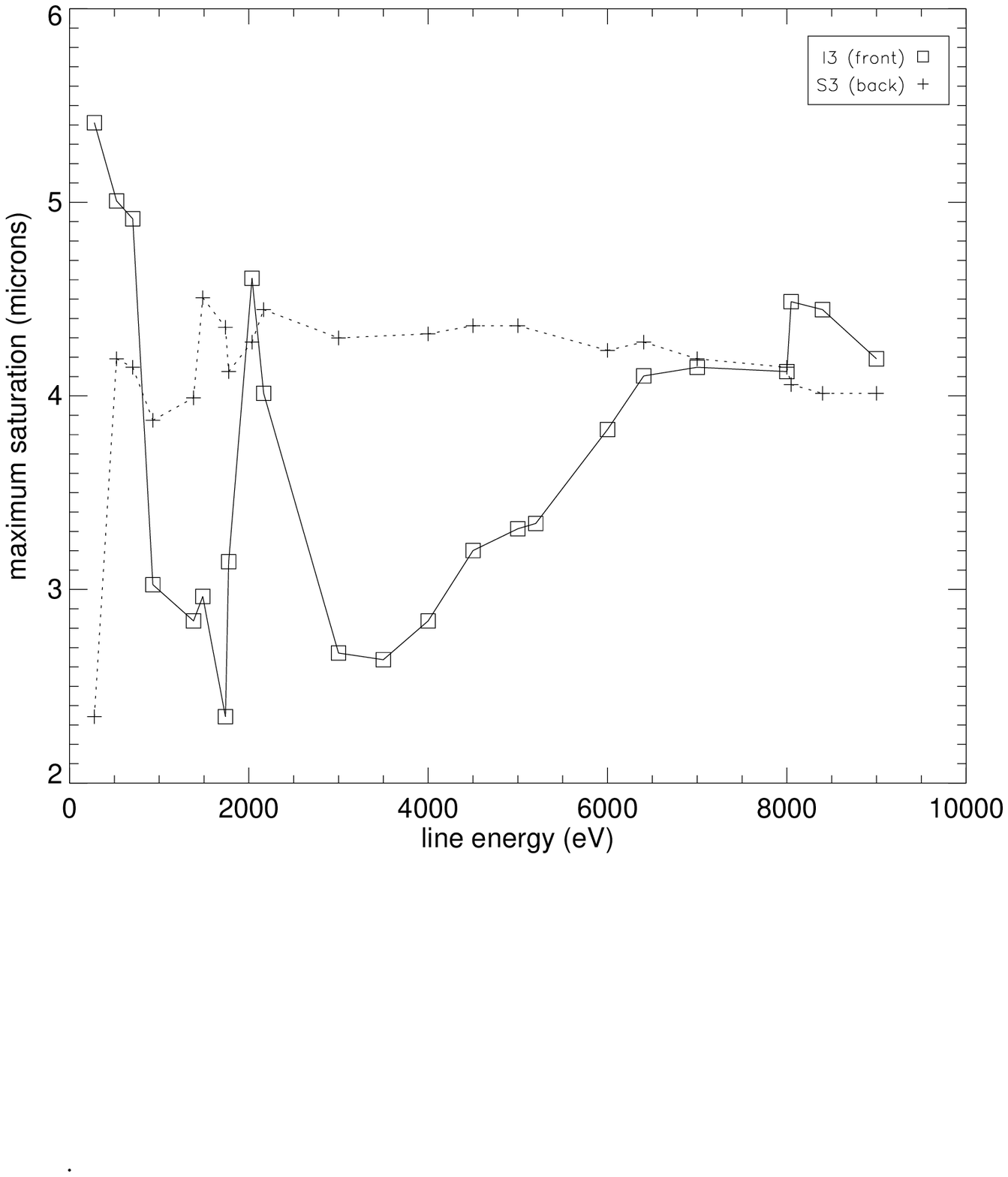,width=3.0in}
         \hspace{0.5in}
         \epsfig{file=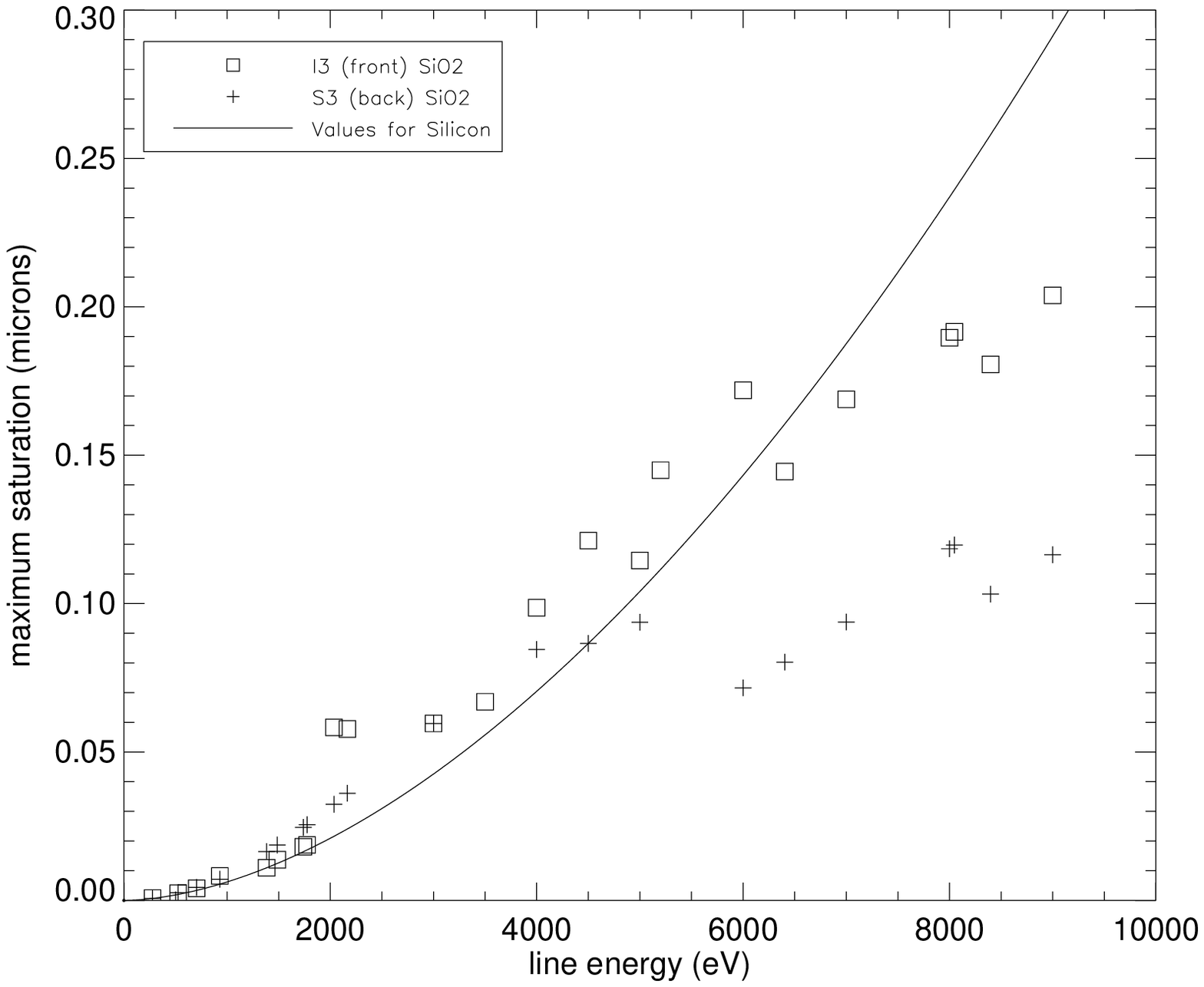,width=3.0in}}}
                  
\caption{\protect \small Charge cloud radius tuning parameters for the
I3 and S3 chips, based on XRCF data.  The left panel shows the maximum
value of the square root of the second term in Equation~\req{rdsat} (we
have assumed $\frac{z_0}{\zd} = 1$).  Thus actual values of this term
vary between zero and the maximum plotted here, depending on the
interaction depth of the photon.  The right panel shows a plot of
Equation~\req{rinitial}, the initial charge cloud radius $r_{i}$ as a
function of photon energy for interactions in $Si$, and modified values
$r_{i} \cdot {\mathcal I}$ that estimate the initial charge cloud radii
in $SiO_2$.} 

\normalsize
\label{fig:radiustuning}
\end{figure}

The left panel of Figure~\ref{fig:radiustuning} shows the effects of
the charge cloud radius saturation tuning parameter $\mathcal S$.  A
value between zero and this maximum value (depending on the photon
interaction depth) is added in quadrature to the expression for the
charge cloud radius in the depletion region and provides a very
sensitive means of tuning the size of the charge clouds to reproduce
the fraction of grade 0 events seen in the data.  It may represent the
ensemble average degree of drift velocity saturation at energies where
most photon interactions occur in the depletion region ({\em e.g.}
above 300~eV for BI devices and above 2500~eV for FI devices).  At
energies where most photon interactions are shallow, this effect may be
masked by the complexities of charge cloud diffusion in $SiO_2$ or
$Si_{3}N_{4}$ layers.  This is apparent in the I3 curve in this figure;
photons at low energies and those just above the $Si$ absorption edge
(at 1839~eV) interact in the top 1 or 2 microns of the CCD, often in
the gate structure or channel stops, and tend to require larger values
of this saturation term in the simulation.  The BI device S3 behaves
very differently, presumably because the surface layers encountered by
low-energy photons are much different than the gate structure and
because the device is thinned so that most photon interactions occur in
the depletion region, independent of energy.  However at 277~eV (our
lowest datapoint), 80\% of
the events interact in the surface layers of the BI device, so charge
cloud diffusion in $SiO_2$ and in the epitaxial field-free layer dominate
saturation effects.  

The right panel of Figure~\ref{fig:radiustuning} shows the initial
charge cloud radius as a function of energy for interactions in $Si$
(Equation~\req{rinitial}) and how this gets modified by the tuning
parameter $\mathcal I$ for interactions in $SiO_2$.  The size of
$SiO_2$ charge clouds seems to follow the prediction for $Si$ in both
devices up to a limiting energy ($\sim$4~keV in the BI device,
$\sim$6~keV in the FI device), above which the radius increases much
more slowly with energy.  Note that the $SiO_2$ layers considered to
be relevant in both the FI and BI models are only $\sim$0.06$\mu$m thick, 
so the flattening seen in the plot may be due to an insensitivity
to the tuning parameter at high energies.

\begin{figure}[htb] 
 
\centerline{\mbox{
         \epsfig{file=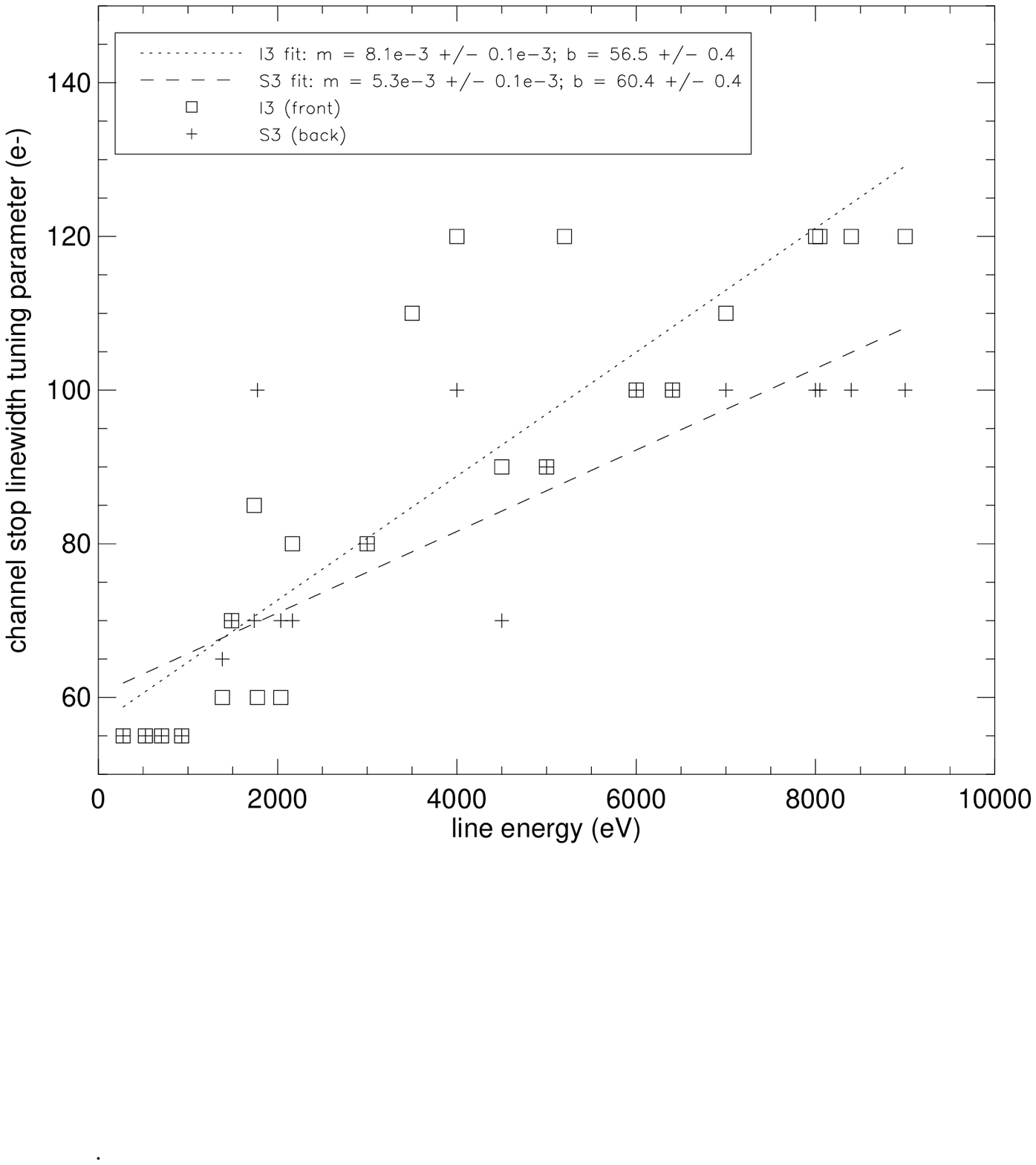,width=3.0in}
         \hspace{0.5in}
         \epsfig{file=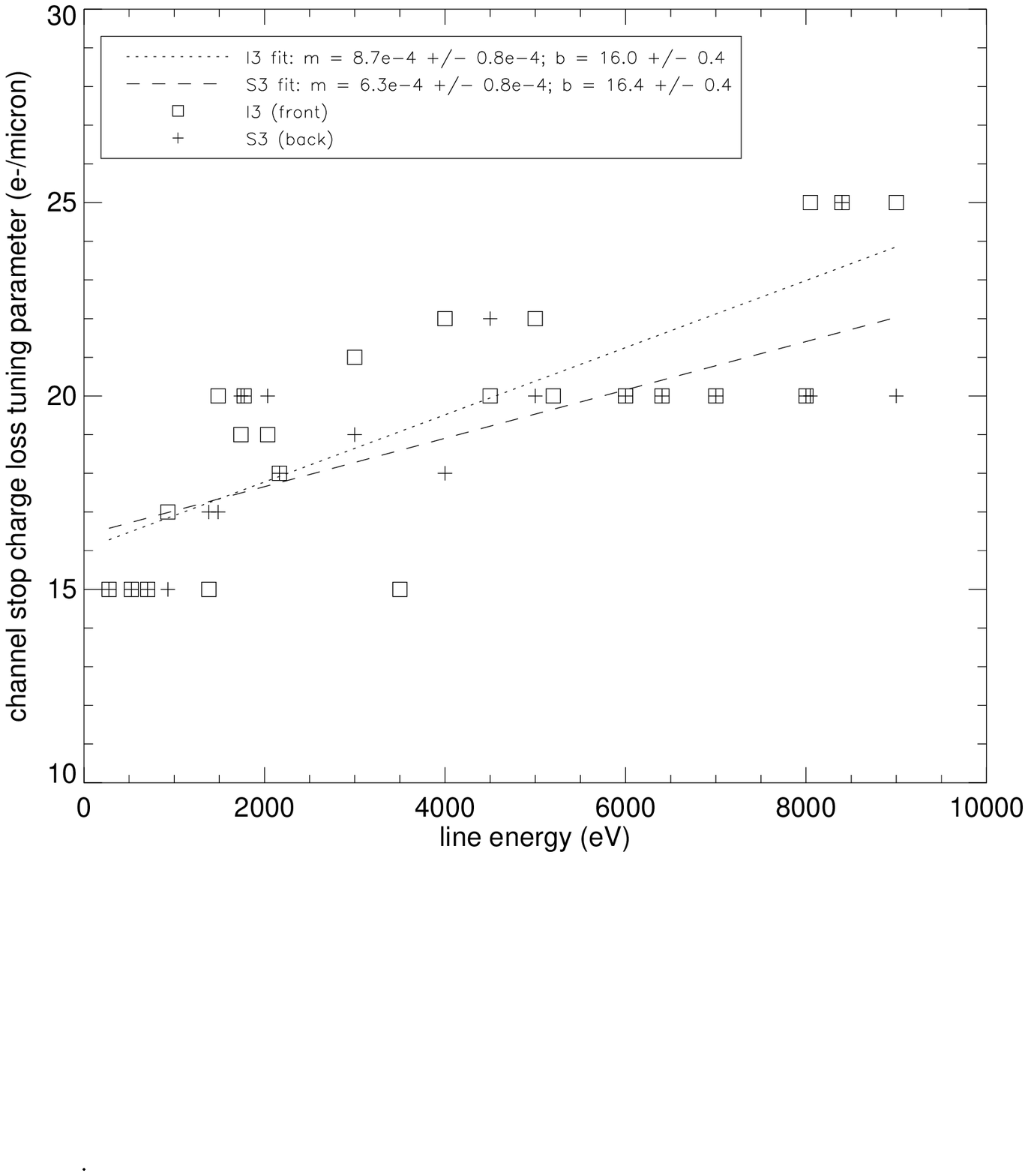,width=3.0in}}}
                  
\caption{\protect \small  Channel stop tuning parameters for the 
I3 and S3 chips, based on XRCF data.  The left panel gives values for
$\mathcal W$, the channel stop linewidth tuning parameter.  This is
a noise parameter that is randomized for each event.  The right
panel shows $\mathcal C$, the channel stop charge loss tuning parameter.
This gives the charge loss per micron that the charge cloud suffers
as it laterally traverses the channel stop.} 

\normalsize
\label{fig:cstuning}
\end{figure}

Figure~\ref{fig:cstuning} shows the channel stop tuning parameters.
The left panel gives $\mathcal W$, the noise amplitude in electrons.
This quantity is modulated by the distance the charge packet must
travel to reach the edges of the channel stop, then randomized and
added to each channel stop event to reproduce the width of the soft
shoulder of the main peak in each of the monochromatic XRCF spectra
(see the last term in Equations~\req{csleft} and \req{csright}).  The
right panel shows $\mathcal C$, the channel stop charge loss in
electrons per micron traversed by the charge packet (see the second
term in Equations~\req{csleft} and \req{csright}).  This parameter
governs the offset from the main peak energy of the soft shoulder.

The channel stop parameters are especially affected by corruption from
the source spectrum in XRCF data, so individual values are highly
uncertain.  There does seem to be a modest trend in these parameters
with input photon energy, however.  Unweighted linear fits are given
for each device in the two panels of Figure~\ref{fig:cstuning}.

As noted above, the simulator currently performs a simple linear
interpolation to estimate tuning parameter values for unsampled
energies.  Given the noise in these data, it may be more appropriate to
substitute a functional form determined from fits to these XRCF
parameter values, as shown above for the channel stop parameters.
Future versions of the code may employ such fits.

\section{Simulator Implementation} \label{sec:machinery}

We implement the physical models described in
Section~\ref{sec:generic}, the ACIS geometries described in
Section~\ref{sec:acisspecs}, and the CTI model described in Townsley
{\em et al.} \cite{townsley01b} using a set of IDL \cite{idl} programs
and FITS parameter files.  For modeling celestial sources, photons are obtained from the MARX
\cite{wise97} simulator which passes an arbitrary spatial and spectral
model of a source through a model of the {\em Chandra} mirrors.  For
simulating uniform-illumination calibration data, we have included a
photon generator in the simulator.  Whatever photon generator is used
produces a list of photons with a Poisson distribution of arrival times that impact the
ACIS detectors.  The properties of this photon list are CCD number,
position in the CCD's X/Y coordinate system, trajectory (direction
cosines), energy, and arrival time.

The user specifies a rectangular region that should be simulated on
each CCD, and the desired frame time.  For each simulated CCD, the
simulator generates a series of CCD frames by performing the following
steps.
\begin{itemize}
\item Determine which photons arrived during the current frame interval
and fall within the specified region of the CCD.  At present, the code
assumes a frame interval for ACIS to be 3.24~seconds; events that fall
on the device during the frame readout (0.041~seconds at the end of the
frame interval) are not modeled separately.  The user may
specify that photon arrival times should be ignored and exactly one photon
should arrive in each CCD frame, eliminating pile-up completely.
\item Add a random offset to each photon's position if desired,
simulating {\em Chandra}'s aspect motion.
\item Simulate the physics of each photon's interaction with the device,
producing a ``charge frame,'' the pattern of charge (in units of electrons)
that was collected at the gates.
\item Simulate the physics of a set of minimally-ionizing particles
interacting with the CCD and adding charge to the charge frame, if modeling
of the particle background is desired.
\item Model CTI by moving and destroying charge in the charge frame.
\item Add Poisson noise to model dark current fluctuations and Gaussian noise
to model electronic noise.
\item Model the analog-to-digital conversion performed in ACIS by applying
an appropriate gain value to the charge frame and rounding to integer values.
\end{itemize}
The resulting digital number (DN) frame represents our best attempt to
reproduce the digital images produced by ACIS. 

Next, the ACIS event detection algorithm is applied to these DN frames,
producing a set of detected events that is written out in a FITS table
format very similar to the event lists produced by ACIS itself.  If
simulated aspect motion was applied to the photon positions, then that
offset and a random error are subtracted from the event positions,
simulating an imperfect spacecraft aspect solution.

Particle events are modeled similarly to photon events; the track of a
particle through the detector is randomly generated, then simulated by
allowing the detector to absorb energy from the particle every time it
traverses one micron in depth through the device.  (Finer sampling
could be implemented but makes the code prohibitively slow.)  The
charge cloud generated at each interaction point is propagated just as
photon charge clouds are propagated.

One advantage of simulating actual large-area CCD frames is that
interference between photon events (pile-up) and between particle
events and photon events may be modeled.  Pile-up is a significant
concern in Chandra due to the very sharp PSF.  Particle events, which
can cover large areas, can corrupt the energies and grades of the X-ray
events they touch.  The disadvantage of this approach is that large
arrays must be used in the code, which necessitates the use of careful
programing techniques to avoid oppressive runtimes.

\begin{figure}[htb]
\centerline{\mbox{
	 \epsfig{file=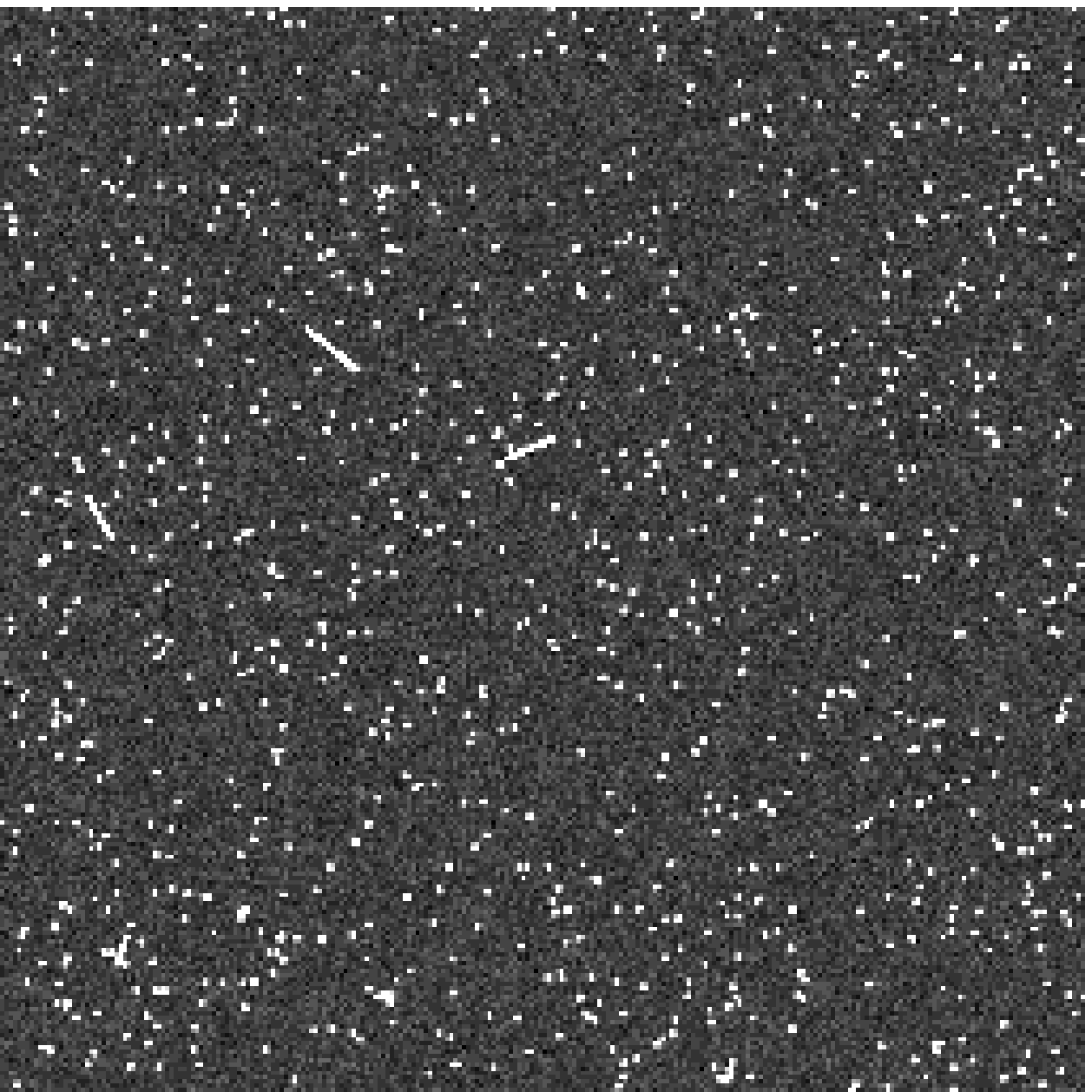,width=3.0in}
	 \epsfig{file=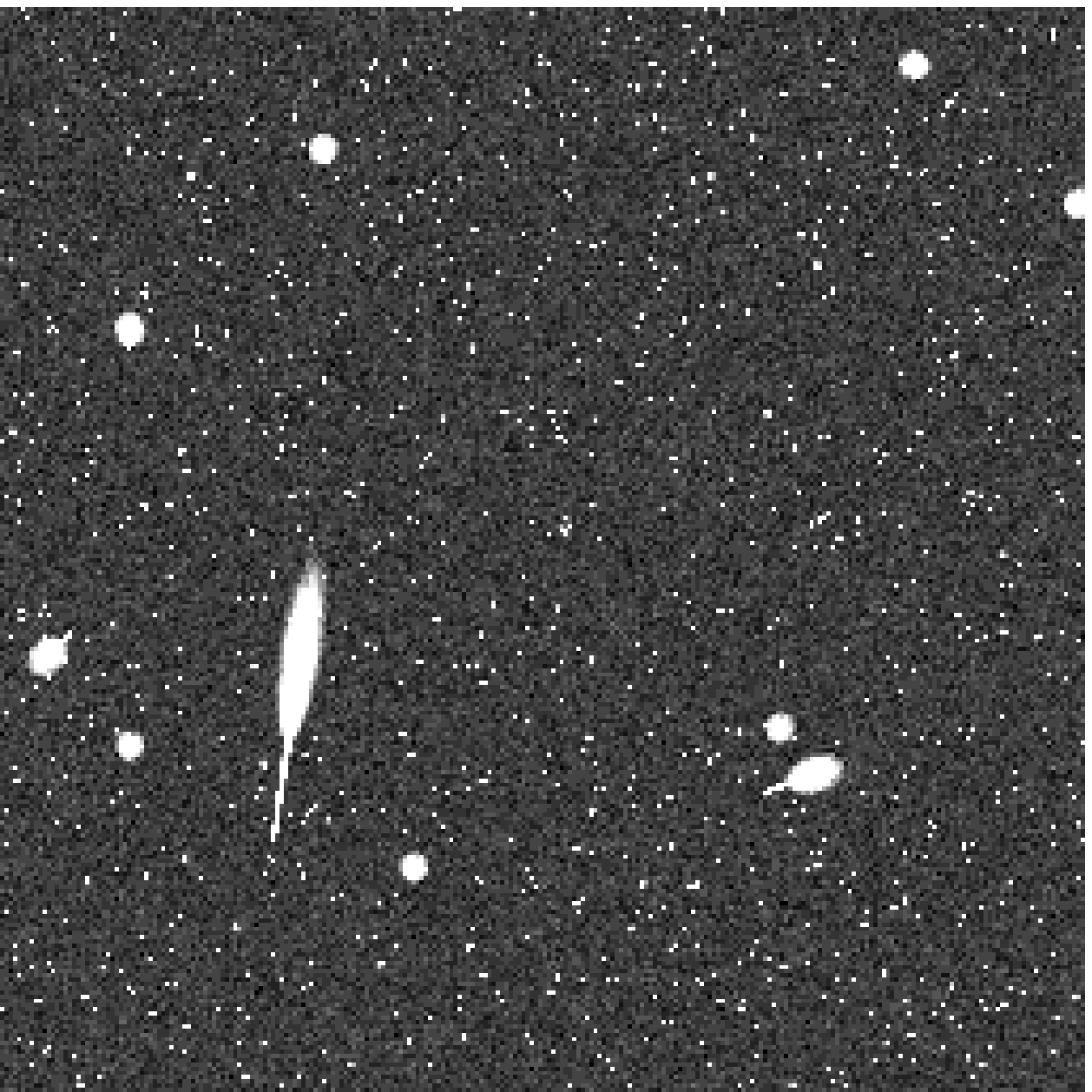,width=3.0in}}}
\caption{\protect \small Examples of simulated generic CCD frames, 
containing 1000 1-keV photon
events and 10 particle events.  A BI device is shown on
the left; a bulk FI device is shown on the right.  The
particle events bloom more in the FI device due to charge diffusion in
the bulk field-free region.} 
\label{fig:sim_images}
\end{figure}

Figure~\ref{fig:sim_images} (left panel) shows a simulated frame for a
generic (CTI-free) BI device with 1000 1-keV photon
events and 10 particle events, assuming the uniform illumination one
would expect in subassembly calibrations.  Note that most of the photon
events are spread among several pixels at this energy because these
low-energy photons interact close to the back surface and their charge
clouds undergo substantial spreading as they travel many microns
through the depletion region toward the gates.  For comparison,
consider the right panel of Figure~\ref{fig:sim_images}, which also
contains 1000 1-keV photon events and 10 particle events, but detected
by a generic FI device.  Most of the photon events are
contained in a single pixel because shallow interactions in an FI
device generate charge clouds that only travel a short distance, thus
suffering minimal spreading.  Note how much more the particle events
have bloomed -- this is a result of the thick substrate assumed for
this device.

\begin{figure}[htb]
\centerline{\mbox{
	\epsfig{file=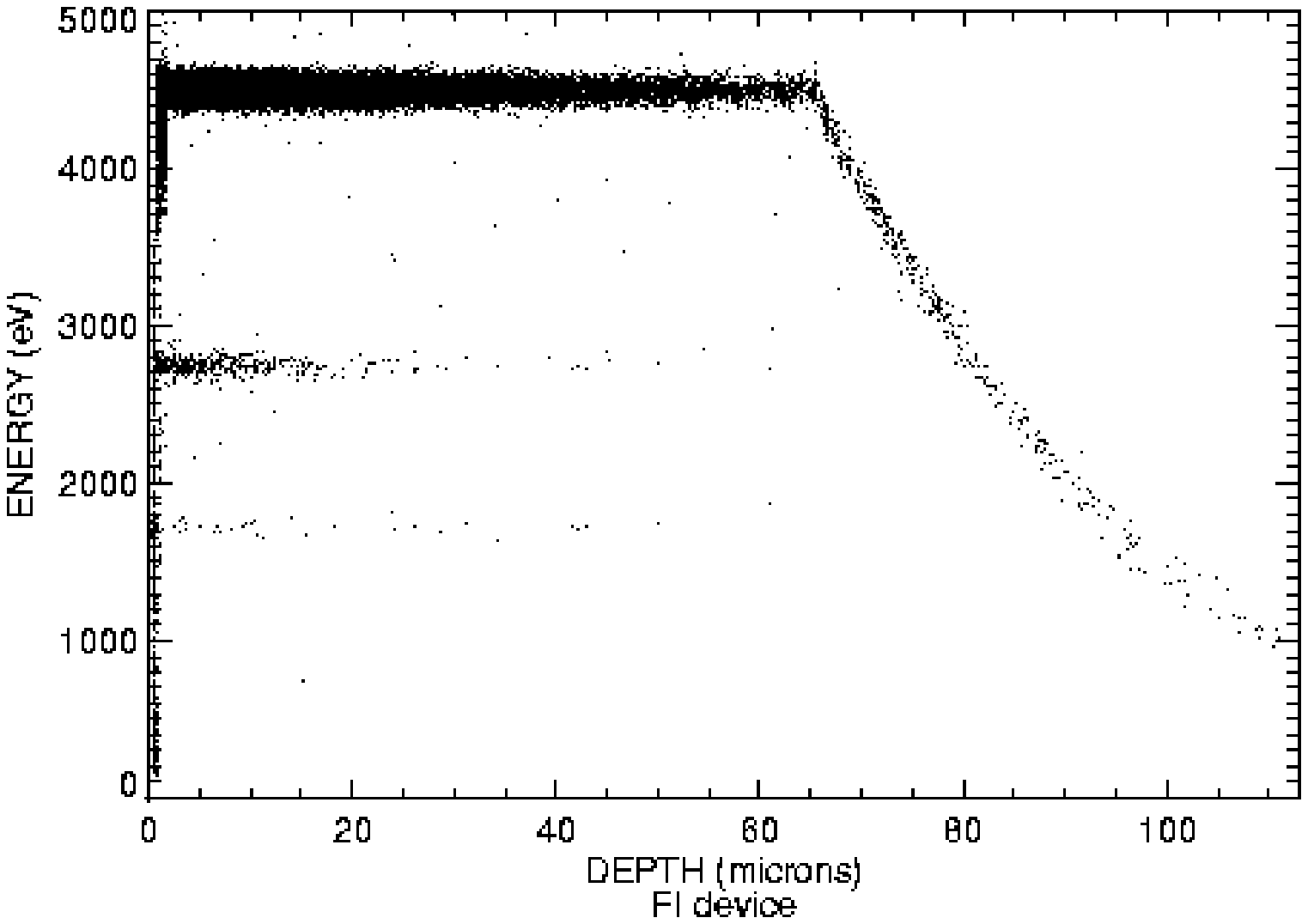,width=3.0in}
	\hspace{0.5in}
	\epsfig{file=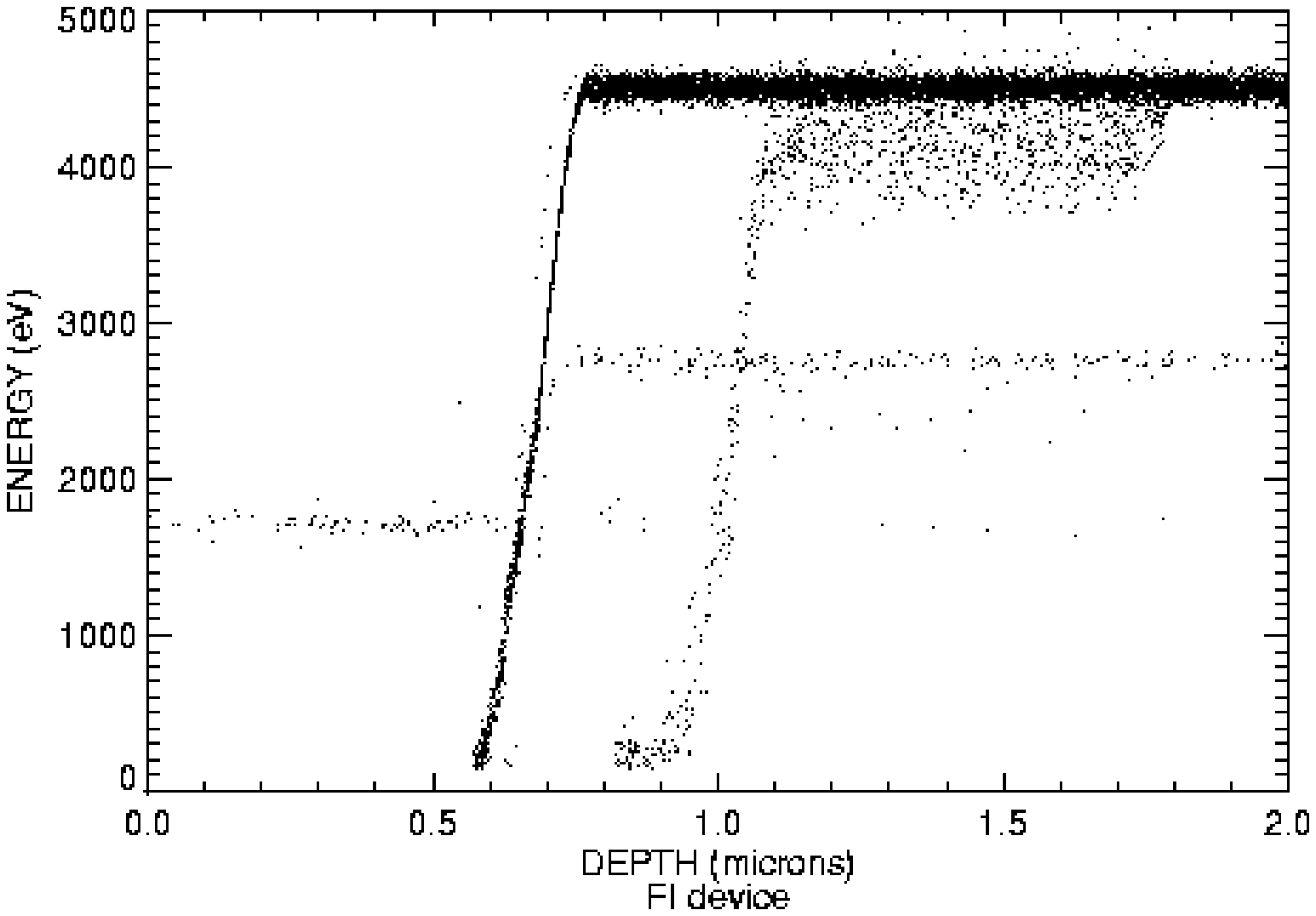,width=3.0in}}}
\centerline{\mbox{
	\epsfig{file=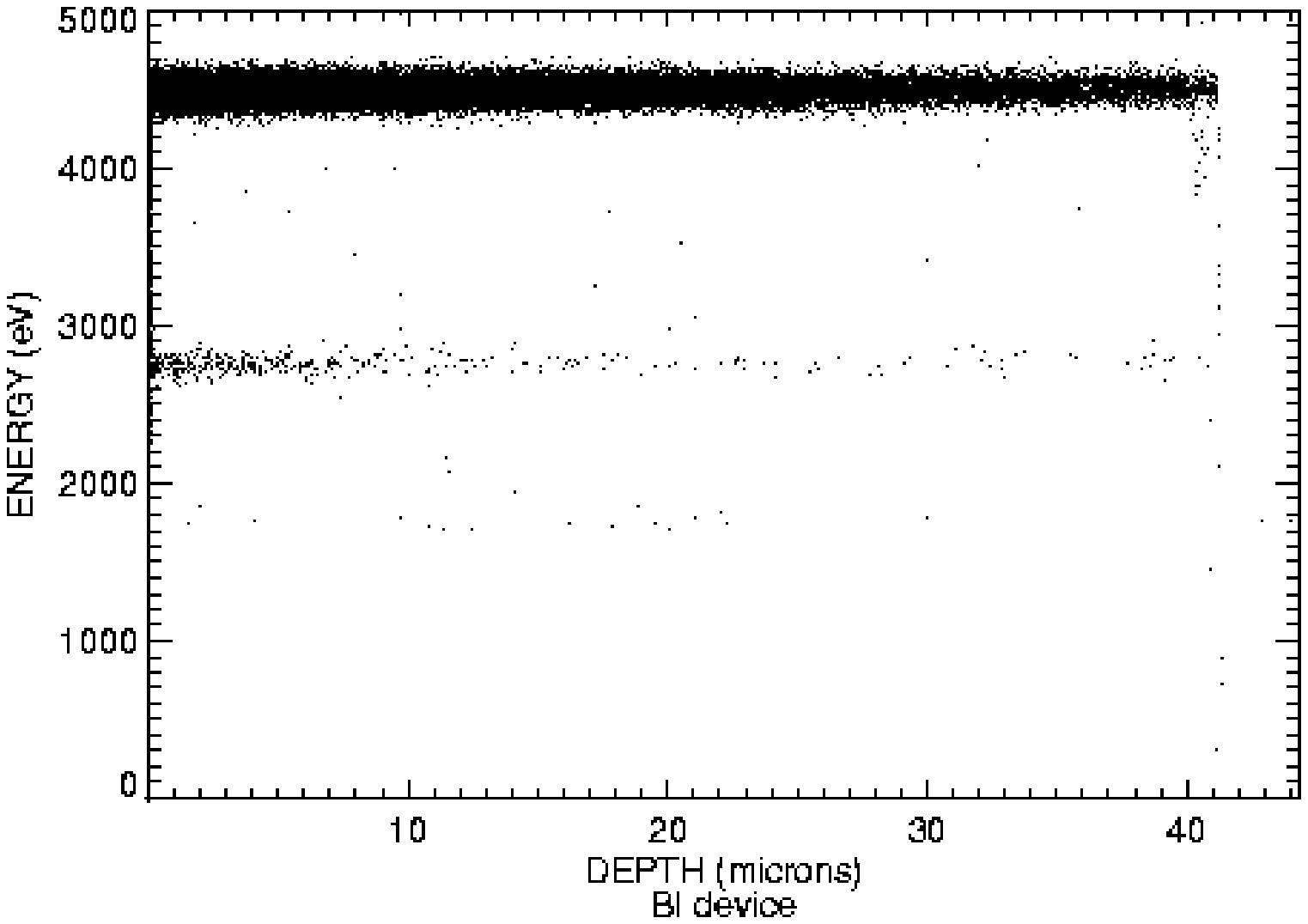,width=3.0in}
	\hspace{0.5in}
	\epsfig{file=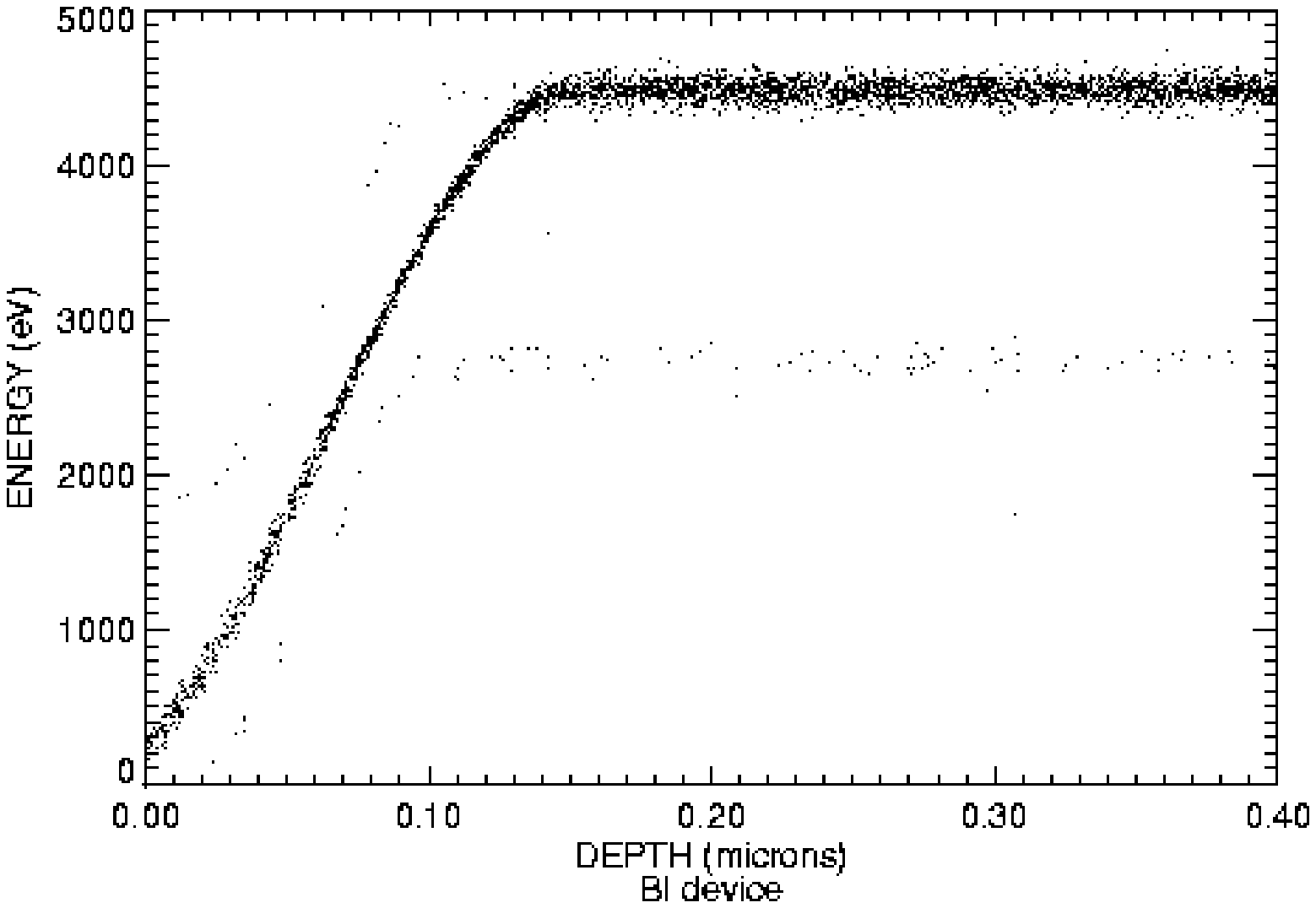,width=3.0in}}}
\caption{\protect \small Photon interaction depths and detected event energies from 
simulations of 4.5~keV photons interacting with ACIS CCDs; upper panels are for
an FI device, lower panels are for a BI device.  Such depth 
information is unfortunately unknowable
for physical CCDs.  The left panels show the full range of depths; the right
panels zoom in on the top surfaces of each device.  The simulated spectra
that correspond to these depth plots are shown in Figures~\ref{fig:specfi4}
and \ref{fig:specbi4}. } 
\label{fig:depths}
\end{figure}

The simulator may optionally attempt to match up photons and events --
there is not a one-to-one correspondence due to fluorescence and other
effects -- and then save for each event the photon's interaction
position in three dimensions.  The interaction depths, plotted against
event energy, have been very helpful in diagnosing what layer of the
device is responsible for certain spectral features in the
redistribution function.  Figure~\ref{fig:depths} illustrates this for
both FI and BI devices with a simulation of 4.5~keV photons impacting
ACIS.  The simulated spectra that correspond to these plots are shown
in the next section.

The left panels of Figure~\ref{fig:depths} show the full range of
photon interaction depths.  Although most photons interact in the
depletion regions of both devices (as designed) and generate events
that populate the main photopeak (dark locus of points at 4500~eV in
the figure), these plots clearly illustrate the origins of other
features in the spectral redistribution function.  For example, the
locus at $\sim$2700~eV comprises the escape peak; the locus at 1740~eV
represents fluorescent photon events that populate the $Si$
fluorescence peak.

In the FI device (top), events lose charge in the bulk field-free
substrate (65$\mu$m and deeper) due to charge spreading and add to the
continuum; many of these events will be removed by grade filtering, but
some remain in the standard g02346 grades and contribute to the soft
shoulder in the spectrum of high-energy photons.  The thinned BI device
(bottom) has no field-free substrate to contribute to the soft shoulder
and the data confirm that this feature is much reduced in BI spectra.

The right panels of Figure~\ref{fig:depths} show the importance of the
gate structure and channel stops in the FI spectral redistribution
function and that of the damage and epitaxial layers in the BI spectral
redistribution function.  Events partially interacting in these surface
layers populate the continuum in the spectra and are shown in the depth
plots as the steep-sloped components rising from the event detection
threshold up to the main peak at 4500~eV.  In FI devices, the channel
stops are near the illuminated surface of the CCD, so a nontrivial
fraction of events interact there and populate the soft shoulder just
below the main peak in the spectrum; the top right panel shows these
events at interaction depths of 1.1--1.8$\mu$m.  In BI devices, fewer
events make it through the entire depth of the device to encounter the
channel stops; the lower left panel of Figure~\ref{fig:depths} shows
them at depths of 40--41$\mu$m.  Plots such as these were examined for
simulated datasets at a wide range of energies and used to confirm how
$SiO_2$ and channel stop tuning parameters affected the spectral
response.

\section{Comparing Simulations to Data} \label{sec:comparisons}

To illustrate the performance of the model, below are various examples
of simulated output compared to XRCF data.  The main
comparitors are spectra, quantum efficiency (QE) and grade
distributions.  In addition, the BI device requires a comparison of
simulated and real CTI effects (see the companion paper).  Since the FI
chips showed minimal CTI at XRCF, we can eliminate this factor when
comparing to XRCF data.  This allows us to instantiate the CCD model
without the confusing effects of CTI, then separately add in CTI effects
to reproduce on-orbit data.  These FI CTI effects will also be described in 
detail in the companion CTI paper.


\subsection{An ACIS FI Device}

Here we compare simulations and XRCF data for the ACIS I3 chip.  Data
from all four amplifiers are combined.  Note that all spectra are
presented using log-linear scales; most of the events are contained in
the main spectral peak, but most of the insight into the device physics is
to be gained by studying the relatively few events that appear at
lower energies.

\begin{figure}[htb] 
 
\centerline{\mbox{
	\epsfig{file=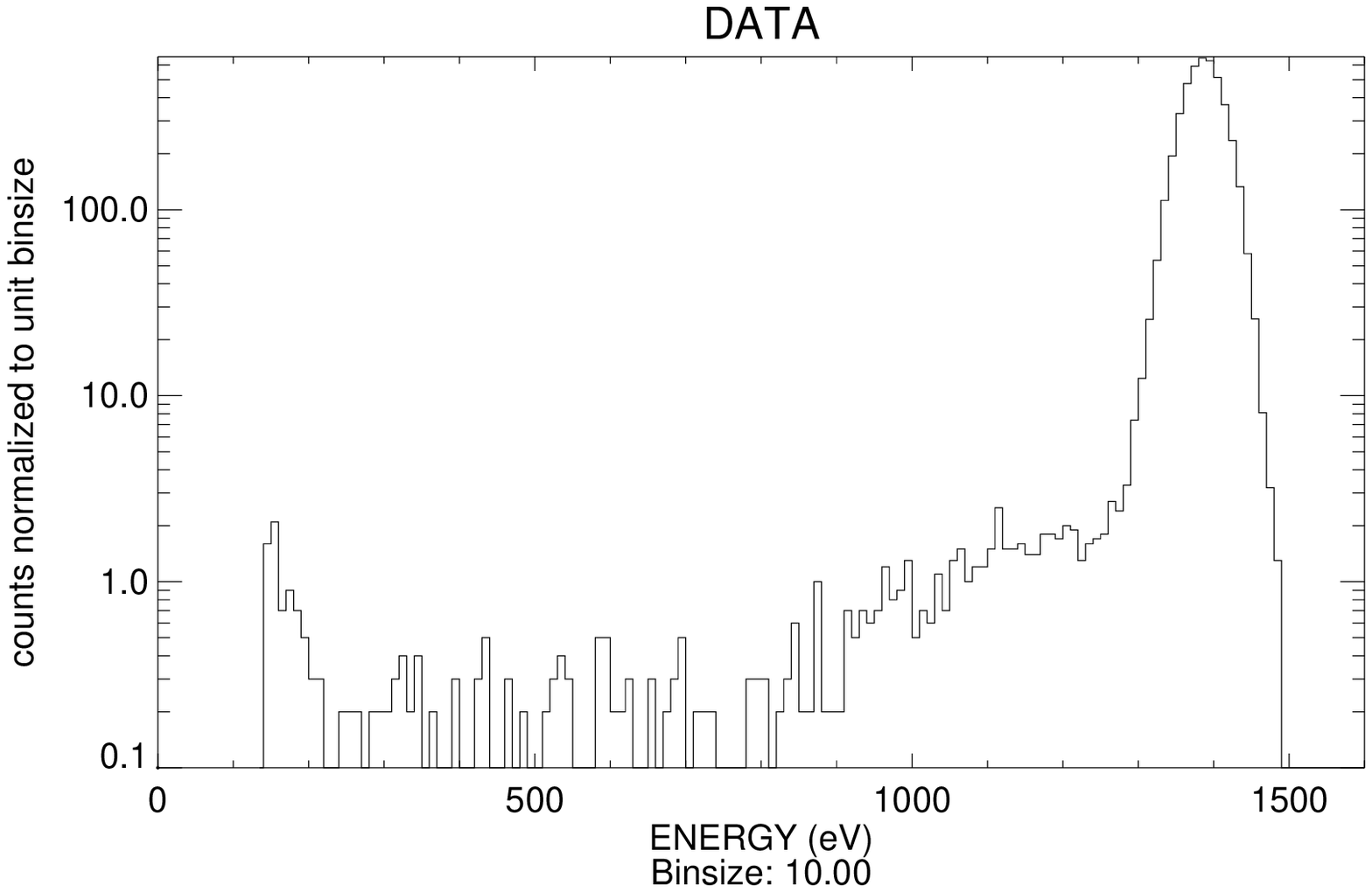,width=3.0in}
	\hspace{0.5in}
	\epsfig{file=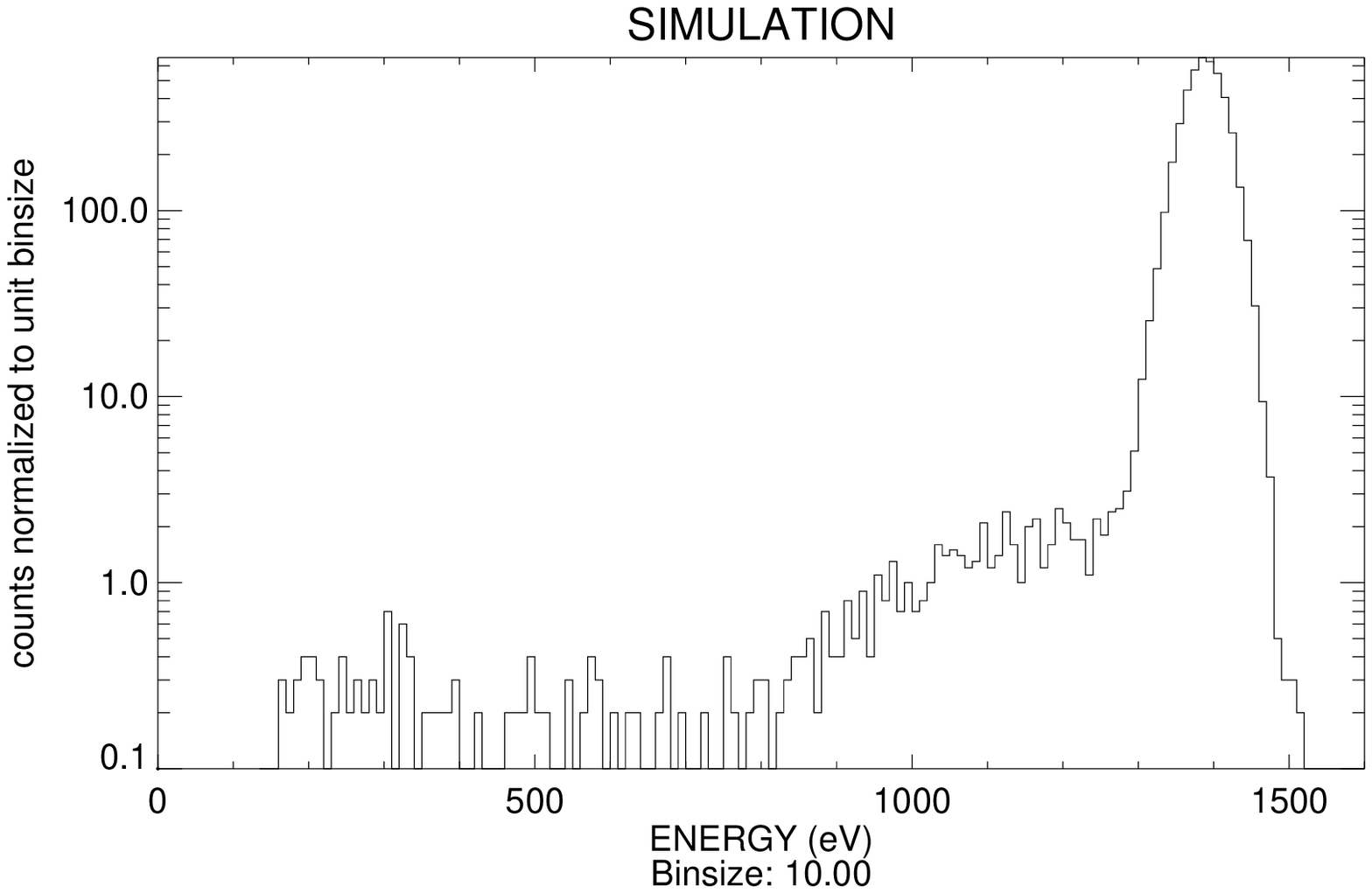,width=3.0in}}}
 
\caption{\protect \small Spectrum of XRCF Phase I data for an FI chip (left) and the 
corresponding simulation (right).  Photon energy
is 1.383~keV; event lists were grade filtered to keep only ASCA g02346, resulting
in $\sim 45100$ events in each spectrum. }
 
\normalsize
\label{fig:specfi1}
\end{figure}

Figure~\ref{fig:specfi1} shows a spectrum of 1.383~keV Phase I data
with uniform illumination taken on the I3 chip and its corresponding
simulation.  The low-energy peak is larger in the data than in the
simulation because of high-energy continuum photons from the source
interacting in the gate structure of the CCD.  The simulation models a
monochromatic input spectrum at 1.383~keV rather than the actual source
spectrum.
  
\begin{figure}[htb] 
 
\centerline{\mbox{
	\epsfig{file=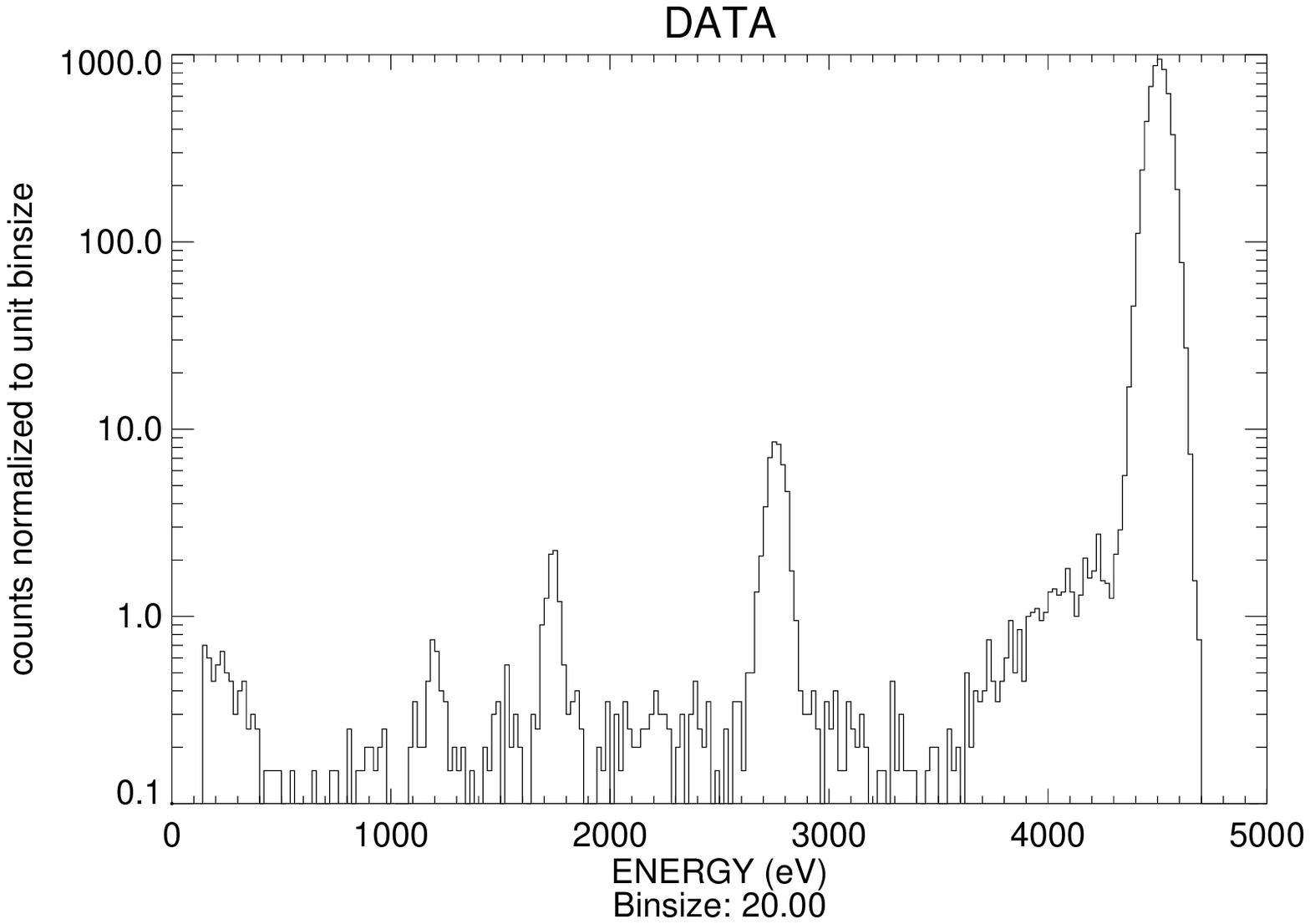,width=3.0in}
	\hspace{0.5in}
	\epsfig{file=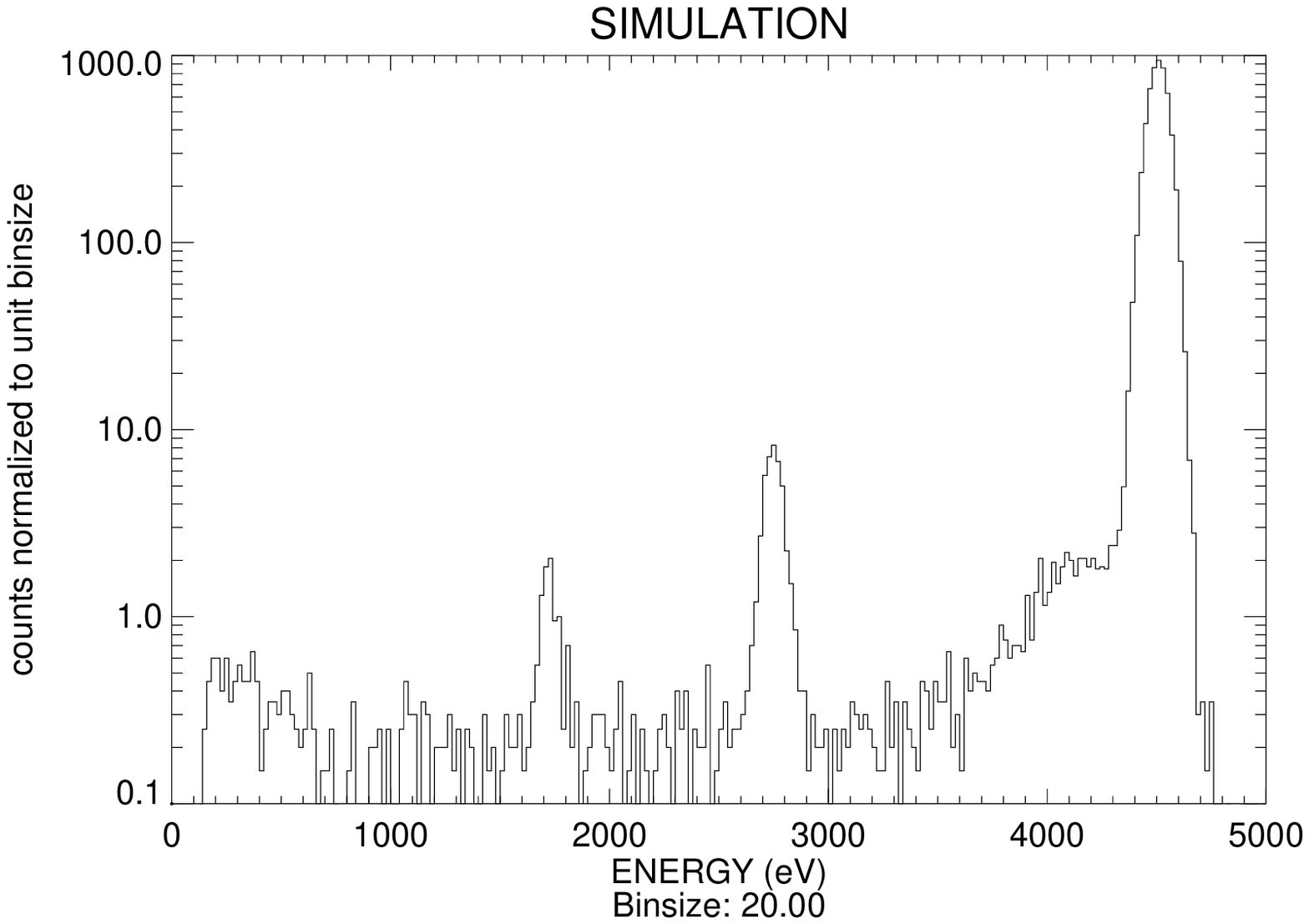,width=3.0in}}}
 
\caption{\protect \small Spectrum of XRCF Phase I data for an FI chip (left) and the 
corresponding simulation (right).  Photon energy
is 4.5~keV; event lists were grade filtered to keep only ASCA g02346, resulting
in $\sim 112500$ events in each spectrum.  Note the Ge~L complex near 1.2~keV
that is present in the data but not modeled in the simulation.}
 
\normalsize
\label{fig:specfi4}
\end{figure}

Figure~\ref{fig:specfi4} gives a spectrum of 4.5~keV Phase I data with
uniform illumination taken on the I3 chip and its corresponding
simulation.  The peak at about 1.2~keV in the data is not reproduced by
the simulator because it is due to a Ge~L line complex in the source
rather than spectral redistribution by the CCD.

\begin{figure}[htb] 
 
\centerline{\mbox{
	\epsfig{file=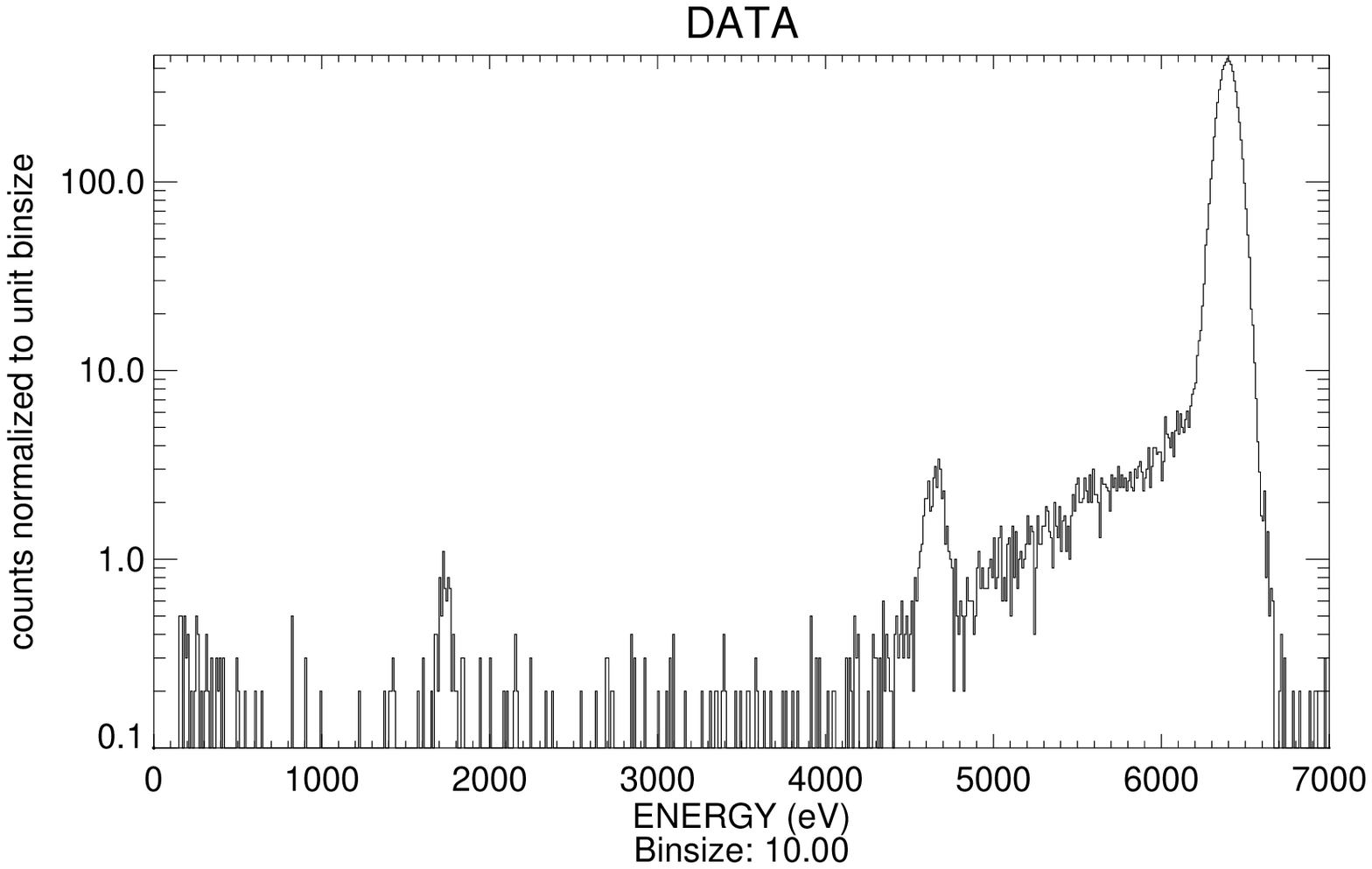,width=3.0in}
	\hspace{0.5in}
	\epsfig{file=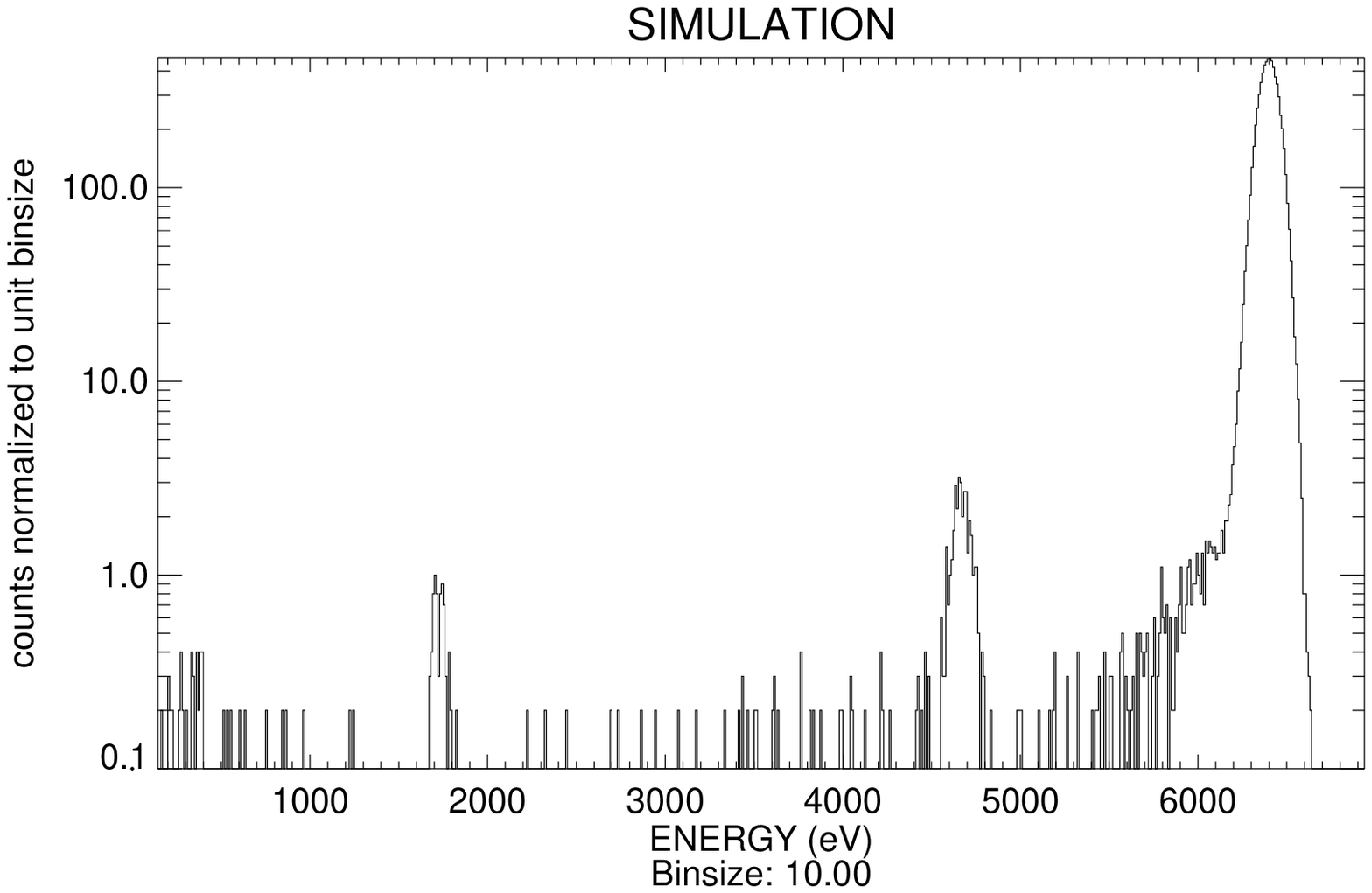,width=3.0in}}}
 
\caption{\protect \small Spectrum of XRCF Phase I data for an FI chip
(left) and the corresponding simulation (right).  Photon energy is
6.404~keV; event lists were grade filtered to keep only ASCA g02346,
resulting in $\sim 69100$ events in each spectrum.}
 
\normalsize
\label{fig:specfi6}
\end{figure}

Figure~\ref{fig:specfi6} illustrates how continuum emission from the
XRCF source can contaminate the calibration spectrum.  The continuum
emission is due to leakage of a bremmsstrahlung continuum through the
source filters, which explains why the contamination is preferentially
to the soft (low energy) side of the line.  The 6.404~keV data show a
much larger soft shoulder than the simulation, due to this
contamination.  Such effects limit our ability to tune the simulator to
reproduce subtle spectral redistribution features.

Table~\ref{table:gradesfi} compares the grade distributions of the data to
the simulated event list for the three example energies given above.
Only the standard ASCA grades are listed because contamination from the
source continuum causes grades 1, 5, and 7 to be overpopulated in the
data.  Again, the tuning metric used was the number of g0 events only.
The simulator keeps symmetry in the number of right- and
left-hand split events (Grades 3 and 4) with respect to the number
of up and down split events (Grade 2): g2 $\simeq$ g3 + g4.  The asymmetry 
in the data (g2 $>$ g3 + g4) could be due to the complicated geometry of the
gate structure or channel stops or other effects not contained in the
simulations.

\begin{table}[htb] \centering

\begin{tabular}{||c|c|c|c|c|c|c||} \hline

ASCA	& \multicolumn{2}{c|}{1.383~keV} & \multicolumn{2}{c|}{4.5~keV}  & \multicolumn{2}{c||}{6.404~keV}\\    
grade   & Data (\%) & Simulation (\%) & Data (\%) & Simulation (\%) & Data (\%) & Simulation (\%)\\ \hline \hline
 
0  &  81.6 $\pm$0.4  &  81.3 $\pm$0.4 &  63.9 $\pm$0.3  &  63.9 $\pm$0.3 &  41.6 $\pm$0.2  &  41.9 $\pm$0.2 \\ \hline
2  &  10.3 $\pm$0.2  &  9.2 $\pm$0.1 &  16.6 $\pm$0.1  &  15.7 $\pm$0.1 &  18.8 $\pm$0.2  &  20.7 $\pm$0.2 \\ \hline
3  &   3.3 $\pm$0.1  &   4.2 $\pm$0.1 &   5.9 $\pm$0.1  &   7.6 $\pm$0.1 &   7.3 $\pm$0.1  &   10.4 $\pm$0.1 \\ \hline
4  &   3.2 $\pm$0.1  &   4.3 $\pm$0.1 &   5.8 $\pm$0.1  &   7.8 $\pm$0.1 &   7.3 $\pm$0.1  &   10.2 $\pm$0.1 \\ \hline
6  &   1.6 $\pm$0.1  &   1.1 $\pm$0.1 &   7.8 $\pm$0.1  &   5.1 $\pm$0.1 &   25.0 $\pm$0.2  &   16.8 $\pm$0.2 \\ \hline
 
\end{tabular}
\caption{\protect \small A comparison of real and simulated grade distributions for the I3 chip at sample energies.  The simulator was tuned
to reproduce the number of Grade 0 events in the main spectral peak.}
\normalsize
\label{table:gradesfi}
\end{table}

The quantum efficiency of the simulator can be measured with vastly
less effort than that required to measure the QE of real ACIS devices.
Here we use the predicted QE as a way to compare the PSU CCD simulator to the
MIT/CSR simulator.  For the PSU simulation, a known number of photons
are simulated, including CTI and excluding events with grades that are
not telemetered by the ACIS instrument, then corrected using our CTI
corrector.  The resulting events are grade filtered (g02346).  Then the
number of g02346 events is computed and divided by the number of input
photons.  This simple definition of QE thus includes events that have
been spectrally redistributed out of the main photopeak.

\begin{figure}[htb]
\centerline{\epsfig{file=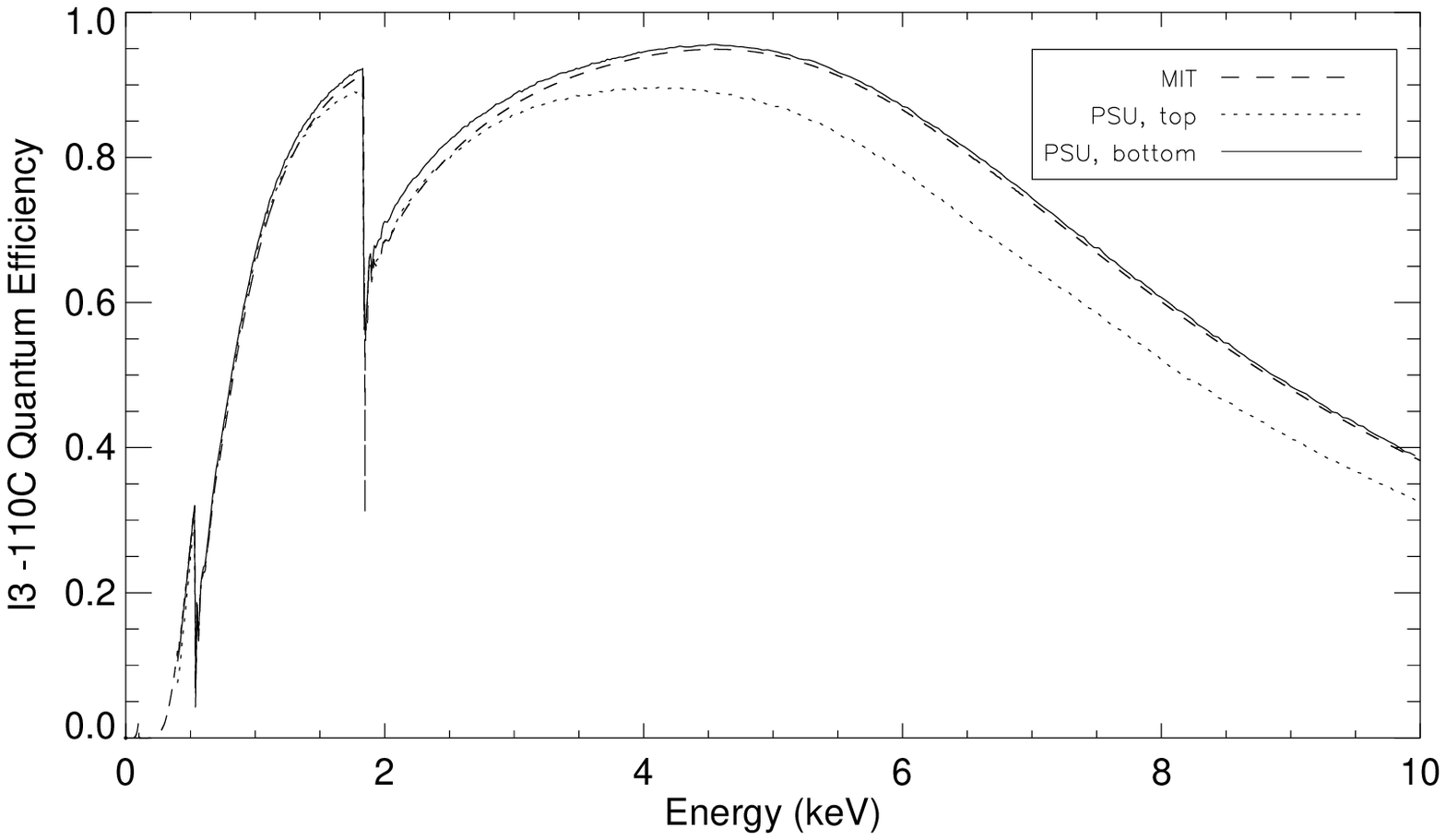, height=3in}}
\centerline{\epsfig{file=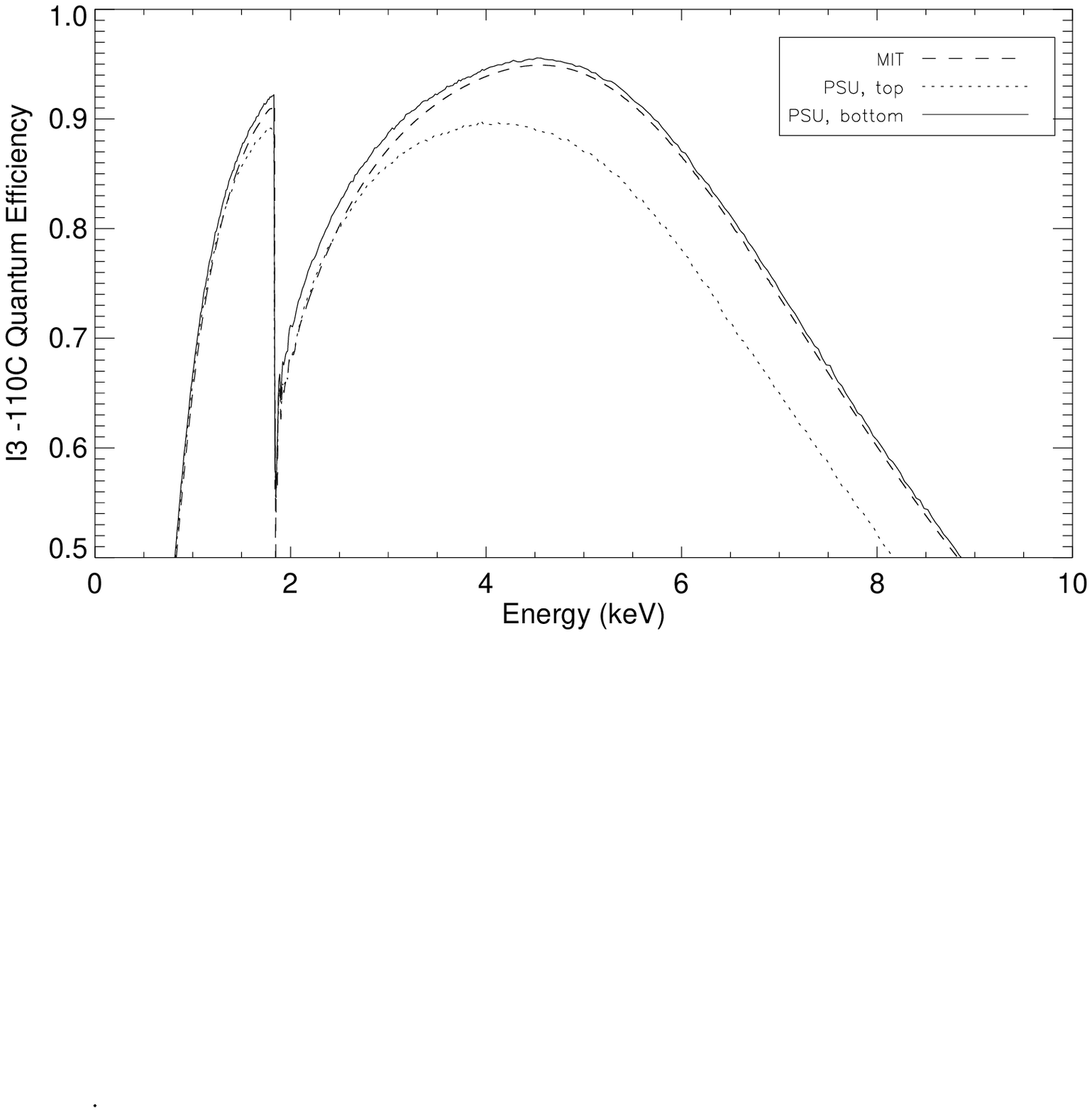, height=3in}}

\caption{\protect \small $-$110C QE curves from the PSU FI simulator
compared to the standard data product for the ACIS I3 FI device (based
on the MIT FI simulator and calibration data).  The dashed line is the
standard QE curve.  The dotted line shows the QE predicted by the PSU
simulator for the top 128 rows of the I3 device (near the aimpoint).
The solid line shows the QE predicted by the PSU simulator for the
bottom 128 rows of the I3 device (those least affected by CTI).  The
top figure panel shows the full range; the bottom panel highlights the
areas where the QE is above 50\%.}

\normalsize
\label{fig:qe3}
\end{figure}

The resulting QE curves, for the bottom and top 128 rows of the I3 FI
device at $-$110C, are shown in Figure~\ref{fig:qe3} along with that
reported for the spatially averaged (full-CCD) pre-launch ACIS I3
device by MIT/CSR \cite{calreport}.  The MIT QE curve is calculated
from their CCD simulator results, based on the detailed absolute and
relative QE measurements made on the ACIS reference and flight CCDs at
the BESSY synchrotron and MIT labs \cite{calreport}.  The pointwise
deviations in the PSU estimates reflect counting statistics; each point
is obtained from simulations of $\sim 1 \times 10^5$ events.  The MIT
release notes state that the spatial variations of QE were less than
$\pm$5\% for FI devices, but subsequent radiation damage changes this
result.  

The PSU curve for the bottom of the device (the area least affected by
CTI) is useful for comparing to the MIT results; this curve generally
falls within the 5\% deviations expected for pre-launch devices.  We do
not tune the layer thicknesses to try to get a specific QE performance
from the PSU simulator.  The discrepancies between our simulated QE and
the MIT results might be due in part to slight differences in our
methods for calculating QE or our simplified model of the gate
structure.  A detailed error analysis of the MIT QE calculation is
given in the ACIS Calibration Report \cite{calreport}.

The curve for the top of the device (the area most affected by CTI)
clearly deviates from the pre-launch predictions, even after CTI
correction.  This is because CTI smears the charge distributions,
causing events to migrate into grades that are not telemetered to the
ground, thus they are not available for CTI correction.  This effect is
only important for data gathered at a focal plane temperature of
$-$110C (used only in the first few months of the mission).  By
lowering the temperature to $-$120C, FI CTI effects were mitigated
enough that the FI QE does not exhibit this spatial dependence.
We show the $-$110C simulation here as a warning to users of early
ACIS data that the spatial dependence of the QE cannot be ignored.

	

\subsection{An ACIS BI Device}	
	
Inherent in the BI simulation is a model for the CTI.  Since this is
detailed in the companion paper \cite{townsley01b}, we will not dwell
on the CTI model here.  Its fidelity does affect the comparison
diagnostics presented below, however.  Subtle spectral features obvious
in FI XRCF data become apparent in BI data when CTI is accounted for,
although the origins of such features are not always exactly the same.

\begin{figure}[htb] 
 
\centerline{\mbox{
	\epsfig{file=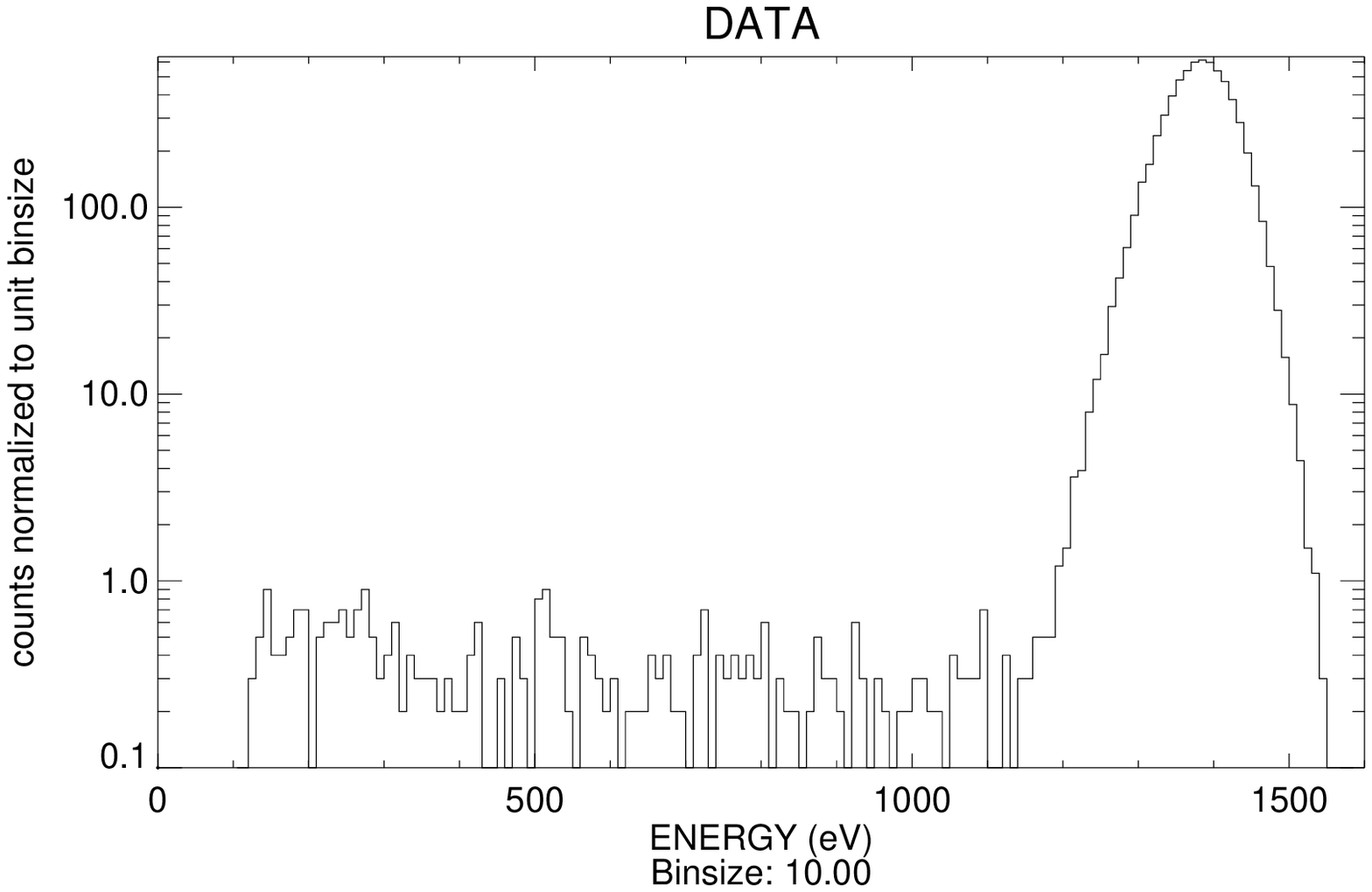,width=3.0in}
	\hspace{0.5in}
	\epsfig{file=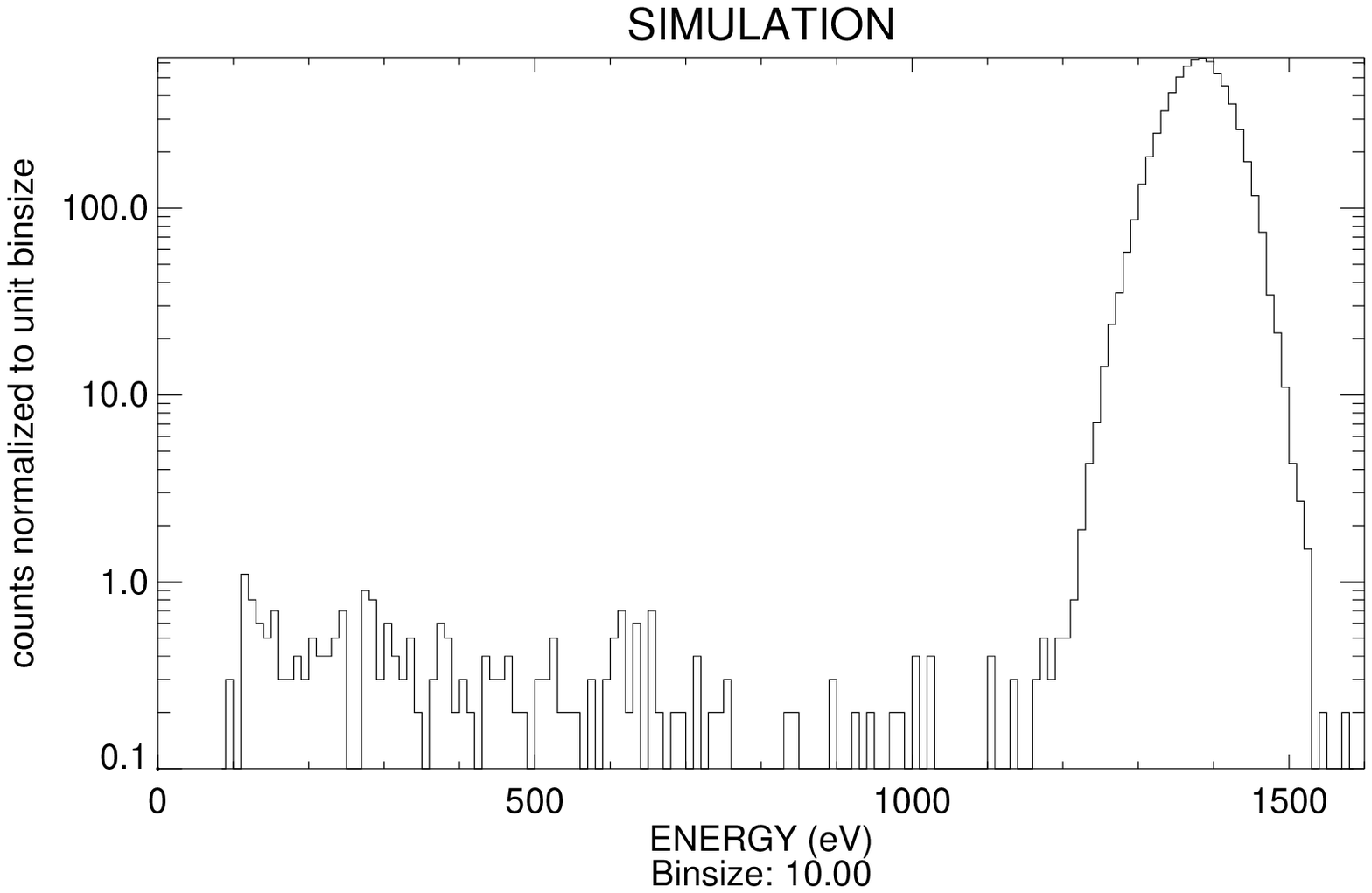,width=3.0in}}}
 
\caption{\protect \small Spectrum of XRCF Phase I data for a BI chip
(left) and the corresponding simulation (right).  Photon energy is
1.383~keV; event lists were grade filtered to keep only ASCA g02346,
resulting in $\sim 65700$ events in each spectrum. }
 
\normalsize
\label{fig:specbi1}
\end{figure}

Figure~\ref{fig:specbi1} (left) gives a spectrum of CTI-corrected
1.383~keV data with uniform illumination taken on a BI chip.  On the
right is the corresponding simulation of a CTI-free device.  Noticeably
absent (compared to Figure~\ref{fig:specfi1}) is the soft shoulder just
below the main peak; since the channel stops are far from the
illuminated surface on BI devices, fewer events interact there thus the
soft shoulder is less populated.

\begin{figure}[htb] 
 
\centerline{\mbox{
	\epsfig{file=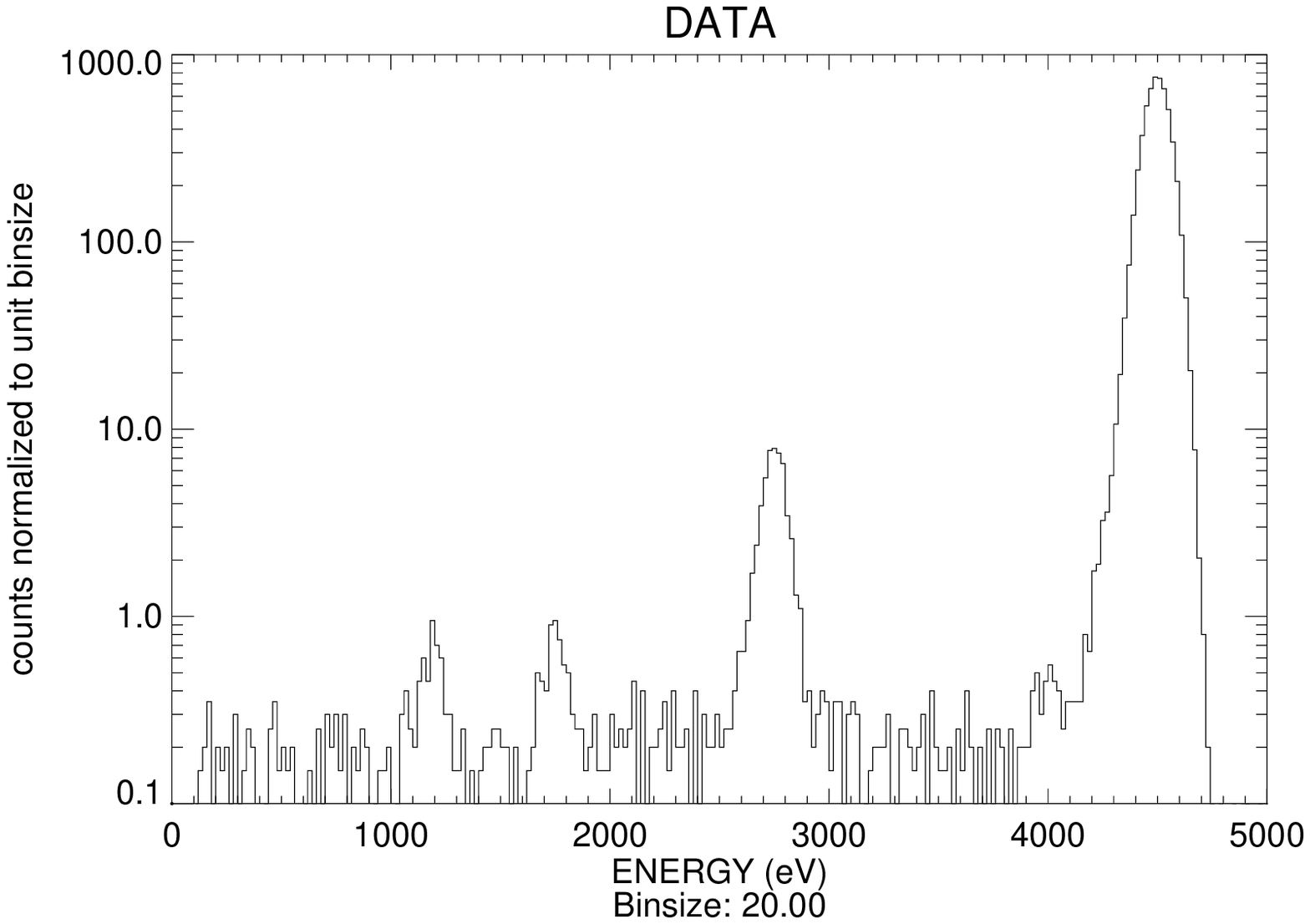,width=3.0in}
	\hspace{0.5in}
	\epsfig{file=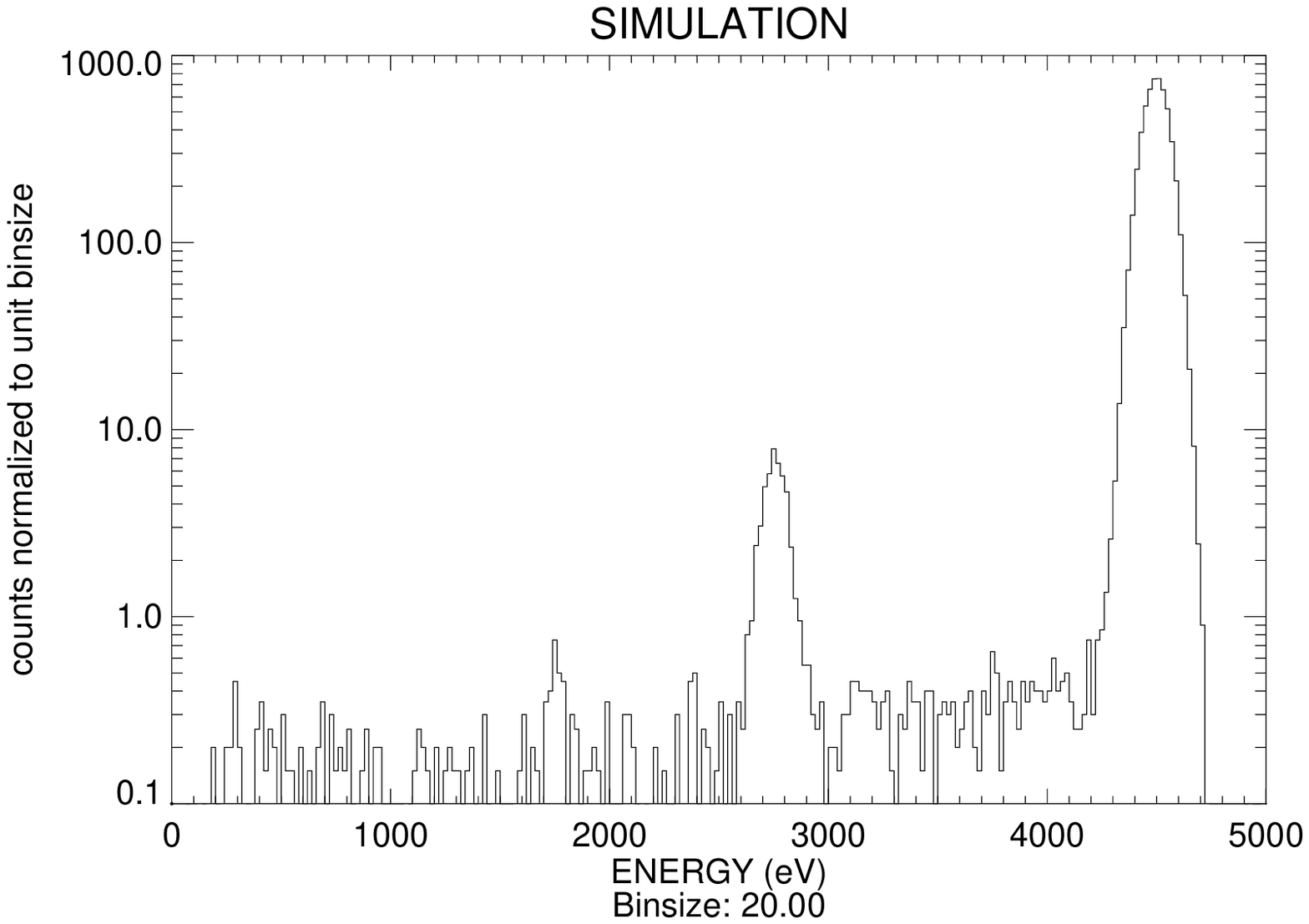,width=3.0in}}}
 
\caption{\protect \small Spectrum of CTI-corrected XRCF Phase I data
for a BI chip (left) and the corresponding simulation (right).  Photon
energy is 4.5~keV; event lists were grade filtered to keep only ASCA
g02346.  We show $\sim 112500$ events in each spectrum for comparison
with the FI spectra in Figure~\ref{fig:specfi4}.}
 
\normalsize
\label{fig:specbi4}
\end{figure}

Figure~\ref{fig:specbi4} (left) shows a spectrum of CTI-corrected
4.5~keV data with uniform illumination taken on a BI chip, with the
CTI-free simulation on the right.  As expected, the Ge~L complex at
about 1.2~keV in the data is not reproduced by the simulator.  Although
the width of the main peak in the simulation is tuned to match that in
the data, the main peak in the data appears to have broader wings than
in the simulation.  This is probably due to residual CTI effects ({\em e.g.}
the non-uniform trap distribution results in column-to-column variations
in the charge loss; incomplete correction for these variations contributes
to wings in the spectral lines).

In spite of the extra $SiO_2$ artificially added below the gate
structure of the device to make a ``virtual dead layer'' for generating
fluorescent photons, we are still underestimating the amplitude of the
$Si$ fluorescence peak at this and higher energies.  This implies that
the source of these fluorescent photons is not some silicon compound
lying below the gate structure in the BI device.  Again the $Si$ rim
around the device could be the source, but finding a spatial signature
of this effect in the XRCF data is difficult because we don't have
full-frame data for most energies and the photon statistics are quite
limited.

\begin{figure}[htb] 
 
\centerline{\mbox{
	\epsfig{file=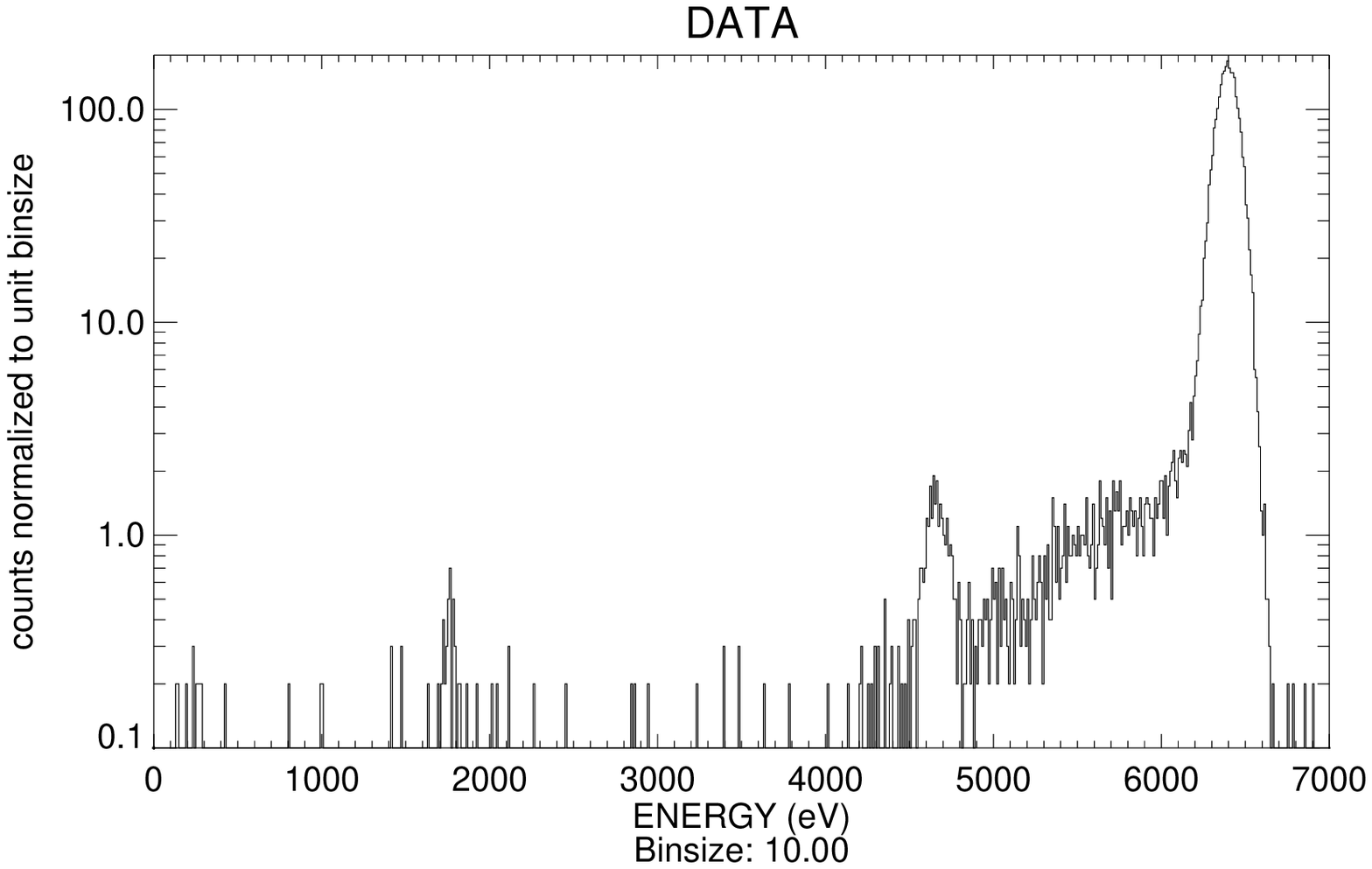,width=3.0in}
	\hspace{0.5in}
	\epsfig{file=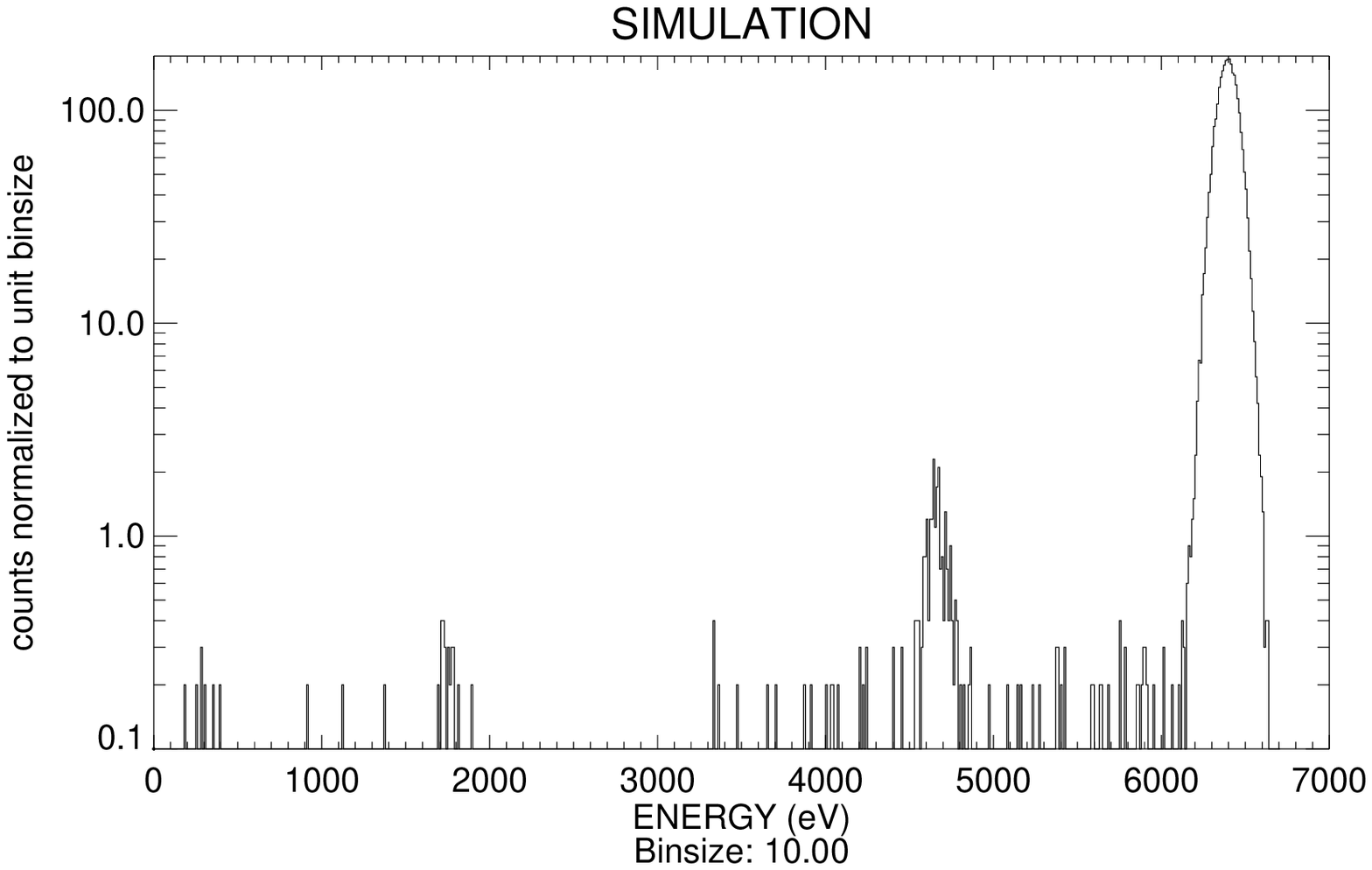,width=3.0in}}}
 
\caption{\protect \small Spectrum of XRCF Phase I data for a BI chip
(left) and the corresponding simulation (right).  Photon energy is
6.404~keV; event lists were grade filtered to keep only ASCA g02346,
resulting in $\sim 28500$ events in each spectrum.}
 
\normalsize
\label{fig:specbi6}
\end{figure}

Figure~\ref{fig:specbi6} confirms our assertion above that continuum
emission from the XRCF source is contaminating the calibration spectrum
at this energy.  The BI 6.404~keV data show a much larger soft shoulder
than the simulation due to this contamination, just as the FI data
showed (Figure~\ref{fig:specfi6}).

Table~\ref{table:gradesbi} compares the grade distributions of the BI data
to the simulated event list.  Again, the tuning metric used was the
number of g0 events in the main spectral peak.  The data show slightly
fewer single-split (ASCA grades 2, 3, or 4) events than the simulation
and correspondingly more g6 events; the overall match is comparable to
the FI examples above.   

\begin{table}[htb] \centering

\begin{tabular}{||c|c|c|c|c|c|c||} \hline

ASCA	& \multicolumn{2}{c|}{1.383~keV} & \multicolumn{2}{c|}{4.5~keV}  & \multicolumn{2}{c||}{6.404~keV}\\    
grade   & Data (\%) & Simulation (\%) & Data (\%) & Simulation (\%) & Data (\%) & Simulation (\%)\\ \hline \hline
 
0  &  27.6 $\pm$0.2   &  27.8 $\pm$0.2  &  19.5 $\pm$0.2   &  19.5 $\pm$0.2 &  24.4 $\pm$0.3  &  25.2 $\pm$0.3 \\ \hline
2  &  26.4 $\pm$0.2   &  25.9 $\pm$0.2  &  23.0 $\pm$0.2   &  23.8 $\pm$0.2 &  20.5 $\pm$0.3  &  23.0 $\pm$0.3 \\ \hline
3  &   12.1 $\pm$0.1  &   13.1 $\pm$0.1 &   10.0 $\pm$0.1  &   11.9 $\pm$0.1 &   8.7 $\pm$0.2  &   11.3 $\pm$0.2 \\ \hline
4  &   12.8 $\pm$0.1  &   12.9 $\pm$0.1 &   11.3 $\pm$0.1  &   11.9 $\pm$0.1 &   9.9 $\pm$0.2  &   11.2 $\pm$0.2 \\ \hline
6  &   21.2 $\pm$0.2  &   20.3 $\pm$0.2 &   36.2 $\pm$0.2  &   32.9 $\pm$0.2 &   36.5 $\pm$0.4  &   29.3 $\pm$0.3 \\ \hline
 
\end{tabular}
\caption{\protect \small A comparison of real and simulated grade distributions for the S3 chip at sample energies.  The simulator was tuned
to reproduce the number of Grade 0 events in the main spectral peak.}
\normalsize
\label{table:gradesbi}
\end{table}

\begin{figure}[htb]

\centerline{\epsfig{file=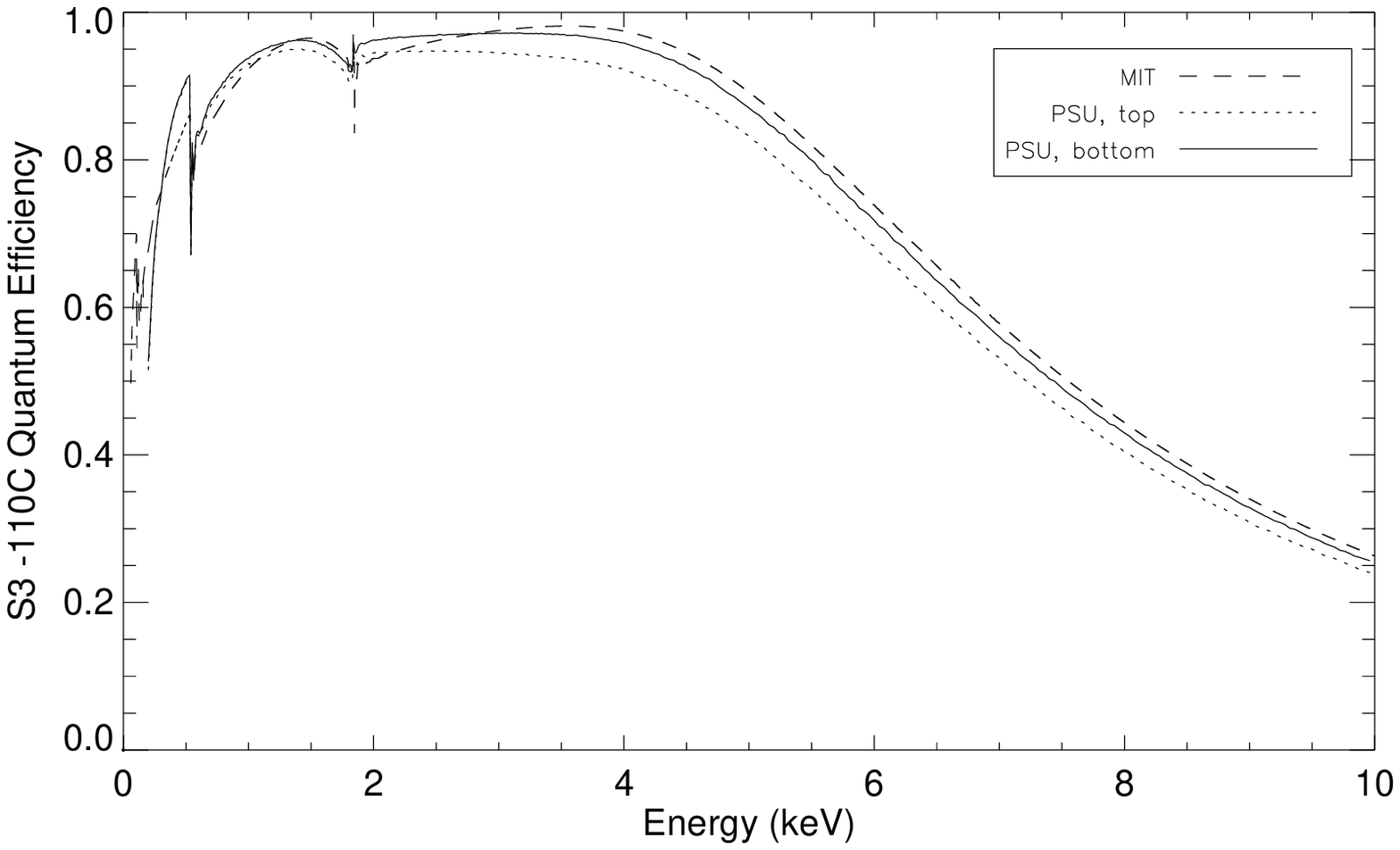, height=2.5in}}
\centerline{\epsfig{file=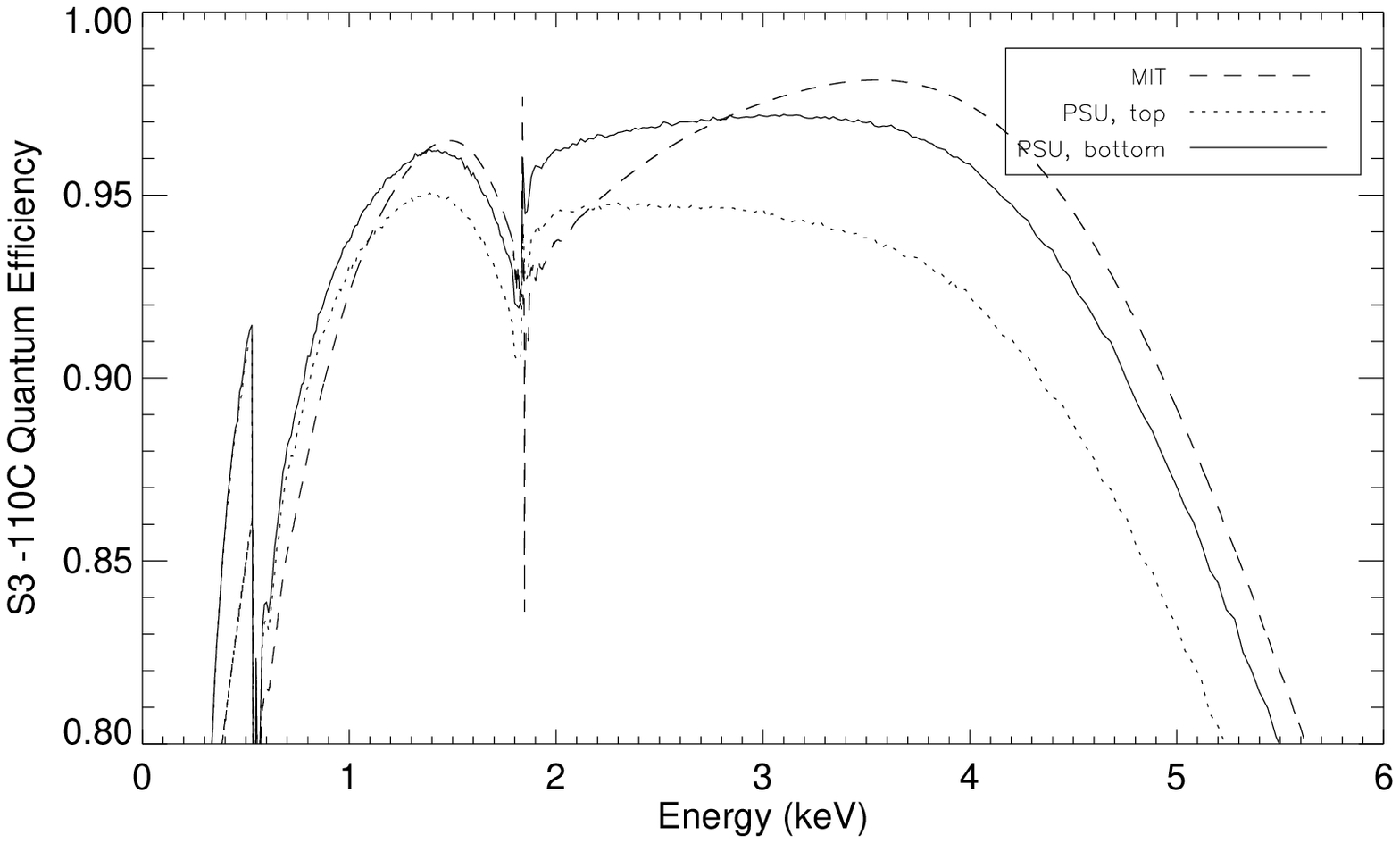, height=2.5in}}
\centerline{\epsfig{file=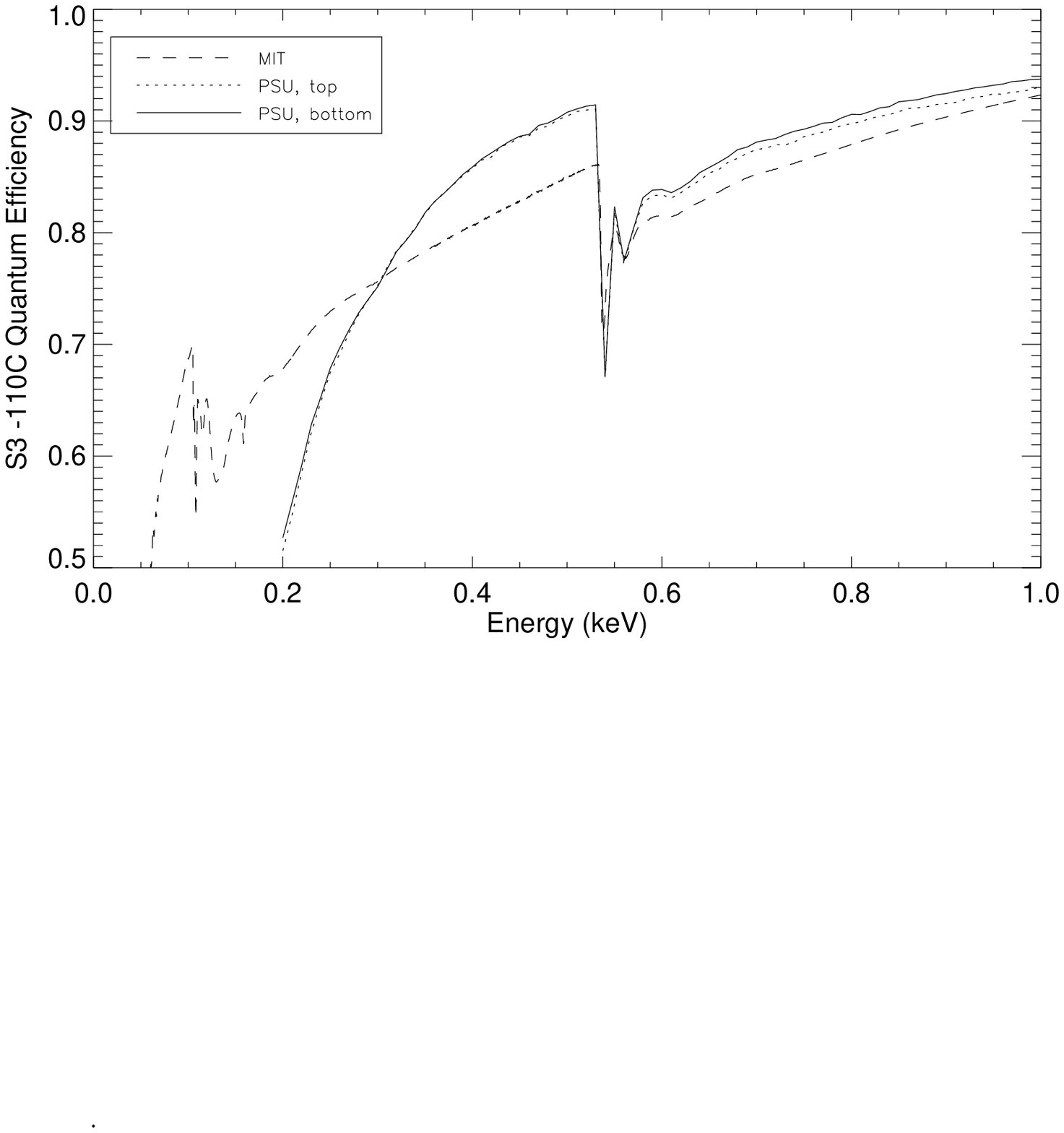, height=2.5in}} 

\caption{\protect \small $-$110C QE curves from the PSU BI simulator
compared to the standard data product for the ACIS S3 BI device (based
on the MIT BI simulator and calibration data).  The dashed line is the
standard QE curve.  The dotted line shows the QE predicted by the PSU
simulator for the top 128 rows of the S3 device (those most affected by
combined parallel and serial CTI).  The solid line shows the QE
predicted by the PSU simulator for the bottom 128 rows of the S3 device
(those least affected by parallel CTI).  The top figure panel shows the
full range; the middle panel focuses on the areas of high QE; the bottom
panel compares the low-energy QE.}

\normalsize
\label{fig:qe7}
\end{figure}

The BI QE curves, calculated for the same energies and rows as the FI
QE curves shown above, are shown in Figure~\ref{fig:qe7} along with
that reported for the spatially-averaged (full-CCD) pre-launch ACIS S3
device (with no CTI correction) by MIT/CSR \cite{calreport}.  Again the
pointwise deviations in the PSU estimates reflect counting statistics;
$\sim 1.2 \times 10^5$ events per energy were simulated to calculate the QE.
The MIT release notes state that there are spatial variations in the BI
QE of up to 20\% \cite{calreport}; this must be due at least in part to
CTI effects.

The PSU curve for the bottom of the device (the area least affected by
parallel CTI) matches the MIT curve above 3.5~keV fairly well, where surface
effects are minimized.  Serial CTI effects dominate in this spatial
region but don't cause enough grade morphing that events are lost to
on-board grade filtering.  The top of the device is the area most
affected by the combined effects of parallel and serial CTI.  The QE
curve here clearly deviates from the pre-launch predictions, even at
high energies and after CTI correction.  This is again because CTI
caused events to morph into grades that are not telemetered to
the ground, thus they are not available for CTI correction.

Our model shows significant deviation from MIT's in the BI QE below the
oxygen absorption edge (543~eV), for both the top and bottom of the
device (note that we only calculate QE down to 0.2~keV but the MIT
curve continues to lower energies).  CTI effects are not large at low
energies on BI devices ({\em e.g.} the two PSU QE curves converge), so
grade morphing can't explain these QE deviations.  Perhaps differences in
the way the two groups model the dead and damage layers first
encountered by photons impinging on the BI device are causing these
variations in the model QE.   

\clearpage

\section{Applications}

\subsection{Response Matrices}

The CCD spectral redistribution function is encapsulated for the
purposes of spectral fitting in a ``response matrix'' \cite{calreport},
where each row corresponds to the spectral response of the device (in
DN) at a given input monochromatic energy (in eV).  We have used our
Monte Carlo CCD simulator to generate response matrices appropriate for
CTI-corrected on-orbit data.  The simulator is used to generate event
lists at a very large number of monochromatic energies spanning the
range of interest for {\em Chandra} observations (0.2 -- 10 keV).
These event lists are filtered to remove events that would not have
been telemetered by the real ACIS instrument and the remaining events
are CTI-corrected using the PSU corrector.  The resulting spectral
redistribution functions form the rows of the response matrix.

The response matrix generator for FI devices accepts a range of CCD row
numbers as input, so the user may create a matrix applicable to
extended structures as well as point sources located anywhere on the
device.  This is necessary because the PSU CTI corrector cannot
compensate for the CTI-induced energy resolution degradation on FI
devices \cite{townsley01b}, so position-dependent response matrices are
still needed to analyze some ACIS celestial spectra.  Since the BI CTI
is less severe, this position-dependent spectral resolution is not seen
in BI data and a single response matrix is adequate for fitting
pointlike or extended structures anywhere on a BI CCD.

\begin{figure}[htb]
\centerline{\epsfig{file=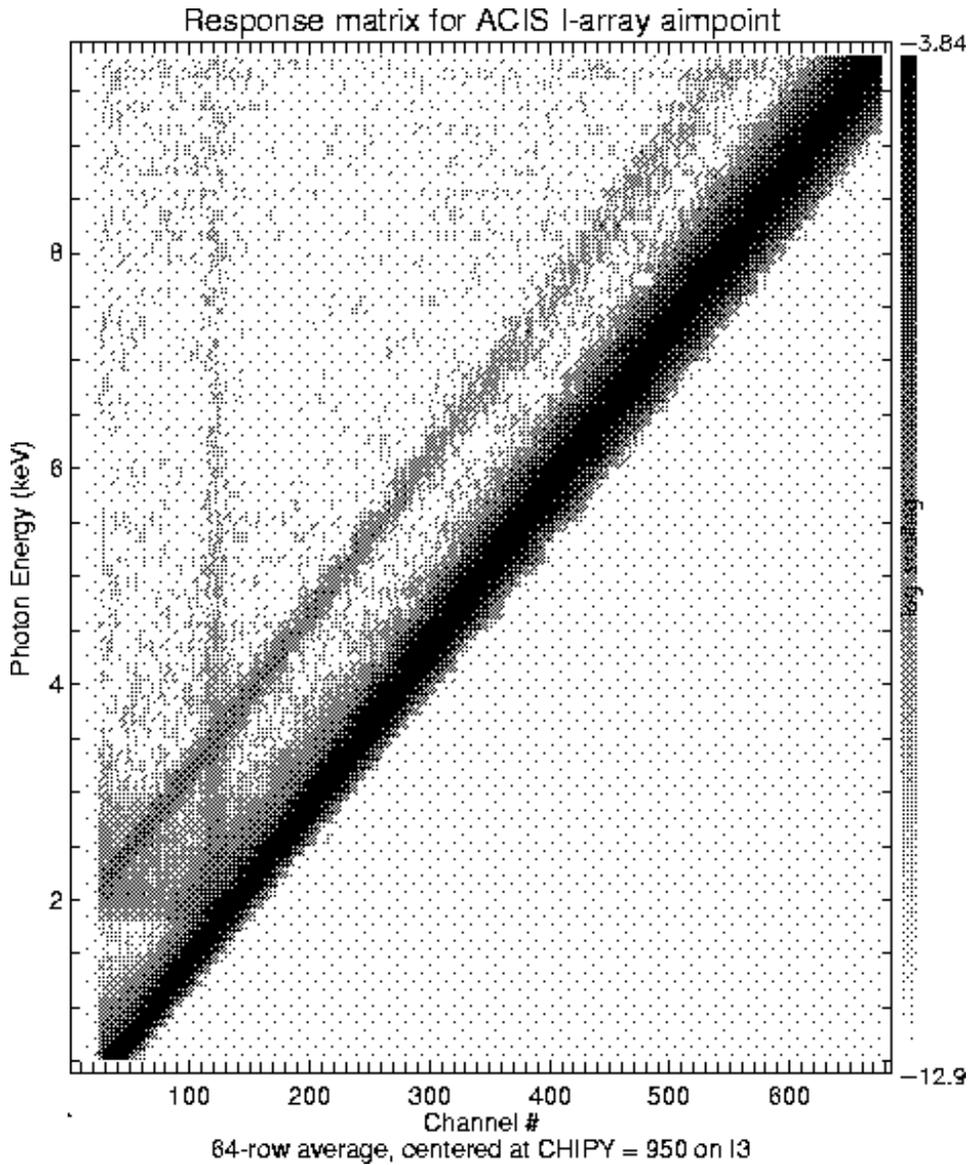, width=5.0in}}
\caption{\protect \small A sample FI response matrix for the I3
aimpoint at $-$110C, normalized to unity.  The dark diagonal locus is the
main energy peak; the fainter, shifted locus is the escape peak; the
faint vertical locus is the silicon fluorescence peak.}

\normalsize
\label{fig:respmat}
\end{figure}

An example FI matrix, generated for the aimpoint of the ACIS imaging
array at a focal plane temperature of $-$110C (where the spectral
broadening is at its worst), is shown in Figure~\ref{fig:respmat}.  The
main peak is represented by the dark diagonal locus.  The grey smear to
the left of the main peak locus is the soft shoulder, produced by
photons interacting in the channel stops.  At low energies photons
interact near the surface of the device and populate the low-energy
tail.  The escape peak tracks the main peak and doesn't appear until
the input photon energy is above the Si absorption edge of 1839~eV.
This edge represents a sharp reduction in the transmissivity of $Si$ so
more events interact near the surface of the device, repopulating the
soft shoulder and low-energy tail; thus the grey smear below the main
peak reappears at 1839~eV and slowly dies off at higher energies, as
photons interact deeper in the device.  The Si fluorescence peak
becomes populated above the Si absorption edge and appears as a
vertical locus of points in the matrix.  Above $\sim$4~keV, the
spectral redistribution features due to the surface layers of the
device have died away and the main and escape peaks are noticeably
broadened by the intrinsic spectral resolution of the device.


This method for generating response matrices involves creating a large
database of simulated events, then accessing a subset of those events
relevant to the spectral analysis task at hand.  A natural result of this
design is that the event database can be filtered for other properties
in addition to location on the CCD, such as event grade, so the user can
tailor a matrix to the specific science goals of the observation.  For
example, higher spectral resolution and background rejection may be obtained
by keeping just ASCA Grade 0 or Grade 0, 2, 3, and 4 events, rather than
the standard ``g02346'' grade set.  Brandt {\em et al.} \cite{brandt01} used
such a restricted grade set to improve the signal-to-noise of faint
source detections in the {\em Chandra} Deep Field North.  See also
Chartas {\em et al.} \cite{chartas98} for a more detailed explanation
of special filtering for optimizing aspects of ACIS data.

Another distinction of this method from typical techniques is that the
spectral redistribution function (the output of the simulator) is
recorded directly as a row in the matrix.  Since the simulator does not
produce an infinite number of events, the resulting matrix is not
perfectly smooth; this is apparent in Figure~\ref{fig:respmat}.  Other
techniques involve parameterizing the simulated spectra by fitting a set
of Gaussians or other functional forms then storing the fit
parameters.  This produces a very compact format for storing the
response matrix and removes the stochastic nature inherent in our
technique, but its fidelity is limited by the quality of the fits to
the simulated data, in addition to any errors introduced by the
simulation itself.

\subsection{Fitting the External Calibration Source Spectrum}

\begin{figure}[htb]
\centerline{\epsfig{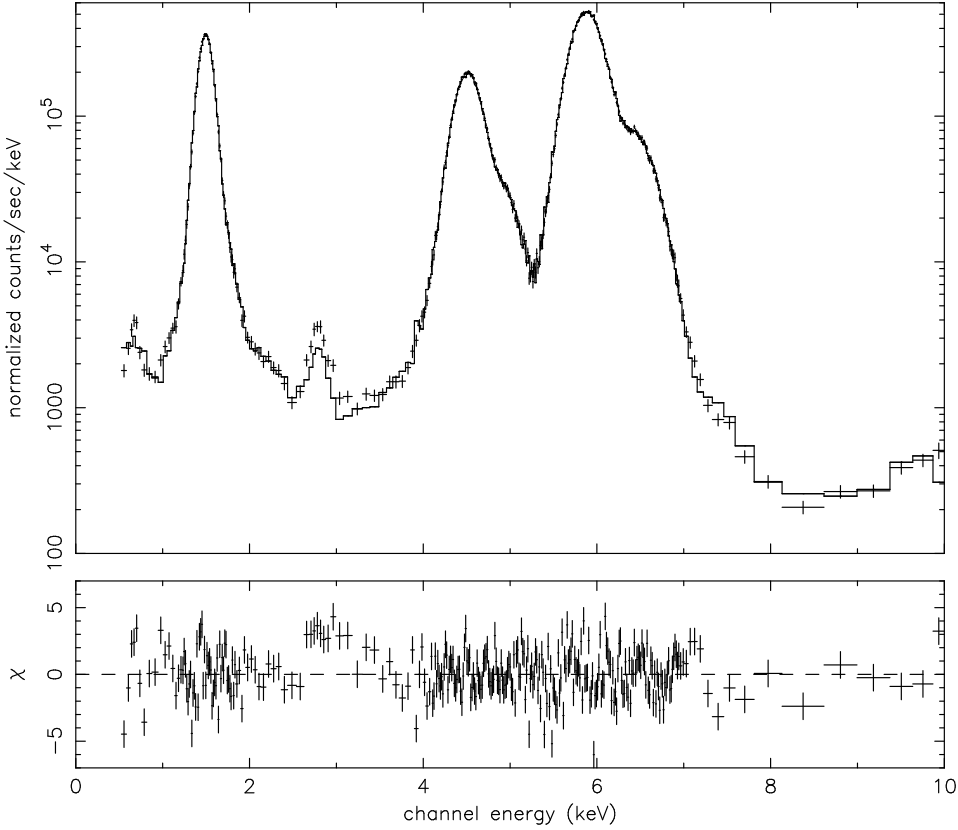}} \caption{\protect \small A spectral fit using the above
response matrix to CTI-corrected ACIS ECS data at $-$110C, isolating
the 100-row region around the Imaging Array aimpoint on the I3 chip.
Approximately $7 \times 10^5$ events were used to create the spectrum.}
\normalsize
\label{fig:calspec}
\end{figure}

We used the above response matrix to fit a spectrum from the External
Calibration Source (ECS), including only events from the 100 rows
surrounding the ACIS Imaging Array aimpoint on the I3 chip.  There are
$\sim 7 \times 10^5$ CTI-corrected events in this spectrum.  Because of
CTI, this region has the worst spectral resolution on the device, but
it is important to do spectral modeling here because it has the best
spatial resolution, so many targets are imaged at this location.  The
main spectral features included in the fit are listed in
Table~\ref{table:calsource} along with the line energies recovered by
the XSPEC \cite{arnaud96} fit.  The actual fit is shown in
Figure~\ref{fig:calspec}.

\begin{table}[htb] \centering

\begin{tabular}{||c|r|c|c|r||} \hline

Spectral line & True energy (eV) & Fit energy (eV) & $\sigma$ & Normalization \\    \hline \hline
 
Mn/Fe L complex &$\sim$680 & fixed & 0.05 (fixed)&   212 $\pm$ 46  \\ 
Al \Ka    &  1486    & 1495 $\pm$ 1 & 0.00        & 75334 $\pm$376  \\ 
Si \Ka    &  1740    & fixed       & 0.06        &  2271 $\pm$106  \\ 
Au M complex    &$\sim$2112& fixed & 0.19        &   869 $\pm$ 71  \\ 
Ti \Ka escape& 2771  & N/A         & N/A         &   N/A 	   \\ 
uncertain & uncertain& 3655 $\pm$27& 0.14	 &   364 $\pm$ 49  \\
Mn \Ka escape& 4155  & N/A         & N/A         &   N/A 	   \\ 
Ti \Ka    &  4511    & 4501 $\pm$ 1 & 0.05        & 73142 $\pm$677   \\ 
Ti \Kb    &  4932    & 4899 $\pm$ 7 & 0.10        & 13415 $\pm$519  \\ 
Mn \Ka    &  5895    & 5874 $\pm$ 1 & 0.05        &233450 $\pm$528   \\ 
Mn \Kb     &  6490   & 6432 $\pm$ 4 & 0.14        & 41599 $\pm$505  \\ 
Mn \Ka + Al \Ka pileup& 7381 & 7330 $\pm$38& 0.17 &  529 $\pm$ 72  \\ 
Au \La    &  9711    & fixed       & 0.11        &   177 $\pm$ 37 \\ \hline
 
\end{tabular}
\caption{\protect \small Gaussian fits to spectral features in the ACIS
ECS at $-$110C, considering only the 100 rows around the I3 aimpoint
and using the response matrix shown in Figure~\ref{fig:respmat} and
CTI-corrected data.  The escape peaks were not independently modeled as
Gaussians.  Energies (and in one case the linewidth) of weak lines were
fixed because they were not well-constrained in the fit.  The errors
represent 90\% confidence intervals.}
\normalsize
\label{table:calsource}
\end{table}

The lines marked as ``uncertain'' may be Ca~\Ka (3.69~keV) or the
escape peak of Cr~\Ka (3.67~keV) (Allyn Tennant, private communication).
The spectrum contains some background as well as the emission lines.
This background has a complicated spectral shape but is modeled here as
a simple power law of photon index 0.72$\pm$0.04 (normalization
1206$\pm$66).  

The fit had a $\chi^{2}=963$ for 281 degrees of freedom; this would be
unacceptably large for a fit to a typical astrophysical spectrum but is
not unreasonable for this calibration spectrum made from such a large
number of events.  We suspect that more sophisticated modeling of the
background would improve $\chi^{2}$; it may also be large because the
spectral lines were modeled as Gaussians when in fact they have
noticeably non-Gaussian wings.   These wings are due to residual CTI effects;
gain variations that were not removed by the CTI corrector are by definition
not incorporated into the CTI model that was used to generate the response
matrix, so to account for this completely the spectral line model would
require broader wings.  

This non-Gaussian line shape may also explain why the fit energies of
the Ti and Mn \Kb lines are low.  The \Kb lines are blended with the
non-Gaussian wings of their \Ka counterparts, which may skew the \Kb
Gaussian fit toward lower energies.  By choosing to fit data taken at
-110C (where CTI is most pronounced) and concentrating on the section of
the device farthest from the readout nodes (where the stochastic nature
of CTI results in the largest gain variations), we have maximized
this effect.

The bright lines contain so many events that the errors on the fit
energies are extremely small.  Conversely, the limited counts in the
weak lines caused those lines to be poorly constrained in the fit.  To
account for this, sometimes their line energies and widths were fixed
before the fit was performed.  The table notes which lines were
constrained.  Escape peaks from Ti~\Ka and Mn~\Ka are prominent in the
spectrum.  They were not modeled as independent Gaussians; rather their
amplitudes, widths, and line centers follow from the spectral
redistribution function instantiated in the response matrix.


\subsection{The LYNX Pile-up Simulator}

Chartas {\em et al.} have developed the spectral fitting tool LYNX
\cite{chartas00} that employs a forward fitting approach to infer
incident astrophysical flux and spectra by interfacing the standard X-ray
spectral fitting package XSPEC \cite{arnaud96} to the raytrace code
MARX \cite{wise97} and this CCD simulator.  Astrophysical spectra
obtained with ACIS are initially fit with XSPEC to provide starting
estimates of the spectral parameters for LYNX. The modeled incident
spectrum is propagated through the {\em Chandra} mirrors with MARX,
through the ACIS Optical Blocking Filters via their X-ray transmission
function \cite{chartas96}, then to the CCD simulator to become
instantiated into events.  A merit function that incorporates the
differences between the observed and simulated spectra is minimized
using a downhill simplex method to yield the best-fit model parameters.

The CCD simulator allows LYNX to account for the possible overlap of
the resulting charge clouds within each exposure, thus the spectra
produced through LYNX will simulate pile-up. The present version of
LYNX also allows spectral fitting using any grade selection, corrects
for telescope vignetting, and accounts for the dither motion of the
source across the CCD.  A simple example is given below, to illustrate
the power of this forward modeling scheme and to show some of the
effects of pile-up on ACIS results.

A set of simulated observations was produced for a point source
placed at the aimpoint of the S3 chip.  The incident spectrum was an
absorbed power law with spectral slope $\Gamma$ = 1.8, modified by
Galactic absorption from neutral hydrogen with column density N$_{H}$ =
$3 \times 10^{20}$ cm$^{-2}$.  The incident count rates (all grades
included) ranged between 0.05 counts per frame and 2 counts per frame.
The standard full-CCD frametime of 3.24 seconds was used.

\begin{figure}[htb]
\centerline{\epsfig{file=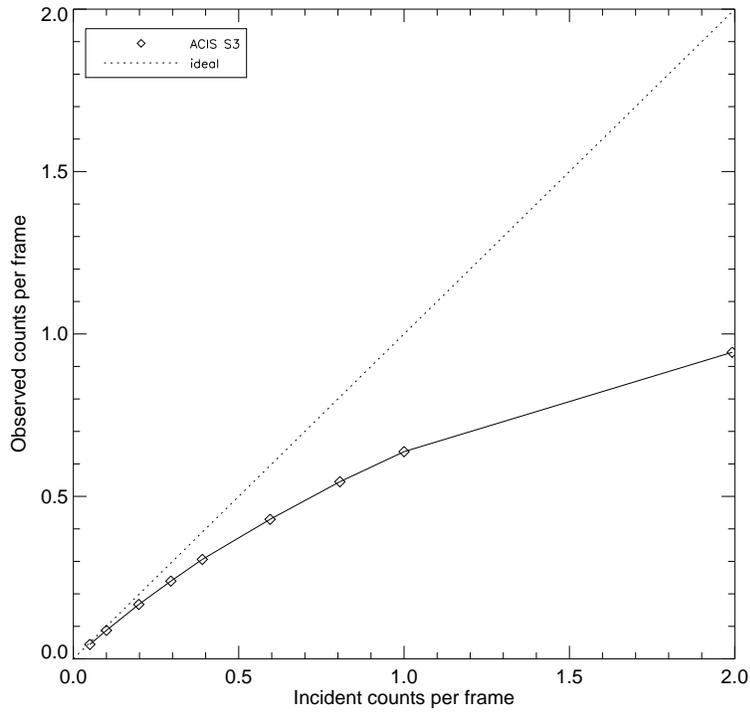,width=4.0in}}
\caption{\protect \small Modeling photon pile-up with LYNX.  Both incident
and observed count rates include all event grades.  An ideal detector
would produce the unit slope shown in the figure.}
\normalsize


\label{fig:lynxrates}
\end{figure}

Figure~\ref{fig:lynxrates} shows the effect of photon pile-up on the
measured event rate.  Even sources of moderate brightness
($\sim$0.2--0.4 counts per frame) will have $>$10\% errors in their
flux estimates if pile-up is not considered.  Naive flux estimates for bright
sources are substantially in error.  Note that this plot assumes
that all events are included; grade filtering makes these flux errors
even worse, as pile-up causes events to migrate into non-standard
grades.

\begin{figure}[htb]
\centerline{\epsfig{file=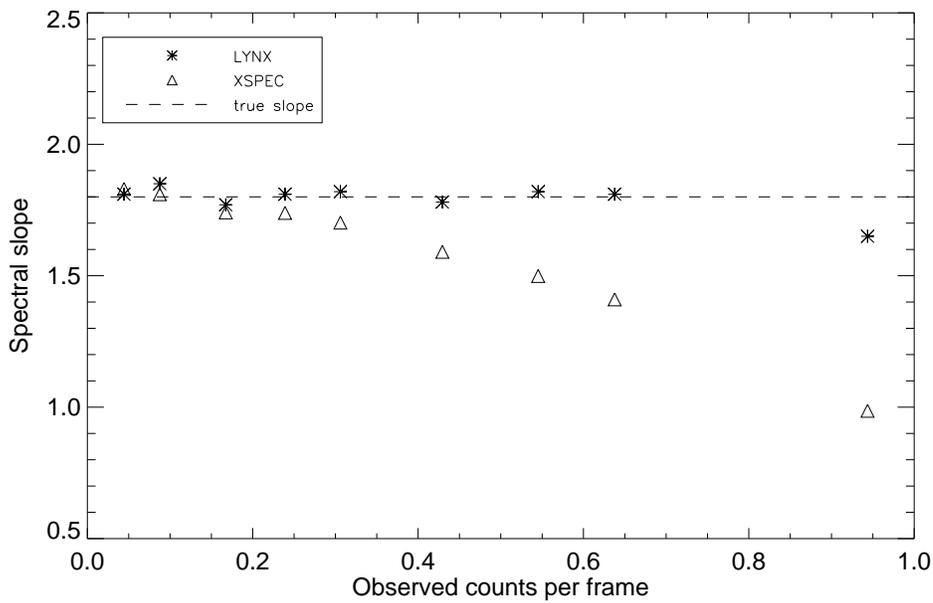,width=5.0in}}
\caption{\protect \small Recovering the incident spectral slope of a
power law model for varying degrees of photon pile-up using LYNX.  The observed
count rate includes events of all grades.  The ``XSPEC'' points show the
inferred slope using standard spectral fitting tools and the standard
response matrix.}


\normalsize
\label{fig:lynxspec}
\end{figure}

Pile-up also compromises spectral fitting, as shown in
Figure~\ref{fig:lynxspec}.  This gives the best fit values of $\Gamma$
obtained with LYNX and XSPEC as a function of the observed counts per
frame.  Here the observed counts per frame include only events with
energies lying between 0.2 and 10~keV and all grades.  For fits
performed with XSPEC we used standard grade selection and the standard
response matrices and ancillary files for the BI device S3 provided by
the CXC.

Above $\sim$ 0.2 counts per frame the pile-up effect produces
noticeably flatter spectra.  LYNX is able to recover the incident
spectral slope for observed rates of up to $\sim$0.7 counts per frame.
This observed rate corresponds to an incident rate of 1 count per
frame.  We are experimenting with improving the performance of LYNX at
even higher count rates by removing the grade filter that was applied
in the spectral fitting (Chartas {\em et al.}, in preparation).

\clearpage
\section{Summary and Further Work}


\subsection{Simulator Status}

The current model reproduces all the main features in ACIS calibration
spectra for both FI and BI devices.  The grade distributions are
reproduced to within a few percent, after the simulator is tuned to
yield the correct percentage of Grade 0 events.  We compare the output
of our simulator to that of the MIT/CSR CCD simulator via the quantum
efficiencies of both FI and BI devices.  They are consistent to
within the errors at most energies, although the errors quoted are
large. 

We must employ tuning parameters to account for physics that we have
not explicitly incorporated in the theory.  The model predicts the
device response at new energies by linearly interpolating the tuning
parameters determined from calibration data.  At some energies, the
quality of the tuning parameters is limited by the quantity and quality
of data available.

The PSU simulator borrows many basic features and physical quantities from
the MIT/CSR simulator, so the QE consistency is not surprising.  The
MIT/CSR simulator employs a much more detailed model of the gate
structure and channel stops, but our simpler slab model seems adequate
to reproduce the spectral features seen in XRCF data.  These two ACIS
simulators share no code and were tuned on independent datasets
(sub-assembly vs.\ XRCF), so they provide useful checks for each other
-- the similarity of their output gives us confidence in their fidelity
and in the stability of the ACIS CCDs.

Our simulator is capable of modeling epitaxial FI devices and some
limited testing has been performed for such devices, but the code
has not been tuned to reproduce an actual laboratory device.
Similarly, the code exists to model minimally-ionizing particle
interactions, but no detailed comparison to actual particle events
has been undertaken.

To facilitate independent assessment of our modeling techniques, the
simulator source code and a Users' Manual are available for use by the
community\footnote{http://www.astro.psu.edu/users/townsley/simulator/}.
Further implementation details and assumptions made are described
there.  We welcome input from other CCD modelers and astronomers
interested in ACIS device calibration and simulation.


\subsection{Future Work}

One clear extension of this work is to replace the point-wise
interpolation of tuning parameters with models.  This should be
straightforward at energies far from the important absorption edges of
O, N, and Si, but care must be taken to avoid oversimplifying the
tuning parameter behavior in these regions.  Our values for the tuning
parameters are affected by the XRCF and flight data used to tune them,
so modeling them might remove some of the simulator's dependence on the
quality of these data.  It is likely that such modeling will yield more
insight into the device physics that is manifested as spectral
redistribution features.

We have generated response matrices for ACIS FI devices I0--I3, S2, and
S4 using the simulator tuning parameters derived for I3 (except for the
CTI-induced, row-dependent spectral resolution, which was tuned for
each device).  This was mandated by lack of time and resources and
justified by MIT's subassembly calibration, which showed that all FI
devices were quite similar prior to launch.  If these matrices are
shown to be inadequate for spectral fitting of celestial sources due to
inaccurate modeling of the spectral redistribution function (rather
than, say, CTI effects), one course of action would be to return to the
XRCF data and tune the simulator for each FI device.  This would be a
labor-intensive undertaking. 

Further refinements to the simulator might be necessary to achieve
adequate modeling of other devices for other missions.  We usually
include physical models in response to the demands of the data rather
than by presupposing the relevance of a physical process, since there
are so many such processes at work in these detectors.  Given that,
CCDs of substantially different design (such as very thin BI devices or
FI devices with radically different gate structures) may require the
invocation of new physics in a future simulator.

Despite its limitations, the current incarnation of the full-frame
simulator is useful for addressing timely issues in CCD
characterization.  The code has been used to develop an algorithm to
determine sub-pixel positions of photons \cite{calreport} and to assess
the problems associated with photon pile-up from bright point sources
\cite{broos98}.  It has been used in conjunction with SAOSAC and MARX
to model the {\em Chandra} point spread function at XRCF
\cite{calreport}.  Event lists generated by this simulator are being
used to generate ACIS response matrices for X-ray spectral analysis of
celestial sources.  

A product of our work on the CTI corrector is the ability to include
some of the effects of CTI in the CCD simulator, namely grade migration
and event energy degradation as a function of position on the device.
We have additionally included another noise term in the simulator to
account for the spectral line broadening caused by charge traps (see
the accompanying paper \cite{townsley01b} for details).  These features
of the model allow us to reproduce on-orbit ACIS calibration data for
both FI and BI CCDs.  Second-generation tools such as LYNX are
combining the power of XSPEC, MARX, and the CCD simulator to model
astrophysical spectra affected by pile-up.  Armed with these simulation
tools, we are prepared to face the challenge of modeling {\em
Chandra}/ACIS celestial data.


\ack
Financial support for this effort was provided by Professor Gordon
Garmire, Principal Investigator of the ACIS project, via NASA contract
NAS8-38252.  This work made use of the NASA Astrophysics Data System.

We very much appreciate the hard work of Konstantin~Getman,
Lisa~Rabban, and Anna~Nousek for their long hours spent tuning and
running the simulator.  We thank David~Lumb for the original simulator
code and for his continuing advice on CCD matters, real and virtual.
We appreciate the loan of CXC CPU cycles during our efforts to generate
the 5.4 billion simulated events necessary to make the response matrices.
We thank the anonymous referee for excellent comments and suggestions.

Special gratitude is owed to our MIT/CSR ACIS colleagues, especially
Gregory~Prigozhin and Mark~Bautz, for their continuing collaborative
spirit and good humor in providing calibration data and ideas in
advance of publication.  Many of our simulation successes are simply
reflections of their own.


\bibliography{xraysim}
\bibliographystyle{elsart-num}

\end{document}